%&plain
%

%
\input harvmac.tex
\input amssym.def
\def\CA{{\cal A}}

\def\CD{{\cal D}}
\def\CE{{\cal E}}
\def\CF{{\cal F}}

\def\CL{{\cal L}}
\def\CM{{\cal M}}
\def\CN{{\cal N}}
\def\CO{{\cal O}}
\def\CP{{\cal P}}

\def\CS{{\cal S}}
\def\CV{{\cal V}}

\def\CJ{{\cal J}}

\def\fG{{\frak G}}

\def\fU{{\frak U}}
\def\fH{{\frak H}}

\def\tA{{\widetilde A}}
\def\tD{{\widetilde D}}
\def\tJ{{\widetilde J}}
\def\tK{{\widetilde K}}

\def\tN{{\widetilde N}}
\def\tP{{\widetilde P}}
\def\tQ{{\widetilde Q}}
\def\tT{{\widetilde T}}

\def\hl{{\hat\lambda}}

\def\Ad{{\dot A}}
\def\Bd{{\dot B}}
\def\Cd{{\dot C}}
\def\a{{\alpha}}
\def\ad{{\dot{\alpha}}}
\def\b{{\beta}}
\def\bd{{\dot{\beta}}}
\def\g{{\gamma}}
\def\gd{{\dot{\gamma}}}
\def\d{{\delta}}
\def\dd{{{\dot\delta}}}
\def\td{{{\tilde\delta}}}
\def\ed{{{\dot\epsilon}}}
\def\l{{\lambda}}
\def\o{{\omega}}

\def\ve{\varepsilon}
\def\i{{\rm i}}

\font\cmss=cmss10 \font\cmsss=cmss10 at 7pt

\def\IR{\relax{\rm I\kern-.18em R}}
\def\inbar{\vrule height1.5ex width.4pt depth0pt}
\def\IC{\relax\,\hbox{$\inbar\kern-.3em{\rm C}$}}
\def\IN{\relax{\rm I\kern-.18em N}}
\def\IF{\relax{\rm I\kern-.18em F}}
\def\IP{\relax{\rm I\kern-.18em P}}
\def\IZ{\relax\ifmmode\mathchoice
{\hbox{\cmss Z\kern-.4em Z}}{\hbox{\cmss Z\kern-.4em Z}}
{\lower.9pt\hbox{\cmsss Z\kern-.4em Z}} {\lower1.2pt\hbox{\cmsss
Z\kern-.4em Z}}\else{\cmss Z\kern-.4em Z}\fi}
\def\sm#1#2{\kern.1em\lower.4ex\hbox{${\scriptstyle #2}$\kern-1em\raise1.6ex
             \hbox{${\scriptstyle #1}$}}}
\def\sfrac#1#2{{\textstyle{#1\over #2}}}
\def\underrightarrow#1{\vbox{\ialign{##\crcr$\hfil\displaystyle
{#1}\hfil$\crcr\noalign{\kern1pt
\nointerlineskip}$\longrightarrow$\crcr}}}
\def\underleftarrow#1{\vbox{\ialign{##\crcr$\hfil\displaystyle
{#1}\hfil$\crcr\noalign{\kern1pt
\nointerlineskip}$\longleftarrow$\crcr}}}
% use of underarrow
%A~~~\underarrow{a}~~~B

\def\fc{\hbox{$f$\kern-0.4em\raise2.25ex\hbox{$_\circ$}}}
\def\Fc{\hbox{$\CF$\kern-0.4em\raise2.35ex\hbox{$_\circ$}}}
\def\Ac{\hbox{$\CA$\kern-0.4em\raise2.35ex\hbox{$_\circ$}}}
\def\nc{\hbox{$\nabla$\kern-.57em\raise2.4ex\hbox{$_\circ$}}}
\def\Wc{\hbox{$W$\kern-.65em\raise2.4ex\hbox{$_\circ$}}}
\def\Gc{\hbox{$G$\kern-.55em\raise2.4ex\hbox{$_\circ$}}}
\def\cc{\hbox{$\chi$\kern-.5em\raise1.8ex\hbox{$_\circ$}}}
\def\pc{\hbox{$\phi$\kern-.4em\raise2.1ex\hbox{$_\circ$}}}
\def\cnab{\hbox{$\nabla$\kern-.6em\raise2.3ex\hbox{$_\circ$}}}
\def\contra#1#2{\,{\buildrel \,\hbox{$
    \vrule height 3pt width 1pt depth 0pt
    \vrule height 3pt width #1pt depth -2pt
    \vrule height 3pt width 1pt depth 0pt $}
    \over {#2} }\,}

%%%%%%%%%%%%%%%%%%%%%%%%%%%%%%%%%%%%%%%%%%%%%%%%%%%%%%%%%%%%%%%%%%%%%
%%%%%%%%%%%%%%%%%%%%%%%%%%%%%%%%%%%%%%%%%%%%%%%%%%%%%%%%%%%%%%%%%%%%%

%\MaldacenaRE
\nref\MaldacenaRE{
      J.~M.~Maldacena,
      {\it The large N limit of superconformal field theories and 
      supergravity},
      Adv.\ Theor.\ Math.\ Phys.\  {\bf 2}, 231 (1998)
      [Int.\ J.\ Theor.\ Phys.\  {\bf 38}, 1113 (1999)]
      [hep-th/9711200];
%%CITATION = HEP-TH 9711200;%%
      S.~S.~Gubser, I.~R.~Klebanov and A.~M.~Polyakov,
      {\it Gauge theory correlators from non-critical string theory},
      Phys.\ Lett.\ B {\bf 428}, 105 (1998)
      [hep-th/9802109];
%%CITATION = HEP-TH 9802109;%%
      E.~Witten,
      {\it Anti-de Sitter space and holography},
      Adv.\ Theor.\ Math.\ Phys.\  {\bf 2}, 253 (1998)
      [hep-th/9802150].
%%CITATION = HEP-TH 9802150;%%
}

%\MinahanVE
\nref\MinahanVE{
      J.~A.~Minahan and K.~Zarembo,
      {\it The Bethe ansatz for $\CN=4$ super Yang-Mills},
      JHEP {\bf 0303}, 013 (2003)
      [hep-th/0212208].
%%CITATION = HEP-TH 0212208;%%
}

%\BerensteinJQ
\nref\BerensteinJQ{
      D.~Berenstein, J.~M.~Maldacena and H.~Nastase,
      {\it Strings in flat space and pp-waves from $\CN=4$ super Yang Mills},
      JHEP {\bf 0204}, 013 (2002)
      [hep-th/0202021].
%%CITATION = HEP-TH 0202021;%%
}

%\LipatovYB
\nref\LipatovYB{
      L.~N.~Lipatov, 
      {\it High-energy asymptotics of multicolor QCD and exactly solvable 
      lattice models},
      JETP Lett.\  {\bf 59}, 596 (1994)
      [hep-th/9311037];
%%CITATION = HEP-TH 9311037;%%
%\FaddeevZG
      L.~D.~Faddeev and G.~P.~Korchemsky,
      {\it High-energy QCD as a completely integrable model},
      Phys.\ Lett.\ B {\bf 342}, 311 (1995)
      [hep-th/9404173];
%%CITATION = HEP-TH 9404173;%%
%\BraunTE
      V.~M.~Braun, S.~E.~Derkachov, G.~P.~Korchemsky and A.~N.~Manashov,
      {\it Baryon distribution amplitudes in QCD},
      Nucl.\ Phys.\ B {\bf 553}, 355 (1999)
      [hep-ph/9902375].
%%CITATION = HEP-PH 9902375;%%
}

%\BeisertRY
\nref\BeisertRY{
      N.~Beisert,
      {\it The dilatation operator of $\CN=4$ super Yang-Mills theory and
      integrability},
      Phys.\ Rept.\  {\bf 405}, 1 (2005)
      [hep-th/0407277];
%%CITATION = HEP-TH 0407277;%%
%\BelitskyCZ
      A.~V.~Belitsky, V.~M.~Braun, A.~S.~Gorsky and G.~P.~Korchemsky,
      {\it Integrability in QCD and beyond},
      Int.\ J.\ Mod.\ Phys.\ A {\bf 19}, 4715 (2004)
      [hep-th/0407232].
%%CITATION = HEP-TH 0407232;%%
}

%\BeisertTQ
\nref\BeisertTQ{
      N.~Beisert, C.~Kristjansen and M.~Staudacher,
      {\it The dilatation operator of $\CN=4$ super Yang-Mills theory},
      Nucl.\ Phys.\ B {\bf 664}, 131 (2003)
      [hep-th/0303060];
%%CITATION = HEP-TH 0303060;%%
%\BelitskySF
      A.~V.~Belitsky, G.~P.~Korchemsky and D.~M\"uller,
      {\it Integrability in Yang-Mills theory on the light cone beyond leading 
      order},
      hep-th/0412054.
%%CITATION = HEP-TH 0412054;%%
}

%\BenaWD
\nref\BenaWD{
      I.~Bena, J.~Polchinski and R.~Roiban,
      {\it Hidden symmetries of the ${\rm AdS}_5\times S^5$ superstring},
      Phys.\ Rev.\ D {\bf 69}, 046002 (2004)
      [hep-th/0305116].
%%CITATION = HEP-TH 0305116;%%
}

%\MetsaevIT
\nref\MetsaevIT{
      R.~R.~Metsaev and A.~A.~Tseytlin,
      {\it Type IIB superstring action in ${\rm AdS}_5\times S^5$ background},
      Nucl.\ Phys.\ B {\bf 533}, 109 (1998)
      [hep-th/9805028];
%%CITATION = HEP-TH 9805028;%%
      R.~Kallosh, J.~Rahmfeld and A.~Rajaraman,
      {\it Near horizon superspace},
      JHEP {\bf 9809}, 002 (1998)
      [hep-th/9805217];
%%CITATION = HEP-TH 9805217;%%
      R.~Roiban and W.~Siegel,
      {\it Superstrings on ${\rm AdS}_5\times S^5$ supertwistor space},
      JHEP {\bf 0011}, 024 (2000)
      [hep-th/0010104].
%%CITATION = HEP-TH 0010104;%%
}

%\MandalFS
\nref\MandalFS{
      G.~Mandal, N.~V.~Suryanarayana and S.~R.~Wadia,
      {\it Aspects of semiclassical strings in ${\rm AdS}_5$},
      Phys.\ Lett.\ B {\bf 543}, 81 (2002)
      [hep-th/0206103].
%%CITATION = HEP-TH 0206103;%%
}

%\PolyakovBR
\nref\PolyakovBR{
      A.~M.~Polyakov,
      {\it Conformal fixed points of unidentified gauge theories},
      Mod.\ Phys.\ Lett.\ A {\bf 19}, 1649 (2004)
      [hep-th/0405106].
%%CITATION = HEP-TH 0405106;%%
}

%\LuscherRQ
\nref\LuscherRQ{
      M.~L\"uscher and K.~Pohlmeyer,
      {\it Scattering of massless lumps and nonlocal charges in the 
      two-dimensional classical nonlinear sigma model},
      Nucl.\ Phys.\ B {\bf 137}, 46 (1978);
%%CITATION = NUPHA,B137,46;%%
      M.~L\"uscher,
      {\it Quantum nonlocal charges and absence of particle production in the
      two-dimensional nonlinear sigma model},
      Nucl.\ Phys.\ B {\bf 135}, 1 (1978).
%%CITATION = NUPHA,B135,1;%%
}

%\SchwarzTD
\nref\SchwarzTD{
      J.~H.~Schwarz,
      {\it Classical symmetries of some two-dimensional models},
      Nucl.\ Phys.\ B {\bf 447}, 137 (1995)
      [hep-th/9503078].
%%CITATION = HEP-TH 9503078;%%
}

%\DolanFQ
\nref\DolanFQ{
      L.~Dolan,
      {\it Kac-Moody algebra is hidden symmetry of chiral models},
      Phys.\ Rev.\ Lett.\  {\bf 47}, 1371 (1981);
%%CITATION = PRLTA,47,1371;%%
      {\it Kac-Moody algebras and exact solvability in hadronic physics},
      Phys.\ Rept.\  {\bf 109}, 1 (1984).
%%CITATION = PRPLC,109,1;%%
}

%\DrinfeldRX
\nref\DrinfeldRX{
      V.~G.~Drin'feld,
      {\it Hopf algebras and the quantum Yang-Baxter equation},
      Sov.\ Math.\ Dokl.\  {\bf 32}, 254 (1985);
%%CITATION = SVMDA,32,254;%%
      {\it A new realization of Yangians and quantized affine algebras},
      Sov.\ Math.\ Dokl.\  {\bf 36}, 212 (1988).
%%CITATION = SVMDA,36,212;%%
}

%\BernardJW
\nref\BernardJW{
      D.~Bernard,
      {\it Hidden Yangians in $2D$ massive current algebras},
      Commun.\ Math.\ Phys.\  {\bf 137}, 191 (1991).
%%CITATION = CMPHA,137,191;%%
}

%\BernardYA
\nref\BernardYA{ 
      D.~Bernard,
      {\it An introduction to Yangian symmetries},
      Int.\ J.\ Mod.\ Phys.\ B {\bf 7}, 3517 (1993)
      [hep-th/9211133];
%%CITATION = HEP-TH 9211133;%%
      N.~J.~MacKay,
      {\it Introduction to Yangian symmetry in integrable field theory},
      hep-th/0409183.
%%CITATION = HEP-TH 0409183;%%
}

%\BerkovitsFE
\nref\BerkovitsFE{
      N.~Berkovits,
      {\it Super-Poincare covariant quantization of the superstring},
      JHEP {\bf 0004}, 018 (2000)
      [hep-th/0001035];
%%CITATION = HEP-TH 0001035;%%
      N.~Berkovits and O.~Chandia,
      {\it Superstring vertex operators in an ${\rm AdS}_5\times S^5$
      background},
      Nucl.\ Phys.\ B {\bf 596}, 185 (2001)
      [hep-th/0009168];
%%CITATION = HEP-TH 0009168;%%
      B.~C.~Vallilo,
      {\it One-loop conformal invariance of the superstring in an 
      ${\rm AdS}_5\times S^5$ background},
      JHEP {\bf 0212}, 042 (2002)
      [hep-th/0210064].
%%CITATION = HEP-TH 0210064;%%
}

%\ValliloNX
\nref\ValliloNX{
      B.~C.~Vallilo,
      {\it Flat currents in the classical ${\rm AdS}_5\times S^5$ pure spinor 
      superstring},
      JHEP {\bf 0403}, 037 (2004)
      [hep-th/0307018].
%%CITATION = HEP-TH 0307018;%%
}

%\AldayZB
\nref\AldayZB{
      L.~F.~Alday,
      {\it Nonlocal charges on ${\rm AdS}_5\times S^5$ and pp-waves},
      JHEP {\bf 0312}, 033 (2003)
      [hep-th/0310146];
%%CITATION = HEP-TH 0310146;%%
      G.~Arutyunov and M.~Staudacher,
      {\it Two-loop commuting charges and the string/gauge duality},
      hep-th/0403077;
%%CITATION = HEP-TH 0403077;%%
      B.~Y.~Hou, D.~T.~Peng, C.~H.~Xiong and R.~H.~Yue,
      {\it The affine hidden symmetry and integrability of type IIB 
      superstring in ${\rm AdS}_5\times S^5$},
      hep-th/0406239;
%%CITATION = HEP-TH 0406239;%%
      G.~Arutyunov, S.~Frolov and M.~Staudacher,
      {\it Bethe ansatz for quantum strings},
      JHEP {\bf 0410}, 016 (2004)
      [hep-th/0406256];
%%CITATION = HEP-TH 0406256;%%
      M.~Hatsuda and K.~Yoshida,
      {\it Classical integrability and super Yangian of superstring on 
      ${\rm AdS}_5\times S^5$},
      hep-th/0407044;
%%CITATION = HEP-TH 0407044;%%
      I.~Swanson,
      {\it Quantum string integrability and AdS/CFT},
      Nucl.\ Phys.\ B {\bf 709}, 443 (2005)
      [hep-th/0410282].
%%CITATION = HEP-TH 0410282;%%
}

%\BerkovitsJW
\nref\BerkovitsJW{
      N.~Berkovits,
      {\it BRST cohomology and nonlocal conserved charges},
      hep-th/0409159.
%%CITATION = HEP-TH 0409159;%%
}

%\DolanUH
\nref\DolanUH{
      L.~Dolan, C.~R.~Nappi and E.~Witten,
      {\it A relation between approaches to integrability in superconformal 
      Yang-Mills theory},
      JHEP {\bf 0310}, 017 (2003)
      [hep-th/0308089];
%%CITATION = HEP-TH 0308089;%%
      {\it Yangian symmetry in $D=4$ superconformal Yang-Mills theory},
      hep-th/0401243;
%%CITATION = HEP-TH 0401243;%%
      L.~Dolan and C.~R.~Nappi,
      {\it Spin models and superconformal Yang-Mills theory},
      hep-th/0411020.
%%CITATION = HEP-TH 0411020;%%
}

%\AgarwalSZ
\nref\AgarwalSZ{
      A.~Agarwal and S.~G.~Rajeev,
      {\it Yangian symmetries of matrix models and spin chains: The dilatation 
      operator of $\CN=4$ SYM},
      hep-th/0409180;
%%CITATION = HEP-TH 0409180;%%
%\ArutyunovRG
      G.~Arutyunov and M.~Staudacher,
      {\it Matching higher conserved charges for strings and spins},
      JHEP {\bf 0403}, 004 (2004)
      [hep-th/0310182];
%%CITATION = HEP-TH 0310182;%%
      A.~Mikhailov,
      {\it Anomalous dimension and local charges},
      hep-th/0411178.
%%CITATION = HEP-TH 0411178;%%
}

%\WittenNN
\nref\WittenNN{
      E.~Witten,
      {\it Perturbative gauge theory as a string theory in twistor space},
      Commun.\ Math.\ Phys.\  {\bf 252}, 189 (2004)
      [hep-th/0312171].
%%CITATION = HEP-TH 0312171;%%
}

%\FerberQX
\nref\FerberQX{
      A.~Ferber,
      {\it Supertwistors and conformal supersymmetry},
      Nucl.\ Phys.\ B {\bf 132}, 55 (1978);
%%CITATION = NUPHA,B132,55;%%
}

%\WittenXX
\nref\WittenXX{
      E.~Witten,
      {\it An interpretation of classical Yang-Mills theory},
      Phys.\ Lett.\ B {\bf 77}, 394 (1978).
%%CITATION = PHLTA,B77,394;%%
}

%\RoibanVT
\nref\RoibanVT{
      R.~Roiban, M.~Spradlin and A.~Volovich,
      {\it A googly amplitude from the $B$-model in twistor space},
      JHEP {\bf 0404}, 012 (2004)
      [hep-th/0402016];
%%CITATION = HEP-TH 0402016;%%
      F.~Cachazo, P.~Svr\v cek and E.~Witten,
      {\it MHV vertices and tree amplitudes in gauge theory},
      JHEP {\bf 0409}, 006 (2004)
      [hep-th/0403047];
%%CITATION = HEP-TH 0403047;%%
      C.~J.~Zhu,
      {\it The googly amplitudes in gauge theory},
      JHEP {\bf 0404}, 032 (2004)
      [hep-th/0403115];
%%CITATION = HEP-TH 0403115;%%
      R.~Roiban, M.~Spradlin and A.~Volovich,
      {\it On the tree-level $S$-matrix of Yang-Mills theory},
      Phys.\ Rev.\ D {\bf 70}, 026009 (2004)
      [hep-th/0403190];
%%CITATION = HEP-TH 0403190;%%
      G.~Georgiou and V.~V.~Khoze,
      {\it Tree amplitudes in gauge theory as scalar MHV diagrams},
      JHEP {\bf 0405}, 070 (2004)
      [hep-th/0404072];
%%CITATION = HEP-TH 0404072;%%
      S.~Gukov, L.~Motl and A.~Neitzke,
      {\it Equivalence of twistor prescriptions for super Yang-Mills},
      hep-th/0404085;
%%CITATION = HEP-TH 0404085;%%
      J.~B.~Wu and C.~J.~Zhu,
      {\it MHV vertices and scattering amplitudes in gauge theory},
      JHEP {\bf 0407}, 032 (2004)
      [hep-th/0406085];
%%CITATION = HEP-TH 0406085;%%
      I.~Bena, Z.~Bern and D.~A.~Kosower,
      {\it Twistor-space recursive formulation of gauge theory amplitudes},
      hep-th/0406133;
%%CITATION = HEP-TH 0406133;%%
      J.~B.~Wu and C.~J.~Zhu,
      {\it MHV vertices and fermionic scattering amplitudes in gauge theory 
      with quarks and gluinos},
      JHEP {\bf 0409}, 063 (2004)
      [hep-th/0406146];
%%CITATION = HEP-TH 0406146;%%
      D.~A.~Kosower,
      {\it Next-to-maximal helicity violating amplitudes in gauge theory},
      Phys.\ Rev.\ D {\bf 71}, 045007 (2005)
      [hep-th/0406175];
%%CITATION = HEP-TH 0406175;%%
      F.~Cachazo, P.~Svr\v cek and E.~Witten,
      {\it Twistor space structure of one-loop amplitudes in gauge theory},
      JHEP {\bf 0410}, 074 (2004)
      [hep-th/0406177];
%%CITATION = HEP-TH 0406177;%%
      G.~Georgiou, E.~W.~N.~Glover and V.~V.~Khoze,
      {\it Non-MHV tree amplitudes in gauge theory},
      JHEP {\bf 0407}, 048 (2004)
      [hep-th/0407027];
%%CITATION = HEP-TH 0407027;%%
      A.~Brandhuber, B.~Spence and G.~Travaglini,
      {\it One-loop gauge theory amplitudes in $\CN=4$ super Yang-Mills from 
      MHV vertices},
      Nucl.\ Phys.\ B {\bf 706}, 150 (2005)
      [hep-th/0407214];
%%CITATION = HEP-TH 0407214;%%
      Y.~Abe, V.~P.~Nair and M.~I.~Park,
      {\it Multigluon amplitudes, $\CN=4$ constraints and the WZW model},
      Phys.\ Rev.\ D {\bf 71}, 025002 (2005)
      [hep-th/0408191];
%%CITATION = HEP-TH 0408191;%%
      V.~V.~Khoze,
      {\it Gauge theory amplitudes, scalar graphs and twistor space},
      hep-th/0408233;
%%CITATION = HEP-TH 0408233;%%
      X.~Su and J.~B.~Wu,
      {\it Six-quark amplitudes from fermionic MHV vertices},
      hep-th/0409228;
%%CITATION = HEP-TH 0409228;%%
      F.~Cachazo, P.~Svr\v cek and E.~Witten,
      {\it Gauge theory amplitudes in twistor space and holomorphic anomaly},
      JHEP {\bf 0410}, 077 (2004)
      [hep-th/0409245].
%%CITATION = HEP-TH 0409245;%%
}

%\NairBQ
\nref\NairBQ{
      V.~P.~Nair,
      {\it A current algebra for some gauge theory amplitudes},
      Phys.\ Lett.\ B {\bf 214}, 215 (1988).
%%CITATION = PHLTA,B214,215;%%
}

%\GiombiIX
\nref\GiombiIX{
      S.~Giombi, R.~Ricci, D.~Robles-Llana and D.~Trancanelli,
      {\it A note on twistor gravity amplitudes},
      JHEP {\bf 0407}, 059 (2004)
      [hep-th/0405086].
%%CITATION = HEP-TH 0405086;%%
}

%\BerkovitsJJ
\nref\BerkovitsJJ{
      N.~Berkovits and E.~Witten,
      {\it Conformal supergravity in twistor-string theory},
      JHEP {\bf 0408}, 009 (2004)
      [hep-th/0406051].
%%CITATION = HEP-TH 0406051;%%
}

%\BerkovitsHG
\nref\BerkovitsHG{
      N.~Berkovits,
      {\it An alternative string theory in twistor space for $\CN=4$ 
      super-Yang-Mills},
      Phys.\ Rev.\ Lett.\  {\bf 93}, 011601 (2004)
      [hep-th/0402045];
%%CITATION = HEP-TH 0402045;%%
      N.~Berkovits and L.~Motl,
      {\it Cubic twistorial string field theory},
      JHEP {\bf 0404}, 056 (2004)
      [hep-th/0403187];
%%CITATION = HEP-TH 0403187;%%
%\LechtenfeldCC
      O.~Lechtenfeld and A.~D.~Popov,
      {\it Supertwistors and cubic string field theory for open $\CN=2$ 
      strings},
      Phys.\ Lett.\ B {\bf 598}, 113 (2004)
      [hep-th/0406179].
%%CITATION = HEP-TH 0406179;%%
}

%\NeitzkePF
\nref\NeitzkePF{
      A.~Neitzke and C.~Vafa,
      {\it $\CN=2$ strings and the twistorial Calabi-Yau},
      hep-th/0402128;
%%CITATION = HEP-TH 0402128;%%
      N.~Nekrasov, H.~Ooguri and C.~Vafa,
      {\it $S$-duality and topological strings},
      JHEP {\bf 0410}, 009 (2004)
      [hep-th/0403167].
%%CITATION = HEP-TH 0403167;%%
}

%\AganagicYH
\nref\AganagicYH{
      M.~Aganagic and C.~Vafa,
      {\it Mirror symmetry and supermanifolds},
      hep-th/0403192;
%%CITATION = HEP-TH 0403192;%%
      A.~Belhaj, L.~B.~Drissi, J.~Rasmussen, E.~H.~Saidi and A.~Sebbar,
      {\it Toric Calabi-Yau supermanifolds and mirror symmetry},
      hep-th/0410291.
%%CITATION = HEP-TH 0410291;%%
}

%\SiegelDJ
\nref\SiegelDJ{
      W.~Siegel,
      {\it Untwisting the twistor superstring},
      hep-th/0404255.
%%CITATION = HEP-TH 0404255;%%
}

%\PopovRB
\nref\PopovRB{
      A.~D.~Popov and C.~S\"amann,
      {\it On supertwistors, the Penrose-Ward transform and $\CN=4$ super 
      Yang-Mills theory},
      hep-th/0405123.
%%CITATION = HEP-TH 0405123;%%
}

%\PopovNK
\nref\PopovNK{
      A.~D.~Popov and M.~Wolf,
      {\it Topological $B$-model on weighted projective spaces and self-dual 
      models in four dimensions}, JHEP {\bf 0409}, 007 (2004)
      [hep-th/0406224].
%%CITATION = HEP-TH 0406224;%%
}

%\AhnXS
\nref\AhnXS{
      C.~H.~Ahn,
      {\it Mirror symmetry of Calabi-Yau supermanifolds},
      hep-th/0407009;
%%CITATION = HEP-TH 0407009;%%
}

%\AhnYU
\nref\AhnYU{
      C.~H.~Ahn,
      {\it $N=1$ conformal supergravity and twistor-string theory},
      JHEP {\bf 0410}, 064 (2004)
      [hep-th/0409195].
%%CITATION = HEP-TH 0409195;%%
}

%\RocekBI
\nref\RocekBI{
      M.~Ro\v cek and N.~Wadhwa,
      {\it On Calabi-Yau supermanifolds},
      hep-th/0408188;
%%CITATION = HEP-TH 0408188;%%
      {\it On Calabi-Yau supermanifolds II},
      hep-th/0410081;
%%CITATION = HEP-TH 0410081;%%
      C.~G.~Zhou,
      {\it On Ricci flat supermanifolds},
      hep-th/0410047.
%%CITATION = HEP-TH 0410047;%%
}

%\SinkovicsFM
\nref\SinkovicsFM{
      A.~Sinkovics and E.~Verlinde,
      {\it A six-dimensional view on twistors},
      hep-th/0410014.
%%CITATION = HEP-TH 0410014;%%
}

\nref\KULAXIZI{
      M.~Kulaxizi and K.~Zoubos,
      {\it Marginal deformations of $\CN=4$ SYM from open/closed twistor 
      strings},
      hep-th/0410122;
%%CITATION = HEP-TH 0410122;%%
      J.~Park and S.~J.~Rey,
      {\it Supertwistor orbifolds: Gauge theory amplitudes and topological 
      strings},
      JHEP {\bf 0412}, 017 (2004)
      [hep-th/0411123];
%%CITATION = HEP-TH 0411123;%%
      S.~Giombi, M.~Kulaxizi, R.~Ricci, D.~Robles-Llana, D.~Trancanelli and 
      K.~Zoubos,
      {\it Orbifolding the twistor string},
      hep-th/0411171.
%%CITATION = HEP-TH 0411171;%%
}

%\SaemannTT
\nref\SaemannTT{
      C.~S\"amann,
      {\it The topological B-model on fattened complex manifolds and 
      subsectors of $\CN=4$ self-dual Yang-Mills theory},
      JHEP {\bf 0501}, 042 (2005)
      [hep-th/0410292].
%%CITATION = HEP-TH 0410292;%%
}

%\WittenFB
\nref\WittenFB{
      E.~Witten,
      {\it Chern-Simons gauge theory as a string theory},
      Prog.\ Math.\  {\bf 133}, 637 (1995)
      [hep-th/9207094].
%%CITATION = HEP-TH 9207094;%%
}

%\PohlmeyerYA
\nref\PohlmeyerYA{
      K.~Pohlmeyer,
      {\it On the Lagrangian theory of anti-(self)-dual fields in 
      four-dimensional Euclidean space},
      Commun.\ Math.\ Phys.\  {\bf 72}, 37 (1980).
%%CITATION = CMPHA,72,37;%%
}

%\ChauGI
\nref\ChauGI{
      L.~L.~Chau, M.~L.~Ge and Y.~S.~Wu,
      {\it The Kac-Moody algebra in the self-dual Yang-Mills equation},
      Phys.\ Rev.\ D {\bf 25}, 1086 (1982);
%%CITATION = PHRVA,D25,1086;%%
      L.~L.~Chau and Wu Yong-Shi,
      {\it More about hidden symmetry algebra for the self-dual Yang-Mills 
      system},
      Phys.\ Rev.\ D {\bf 26}, 3581 (1982);
%%CITATION = PHRVA,D26,3581;%%
      L.~L.~Chau, M.~L.~Ge, A.~Sinha and Y.~S.~Wu,
      {\it Hidden symmetry algebra for the self-dual Yang-Mills equation},
      Phys.\ Lett.\ B {\bf 121}, 391 (1983).
%%CITATION = PHLTA,B121,391;%%
}

\nref\UENO{
      K.~Ueno and Y.~Nakamura,
      {\it Transformation theory for anti-(self)-dual equations and the 
      Riemann-Hilbert problem},
      Phys.\ Lett.\ B {\bf 109}, 273 (1982);
%%CITATION = PHLTA,B109,273;%%
      L.~Dolan,
      {\it A new symmetry group of real self-dual Yang-Mills},
      Phys.\ Lett.\ B {\bf 113}, 387 (1982);
%%CITATION = PHLTA,B113,387;%%
      L.~Crane,
      {\it Action of the loop group on the self-dual Yang-Mills equation},
      Commun.\ Math.\ Phys.\  {\bf 110}, 391 (1987).
%%CITATION = CMPHA,110,391;%%
}

%\ChauRP
\nref\ChauRP{
      L.~L.~Chau and I.~Yamanaka,
      {\it Quantization of the self-dual Yang-Mills system: Algebras and 
      hierarchy},
      Phys.\ Rev.\ Lett.\  {\bf 68}, 1807 (1992);
%%CITATION = PRLTA,68,1807;%%
      {\it Quantization of the self-dual Yang-Mills system: Exchange algebras 
      and local quantum group in four-dimensional quantum field theories},
      Phys.\ Rev.\ Lett.\  {\bf 70}, 1916 (1993).
%%CITATION = PRLTA,70,1916;%%
}

%\PopovQB
\nref\PopovQB{
      A.~D.~Popov and C.~R.~Preitschopf,
      {\it Extended conformal symmetries of the self-dual Yang-Mills 
      equations},
      Phys.\ Lett.\ B {\bf 374}, 71 (1996)
      [hep-th/9512130];
%%CITATION = HEP-TH 9512130;%%
      A.~D.~Popov,
      {\it Self-dual Yang-Mills: Symmetries and moduli space},
      Rev.\ Math.\ Phys.\  {\bf 11}, 1091 (1999)
      [hep-th/9803183];
%%CITATION = HEP-TH 9803183;%%
      {\it Holomorphic Chern-Simons-Witten theory: From $2D$ to $4D$ conformal 
      field theories},
      Nucl.\ Phys.\ B {\bf 550}, 585 (1999)
      [hep-th/9806239];
%%CITATION = HEP-TH 9806239;%%
      T.~A.~Ivanova,
      {\it On infinite-dimensional algebras of symmetries of the self-dual 
      Yang-Mills equations},
      J.\ Math.\ Phys.\  {\bf 39}, 79 (1998)
      [hep-th/9702144].
%%CITATION = HEP-TH 9702144;%%
}

%\MasonRF
\nref\MasonRF{
      L.~J.~Mason and N.~M.~J.~Woodhouse,
      {\it Integrability, self-duality, and twistor theory},
      Clarendon Press, Oxford, 1996.
%\href{http://www.slac.stanford.edu/spires/find/hep/www?irn=3540308}{SPIRES entry}
}

%\PenroseIN
\nref\PenroseIN{
      R.~Penrose,
      {\it The twistor program},
      Rept.\ Math.\ Phys.\  {\bf 12}, 65 (1977);
%%CITATION = RMHPB,12,65;%%
      R.~S.~Ward,
      {\it On self-dual gauge fields},
      Phys.\ Lett.\ A {\bf 61}, 81 (1977).
%%CITATION = PHLTA,A61,81;%%
}

%\PenroseJW
\nref\PenroseJW{
      R.~Penrose and W.~Rindler,
      {\it Spinors and space-time. Vol. 1: Two spinor calculus and relativistic
      fields},
      Cambridge University Press, Cambridge, 1984;
%\href{http://www.slac.stanford.edu/spires/find/hep/www?irn=1396293}{SPIRES entry}
      {\it Spinors and Space-Time. Vol. 2: Spinor and twistor methods in 
      space-time geometry},
      Cambridge University Press, Cambridge, 1985.
%\href{http://www.slac.stanford.edu/spires/find/hep/www?irn=1653610}{SPIRES entry}
}

\nref\Ward{      
      R.~S.~Ward and R.~O.~Wells,
      {\it Twistor geometry and field theory},
      Cambridge University Press, Cambridge, 1990.
%\href{http://www.slac.stanford.edu/spires/find/hep/www?irn=2256894}{SPIRES entry}
}

%\TakasakiGW
\nref\TakasakiGW{
      L.~J.~Mason and G.~A.~J.~Sparling,
      {\it Nonlinear Schr\"odinger and Korteweg-De Vries are reductions of 
      self-dual Yang-Mills},
      Phys.\ Lett.\ A {\bf 137}, 29 (1989);
%%CITATION = PHLTA,A137,29;%%
      K.~Takasaki,
      {\it Hierarchy structure in integrable systems of gauge fields and 
      underlying Lie algebras},
      Commun.\ Math.\ Phys.\  {\bf 127}, 225 (1990);
%%CITATION = CMPHA,127,225;%%
%\AblowitzEC
      M.~J.~Ablowitz, S.~Chakravarty and L.~A.~Takhtajan,
      {\it A self-dual Yang-Mills hierarchy and its reductions to integrable 
      systems in $1+1$ dimensions and $2+1$ dimensions},
      Commun.\ Math.\ Phys.\  {\bf 158}, 289 (1993).
%%CITATION = CMPHA,158,289;%%
%\IvanovaZT
      T.~A.~Ivanova and O.~Lechtenfeld,
      {\it Hidden symmetries of the open N = 2 string},
      Int.\ J.\ Mod.\ Phys.\ A {\bf 16}, 303 (2001)
      [hep-th/0007049].
%%CITATION = HEP-TH 0007049;%%
}

%\BardeenGK
\nref\BardeenGK{
      W.~A.~Bardeen,
      {\it Self-dual Yang-Mills theory, integrability and multiparton 
      amplitudes},
      Prog.\ Theor.\ Phys.\ Suppl.\  {\bf 123}, 1 (1996);
%%CITATION = PTPSA,123,1;%%
      D.~Cangemi,
      {\it Self-dual Yang-Mills theory and one-loop maximally helicity 
      violating multi-gluon amplitudes},
      Nucl.\ Phys.\ B {\bf 484}, 521 (1997)
      [hep-th/9605208];
%%CITATION = HEP-TH 9605208;%%
      {\it Self-duality and maximally helicity violating QCD amplitudes},
      Int.\ J.\ Mod.\ Phys.\ A {\bf 12}, 1215 (1997)
      [hep-th/9610021].
%%CITATION = HEP-TH 9610021;%%
}

%\LeznovUP
\lref\LeznovUP{
      E.~T.~Newman,
      {\it Source-free Yang-Mills theories},
      Phys.\ Rev.\ D {\bf 18}, 2901 (1978);
%%CITATION = PHRVA,D18,2901;%%
      A.~N.~Leznov,
      {\it On equivalence of four-dimensional self-duality equations to 
      continual analog of the main chiral field problem},
      Theor.\ Math.\ Phys.\  {\bf 73}, 1233 (1988).
%%CITATION = TMPHA,73,1233;%%
}

%\DeWittCY
\lref\DeWittCY{
      B.~DeWitt,
      {\it Supermanifolds},
      Cambridge University Press, Cambridge, 1992.
%\href{http://www.slac.stanford.edu/spires/find/hep/www?irn=2650312}{SPIRES entry}
}

%\HarnadVK
\lref\DevchandGV{
      J.~P.~Harnad, J.~Hurtubise, M.~Legare and S.~Shnider,
      {\it Constraint equations and field equations in supersymmetric $\CN=3$ 
      Yang-Mills theory},
      Nucl.\ Phys.\ B {\bf 256}, 609 (1985);
%%CITATION = NUPHA,B256,609;%%
%\HarnadBC
      J.~P.~Harnad and S.~Shnider,
      {\it Constraints and field equations for ten-dimensional super Yang-Mills
      theory},
      Commun.\ Math.\ Phys.\  {\bf 106}, 183 (1986);
%%CITATION = CMPHA,106,183;%%
%\DevchandGV
      C.~Devchand and V.~Ogievetsky,
      {\it Interacting fields of arbitrary spin and $\CN>4$ supersymmetric  
      self-dual Yang-Mills equations},
      Nucl.\ Phys.\ B {\bf 481}, 188 (1996)
      [hep-th/9606027].
%%CITATION = HEP-TH 9606027;%%
}

%\KotrlaKY
\lref\KotrlaKY{
      M.~Kotrla and J.~Niederle,
      {\it Supertwistors and superspace},
      Czech.\ J.\ Phys.\ B {\bf 35}, 602 (1985);
%%CITATION = CZYPA,B35,602;%%
      J.~Lukierski and W.~J.~Zakrzewski,
      {\it Euclidean supersymmetrization of instantons and self-dual 
      monopoles},
      Phys.\ Lett.\ B {\bf 189}, 99 (1987).
%%CITATION = PHLTA,B189,99;%%
}

%\SiegelZA
\lref\SiegelZA{
      W.~Siegel,
      {\it $N=2$ $(4)$ string theory is self-dual $\CN=4$ Yang-Mills theory},
      Phys.\ Rev.\ D {\bf 46}, R3235 (1992)
      [hep-th/9205075].
%%CITATION = HEP-TH 9205075;%%
}

%\MandelstamCB
\lref\MandelstamCB{
      S.~Mandelstam,
      {\it Light cone superspace and the ultraviolet finiteness of the $\CN=4$ 
      model},
      Nucl.\ Phys.\ B {\bf 213}, 149 (1983);
%%CITATION = NUPHA,B213,149;%%
      {\it Covariant superspace with unconstrained fields},
      Phys.\ Lett.\ B {\bf 121}, 30 (1983).
%%CITATION = PHLTA,B121,30;%%
}

%\WittenZZ
\lref\WittenZZ{
      E.~Witten,
      {\it Mirror manifolds and topological field theory},
      in S.-T.\ Yau, ed., {\it Mirror symmetry},
      [hep-th/9112056].
%%CITATION = HEP-TH 9112056;%%
}

%\Nazarov
\lref\Nazarov{
      M.~L.~Nazarov,
      {\it Quantum Berezinian and the classical Capelli identity},
      Lett.\ Math.\ Phys.\  {\bf 21}, 123 (1991);
%%CITATION = LMPHD,21,123;%%
      R.~B.~Zhang,
      {\it The ${\frak g}{\frak l}(M|N)$ super Yangian and its finite 
      dimensional representations},
      Lett.\ Math.\ Phys.\  {\bf 37}, 419 (1996)
      [q-alg/9507029].
%%CITATION = Q-ALG 9507029;%%
}

%\SchoutensAU
\lref\SchoutensAU{
      K.~Schoutens,
      {\it Yangian symmetry in conformal field theory},
      Phys.\ Lett.\ B {\bf 331}, 335 (1994)
      [hep-th/9401154];
%%CITATION = HEP-TH 9401154;%%
      P.~Bouwknegt, A.~W.~W.~Ludwig and K.~Schoutens,
      {\it Spinon basis for higher level $SU(2)$ WZW models},
      Phys.\ Lett.\ B {\bf 359}, 304 (1995)
      [hep-th/9412108].
%%CITATION = HEP-TH 9412108;%%
}

%\WardGZ
\lref\WardGZ{
      R.~S.~Ward,
      {\it Integrable and solvable systems and relations among them},
      Phil.\ Trans.\ Roy.\ Soc.\ Lond.\ A {\bf 315}, 451 (1985);
%%CITATION = PTRSA,A315,451;%%
      {\it Multidimensional integrable systems},
      Lect.\ Notes\ Phys. {\bf 280}, 106 (1986);
      {\it Integrable systems in twistor theory},
      in: {\it Twistors in mathematics and physics}, p.246, 
      Cambridge University Press, Cambridge, 1990.
}

%\BoyerAJ
\lref\BoyerAJ{
      C.~P.~Boyer and J.~F.~Plebanski,
      {\it An infinite hierarchy of conservation laws and 
      nonlinear superposition principles for self-dual Einstein spaces},
J.\ Math.\ Phys.\  {\bf 26}, 229 (1985);
%%CITATION = JMAPA,26,229;%%
%\TakasakiCG
      K.~Takasaki,
      {\it Symmetries of hyper-K\"ahler (or Poisson gauge 
      field) hierarchy},
      J.\ Math.\ Phys.\  {\bf 31}, 1877 (1990);
%%CITATION = JMAPA,31,1877;%%
%\ParkVI
      Q.~H.~Park,
      {\it $2D$ sigma model approach to $4D$ instantons},
      Int.\ J.\ Mod.\ Phys.\ A {\bf 7}, 1415 (1992);
%%CITATION = IMPAE,A7,1415;%%
%\PopovUU
      A.~D.~Popov, M.~Bordemann and H.~R\"omer,
      {\it Symmetries, currents and conservation laws of 
      self-dual gravity},
      Phys.\ Lett.\ B {\bf 385}, 63 (1996)
      [hep-th/9606077];
%%CITATION = HEP-TH 9606077;%%
%\JunemannHI
      K.~J\"unemann, O.~Lechtenfeld and A.~D.~Popov,
      {\it Nonlocal symmetries of the closed $N=2$ string},
      Nucl.\ Phys.\ B {\bf 548}, 449 (1999)
      [hep-th/9901164];
%%CITATION = HEP-TH 9901164;%%
%\LechtenfeldIK
      O.~Lechtenfeld and A.~D.~Popov,
      {\it Closed $N=2$ strings: Picture-changing, hidden 
      symmetries and SDG hierarchy},
      Int.\ J.\ Mod.\ Phys.\ A {\bf 15}, 4191 (2000)
      [hep-th/9912154];
%%CITATION = HEP-TH 9912154;%%
%\DunajskiIQ
      M.~Dunajski and L.~J.~Mason,
      {\it Hyper-K\"ahler hierarchies and their twistor 
      theory},
      Commun.\ Math.\ Phys.\  {\bf 213}, 641 (2000)
      [math.dg/0001008].
%%CITATION = MATH-DG 0001008;%%
}

%\FradkinAM
\lref\FradkinAM{
      E.~S.~Fradkin and A.~A.~Tseytlin,
      {\it Conformal supergravity},
      Phys.\ Rept.\  {\bf 119}, 233 (1985).
%%CITATION = PRPLC,119,233;%%
}

%\ChauMP
\lref\ChauMP{
      L.~L.~Chau,
      {\it Yang-Mills, Yang-Baxter and Yangian algebra -- Quantum group 
      symmetry, quantum bimodule field, and quanta bimodulons},
      Chin.\ J.\ Phys.\  {\bf 32}, 535 (1994);
%%CITATION = CJOPA,32,535;%%
%\ChauIY
      L.~L.~Chau and I.~Yamanaka,
      {\it Conserved monodromy $R^TTT=TTR^T$ algebra in 
      the quantum self-dual Yang-Mills system},
      hep-th/9504106.
%%CITATION = HEP-TH 9504106;%%
}

%%%%%%%%%%%%%%%%%%%%%%%%%%%%%%%%%%%%%%%%%%%%%%%%%%%%%%%%%%%%%%%%%%%%%
%%%%%%%%%%%%%%%%%%%%%%%%%%%%%%%%%%%%%%%%%%%%%%%%%%%%%%%%%%%%%%%%%%%%%

%
\newbox\tmpbox\setbox\tmpbox\hbox{\abstractfont}
\Title{\vbox{
\rightline{\hbox{hep-th/0412163}}
\rightline{\hbox{ITP--UH--28/04}}}}
{\vbox{\centerline{On Hidden Symmetries of a Super Gauge Theory}
\vskip .2cm
\centerline{and Twistor String Theory}}}
\centerline{Martin Wolf \foot{E-mail: \tt wolf@itp.uni-hannover.de}}
\smallskip\smallskip
\centerline{\it Institut f\"ur Theoretische Physik}
\centerline{\it Universit\"at Hannover}
\centerline{\it Appelstra{\ss}e 2, 30167 Hannover, Germany}
\vskip .8cm
\centerline{{\it To the memory of Gerhard Soff}}
\vskip .8cm
\centerline{\bf Abstract}
\smallskip

We discuss infinite-dimensional hidden symmetry algebras (and hence an 
infinite number of conserved nonlocal charges) of the $\CN$-extented self-dual
super Yang-Mills equations  for general $\CN\leq4$ by using the supertwistor 
correspondence. Furthermore, by enhancing the supertwistor space, we construct 
the $\CN$-extended self-dual super Yang-Mills hierarchies, which describe
infinite sets of graded Abelian symmetries.  We also show that the open 
topological $B$-model with the enhanced supertwistor space as target manifold
will describe the hierarchies. Furthermore, these hierarchies will in turn -- 
by a supersymmetric extension of Ward's conjecture -- reduce to the super 
hierarchies of integrable models in $D<4$ dimensions.

\Date{December, 2004}
\listtoc \writetoc
\vskip-.8cm\centerline{{\vrule height 0.5pt width 13.7cm depth 0pt}}

%--------------------------------------------------------------------------
%--------------------------------------------------------------------------
\newsec{Introduction}

Within the last decades, the investigation of four-dimensional $\CN=4$ 
super Yang-Mills (SYM) theory became a quite important area. In particular, the
interest in this theory was again stimulated by the discovery of the AdS/CFT 
correspondence \MaldacenaRE. This conjecture states the equivalence of type IIB
superstring theory on an ${\rm AdS}_5\times S^5$ background with $\CN=4$ SYM 
theory on $\IR^4$.

One important tool for testing the AdS/CFT conjecture, which has been emerged 
lately, is integrability that appears on both sides of the correspondence.
(Quantum) integrable structures in $SU(N)$ $\CN=4$ SYM theory have first been 
discovered by Minahan and Zarembo \MinahanVE, inspired by the work of 
Berenstein, Maldacena and Nastase \BerensteinJQ, in the large $N$ or planar 
limit of the gauge theory.\foot{See references \LipatovYB\ 
for an ealier account of integrable structures in QCD.} Furthermore, 
it has been shown that it is possible to interpret the dilatation operator, 
which measures the scaling dimension of local operators, at one-loop level as 
Hamiltonian of an integrable quantum spin chain (see, e.g., \BeisertRY\ and 
references therein). For discussions beyond leading order see
also \BeisertTQ. Another development which has
pointed towards integrable structures was triggered by Bena, Polchinski and 
Roiban \BenaWD. Their investigation is based on the observation that the 
Green-Schwarz formulation of the superstring on ${\rm AdS}_5\times S^5$
can be interpreted as a coset theory, where the fields take values in the
supercoset space $PSU(2,2|4)/(SO(4,1)\times SO(5))$ \MetsaevIT.
Although this is not a symmetric space
\MandalFS, they found that the classical Green-Schwarz superstring on 
${\rm AdS}_5\times S^5$ possesses an infinite set of conserved
nonlocal charges, quite similar to those that exist in two-dimensional
field theories.\foot{These charges were also independently found by Polyakov
\PolyakovBR. For the construction of nonlocal conserved charges in 
two-dimensional sigma models see, e.g., \refs{\LuscherRQ,\SchwarzTD}.} 
Such charges are in turn related to Kac-Moody algebras 
\refs{\DolanFQ,\SchwarzTD} and generate Yangian
algebras \DrinfeldRX\ as has been discussed, e.g., in \BernardJW. For a review 
of Yangian algebras see, e.g., \BernardYA. Some time later,  
the construction of an analogous set of nonlocal conserved charges 
using the pure spinor formulation of the superstring \BerkovitsFE\ on 
${\rm AdS}_5\times S^5$ was given in \ValliloNX\ (for further developments 
see \AldayZB). Recently \BerkovitsJW, it has been verified that these 
charges are $\kappa$-symmetric in the Green-Schwarz as well 
as BRST invariant in the pure spinor formulation of the superstring on 
${\rm AdS}_5\times S^5$. Dolan, Nappi and Witten related these nonlocal charges
for the superstring to a corresponding set of nonlocal charges in the gauge 
theory \DolanUH\ (see also \AgarwalSZ).

Besides the whole AdS/CFT business, the studies of $\CN=4$ SYM theory have 
recently received important input from quite another point of view. In
\WittenNN, Witten has shown that perturbative $\CN=4$ SYM theory can be
described by the $D$-instanton expansion of a topological string theory which 
turned out to be the open topological $B$-model whose target space is the 
Calabi-Yau supermanifold $\IC P^{3|4}$. The latter space is the supersymmetric 
version of the twistor space \refs{\FerberQX,\WittenXX}. Witten demonstrated 
that one can actually reproduce (initially at tree level) scattering amplitudes
of the gauge theory in a simplified manner by performing string theory 
calculations. Since then, quite some progress in computing and understanding of
such amplitudes has been made (see, e.g., \RoibanVT\foot{Cf. \NairBQ\ for an 
earlier discussion. For properties of gravity amplitudes see, e.g., 
\refs{\GiombiIX,\BerkovitsJJ}.}). Besides the discussion of scattering 
amplitudes, a variety of other interesting aspects has been examined throughout
the literature \refs{\BerkovitsHG-\SaemannTT}.

The open topological $B$-model on the supertwistor space is equivalent to 
holomorphic Chern-Simons (hCS) theory  on the same space \WittenFB. Therefore, 
the moduli space of solutions to the equations of motion
of hCS theory on the supertwistor space can bijectively be mapped onto the 
moduli space of solutions to the equations of motion of self-dual $\CN=4$ SYM 
theory in four dimensions, as has been shown in \WittenNN\ by analyzing the 
sheaf cohomology interpretation of the linearized field equations on the 
supertwistor space. This correspondence -- by now known as 
the supertwistor correspondence -- has then been pushed further beyond the 
linearized level in \PopovRB. 

The purpose of this paper is to use this correspondence for the studies of the 
supertwistor construction of hidden symmetry algebras -- as those mentioned 
above -- of the $\CN$-extended self-dual SYM equations, 
generalizing the results known in the literature for the bosonic self-dual 
Yang-Mills (SDYM) equations \refs{\PohlmeyerYA-\MasonRF}. In 
particular, we will consider infinitesimal perturbations of the transition 
functions of holomorphic vector bundles over the supertwistor space. Using the 
Penrose-Ward transform \PenroseIN\foot{For reviews of twistor theory, 
we refer to \refs{\MasonRF,\PenroseJW,\Ward}.}, we relate these infinitesimal 
perturbations to infinitesimal symmetries of the $\CN$-extended self-dual SYM 
theory, thereby obtaining infinite sets of conserved nonlocal charges. 
After some general words on hidden symmetry algebras, we exemplify
our discussion by constructing super Kac-Moody symmetries which come from 
affine extensions of some Lie (super)algebra ${\frak g}$. Furthermore, we 
consider affine extensions of the superconformal algebra to obtain super
Kac-Moody-Virasoro type symmetries. Moreover, by considering a certain
Abelian subalgebra of the extended superconformal algebra, we introduce a 
supermanifold which we call the enhanced supertwistor space. Actually, we 
find a whole family of such spaces which is parametrized by a set of discrete
parameters. These enhanced spaces then allow us to introduce the 
$\CN$-extended self-dual SYM hierarchies which describe infinite sets of graded
Abelian symmetries. This generalizes the results known for the purely bosonic 
SDYM equations \TakasakiGW. We remark that such symmetries of the latter 
equations are intimately connected with one-loop maximally helicity violating 
(MHV) amplitudes \BardeenGK. As in certain situations the enhanced supertwistor
space turns out to be a Calabi-Yau space, the open topological $B$-model with 
this space as target manifold will describe certain corners of these 
hierarchies. Finally, we point out that since $\CN=3\ (4)$ SYM theory is 
related to a quadric, the superambitwistor space, in  
$\IC P^{3|3}\times\IC P^{3|3}$ \refs{\WittenXX,\WittenNN,\PopovRB}, 
our discussion presented below can directly be translated to the supertwistor 
construction of hidden symmetry algebras and thus of infinite sets of 
conserved nonlocal charges for the full $\CN=4$ SYM theory.

The paper is organized as follows. After a brief review of the supertwistor
correspondence in the context of the $\CN$-extended self-dual SYM equations in 
section 2, we explain in section 3 how one can construct (hidden) infinitesimal
symmetries of these equations in general. In the appendix A, these symmetries
are described in the context of a more mathematical language -- the sheaf 
cohomology theory. Also in section 3, we give the construction of the 
above-mentioned hidden symmetry algebras. After this, we focus in section 4
on the construction of the $\CN$-extended self-dual SYM hierarchy 
thereby discussing
the enhanced supertwistor space. Starting from certain constraint equations, 
we derive the equations of motion of the hierarchies and give the 
superfield expansions of the accompanied fields, first in a gauge covariant 
manner and second in Leznov gauge (also known as light cone gauge when the 
Minkowski signature is chosen). The latter gauge turns out to be
useful, for instance, for the geometrical interpretation of the hierarchies in
the context of certain dynamical systems. Furthermore, in section 5, we  
describe connections between the hierarchies and the open topological 
$B$-model. After this, we present our conclusions in section 6. In the 
appendix B, we discuss the moduli space of almost complex structures on the 
superspace $\IR^{2m|2n}$ thereby presenting yet another
extension of the ordinary twistor space, which might be of interest in further
investigations. Finally, the appendix C contains the superconformal algebra 
for general $\CN\leq4$.

%--------------------------------------------------------------------------
%--------------------------------------------------------------------------
\newsec{Holomorphy and self-dual super Yang-Mills equations}

In order to establish the formalism needed later on, we first summarize some 
aspects of the supertwistor correspondence between the $\CN$-extended self-dual
SYM equations on $\IR^4$ with Euclidean signature $({+}\,{+}\,{+}\,{+})$ and 
holomorphic vector bundles over the supertwistor space
\eqn\defoftwistor{\CP^{3|\CN}\ \equiv\
   \IC P^{3|\CN}\setminus\IC P^{1|\CN}\ \cong\ \CO(1)\otimes\IC^2
\oplus\Pi\CO(1)\otimes\IC^\CN\ \rightarrow\ \IC P^1,}
where $\Pi$ is the parity changing operator. We will, however, be rather brief 
on this and for details refer the reader to \PopovRB.

%--------------------------------------------------------------------------
\subsec{Holomorphic vector bundles and self-dual super Yang-Mills equations}

The supertwistor correspondence -- for the self-dual subsector\foot{Note that
strictly speaking for $\CN=4$ it is not a subsector.} of the
$\CN$-extended SYM theory -- is based on the double fibration
\eqn\doublefibrationc{\CP^{3|\CN} ~~~\underleftarrow{\pi_2}~~~\CF^{5|2\CN}
                               ~~~\underrightarrow{\pi_1}~~~\IC^{4|2\CN}_R,}
where  $\CP^{3|\CN}$ denotes the supertwistor space as given in \defoftwistor,
$\CF^{5|2\CN}=\IC^{4|2\CN}_R\times\IC P^1$ is the correspondence space and 
$\IC^{4|2\CN}_R$ is the anti-chiral subspace of the complexification of the 
(super)spacetime $\IR^{4|4\CN}$. The mappings $\pi_{1,2}$ are projections 
defined momentarily. The correspondence space $\CF^{5|2\CN}$ can be covered by 
two coordinate patches, say $\{{\tilde\CU}_+,{\tilde\CU}_-\}$, and hence 
we may introduce the (complex) coordinates 
$(x^{\a\ad}_R,\l^\pm_\ad,\eta^\ad_i)$, where $\a,\b,\ldots,\ad,\bd,\ldots=1,2$ 
and $i,j,\ldots=1,\ldots,\CN$. These coordinates are $\IZ_2$-graded with the 
$x_R^{\a\ad}$, $\l^\pm_\ad$ being Gra{\ss}mann even and the $\eta^\ad_i$ 
Gra{\ss}mann odd. Moreover, the $\l_\ad^\pm$ are given by
\eqn\defofl{(\l^+_\ad)\ \equiv \ \left(\matrix{1\cr \l_+}\right)
            \qquad{\rm and}\qquad 
            (\l^-_\ad)\  \equiv \ \left(\matrix{\l_-\cr 1}\right),}
with $\l_\pm\in\IC P^1$, such that $\l_+$ is defined on $U_+$ and $\l_-$ on 
$U_-$, respectively, where $\{U_+,U_-\}$ represents the canonical cover of 
$\IC P^1$. The projections $\pi_{1,2}$ are defined as
\eqn\defofpi{\eqalign{
             \pi_1\,:\,(x^{\a\ad}_R,\l^\pm_\ad,\eta^\ad_i)&\ \to\
             (x^{\a\ad}_R,\eta^\ad_i),\cr
             \pi_2\,:\,(x^{\a\ad}_R,\l^\pm_\ad,\eta^\ad_i)&\ \to\
             (x^{\a\ad}_R\l^\pm_\ad,\l^\pm_\ad,\eta^\ad_i\l^\pm_\ad),\cr
}}
where we call the $x^{\a\ad}_R$ anti-chiral coordinates. 
The supertwistor space can also be covered by two coordinate 
patches which we denote by $\{\CU_+,\CU_-\}$.\foot{Clearly, we have
$U_\pm=\CU_\pm\cap\IC P^1$.} Note that if we are given a real structure on the 
supertwistor space which is induced by an anti-linear involutive 
automorphism $\tau$ acting on the various coordinates (and which
naturally extends to arbitrary functions),
the double fibration \doublefibrationc\ reduces according to
\eqn\doublefibrationi{\pi\,:\,\CP^{3|\CN}\ \cong\ \IR^{4|2\CN}\times\IC P^1
                       \ \to\  \IR^{4|2\CN},}
by requiring reality of $(x^{\a\ad}_R,\eta^\ad_i)$.\foot{Requiring reality
of the fermionic coordinates in the case of Euclidean signature is only 
possible for $\CN=0,2,4$; see below.} Here, the $\l_\pm$ are 
kept complex. One is eventually interested in real self-dual 
YM fields and hence it is enough to take the fibration 
\doublefibrationi. However, it is more convenient to consider the 
complexification \doublefibrationc\ of \doublefibrationi\ first and, when 
necessary, to impose suitable reality conditions later on (cf. subsection 2.3).
As we will also see below in our discussion of the self-dual SYM 
hierarchies, double fibrations are natural -- even in the real setup.

We stress that from the explicit form of the projections \defofpi\ it follows 
that holomorphic sections of the bundle \defoftwistor\ are rational degree one
curves, $\IC P^1_{x_R,\eta}\hookrightarrow\CP^{3|\CN}$, given by the 
expressions
\eqn\ratcurvesone{z_\pm^\a\ =\ x_R^{\a\ad}\l^\pm_\ad
                  \qquad{\rm and}\qquad
                  \eta^\pm_i\ =\ \eta^\ad_i\l^\pm_\ad,
                  \qquad{\rm with}\qquad
                  \l^\pm_\ad\in U_\pm,}
and parametrized by the supermoduli $(x_R,\eta)=(x^{\a\ad}_R,\eta^\ad_i)\in
\IC^{4|2\CN}_R$. In this respect, the complexified (super)spacetime
can be interpreted as the moduli space of rational degree one curves living in
the supertwistor space. By a slight abuse of notation, we shall omit 
the subscript ``$R$'' on all the appearing expressions from now on.

For the discussion of the supertwistor correspondence, holomorphic vector 
bundles over the supertwistor space are needed. It is enough, however, to 
restrict ourselves
to such vector bundles whose fibers are not $\IZ_2$-graded, that is, 
we shall not need super vector bundles. Generally speaking,
a collection consisting of five objects, $(\CE,M,{\rm pr},{\frak U},f)$, is 
called a holomorphic rank $n$ vector bundle, whenever $\CE$ and $M$ are complex
(super)manifolds, the map 
$${\rm pr}\,:\,\CE\ \to\  M$$ 
is a surjective holomorphic projection, ${\frak U}=\{\CU_m\}$ is an open 
covering of $M$ and $f=\{f_{mn}\}$ is a collection of holomorphic transition 
functions defined on nonempty intersections $\CU_m\cap\CU_n$ and being 
$GL(n,\IC)$-valued. 

Let $M$ be either the supertwistor space $\CP^{3|\CN}$ or the correspondence 
space $\CF^{5|2\CN}$. We denote the coverings of $\CP^{3|\CN}$ and  
$\CF^{5|2\CN}$ by ${\frak U}=\{\CU_+,\CU_-\}$ and 
${\tilde{\frak U}}=\{{\tilde\CU}_+,{\tilde\CU}_-\}$, respectively. Consider a 
holomorphic vector bundle $\CE\to\CP^{3|\CN}$ and its pull-pack 
$\pi^*_2\CE\to\CF^{5|2\CN}$. These bundles are defined by the transition 
functions $f=\{f_{+-}\}$ on the intersection $\CU_+\cap\CU_-$ and $\pi^*_2f$
on ${\tilde\CU}_+\cap{\tilde\CU}_-$. For notational simplicity, we shall use
the same letter, $f$, for the transition functions of both bundles in the 
following course of discussion. Then, $f$ is annihilated by the vector fields
\eqn\defofvf{D^\pm_\a\ =\ \l^\ad_\pm\partial_{\a\ad},\qquad
             D^\pm_3\ =\ \partial_{{\bar\l}_\pm}
             \qquad{\rm and}\qquad
             D^i_\pm\ =\ \l^\ad_\pm\partial^i_\ad.}
Here, we have abbreviated
$\partial_{\a\ad}\equiv\partial/\partial x^{\a\ad}$,
$\partial_{{\bar\l}_\pm}\equiv\partial/\partial{\bar\l}_\pm$,
$\partial^i_\ad\equiv\partial/\partial\eta^\ad_i$. 
Moreover, spinor indices are raised and lowered via the $\epsilon$-tensors, 
e.g., $\o^\a=\epsilon^{\a\b}\o_\b$ and $\o^\ad=\epsilon^{\ad\bd}\o_\bd$, where
\eqn\epsdef{ 
  (\epsilon^{\a\b})\ =\ (\epsilon^{\ad\bd})\ =\
   \left(\matrix{0&1\cr -1&0\cr}\right)
   \qquad{\rm and}\qquad
  (\epsilon_{\a\b})\ =\ (\epsilon_{\ad\bd})\ =\
  \left(\matrix{0&-1\cr 1&0\cr}\right),}
with $\epsilon_{\a\b}\epsilon^{\b\g}=\d^\g_\a$ and 
$\epsilon_{\ad\bd}\epsilon^{\bd\gd}=\d^\gd_\ad$. Note that the vector fields
$D^\pm_\a$ and $D^i_\pm$ in \defofvf\ are tangent to the leaves of the 
fibration $\pi_2\,:\,\CF^{5|2\CN}\to\CP^{3|\CN}$.

We assume that the bundle $\CE$ is topologically trivial and moreover
holomorphically trivial when restricted to any projective line 
$\IC P^1_{x,\eta}\hookrightarrow\CP^{3|\CN}$. These conditions imply that 
there exist some smooth $GL(n,\IC)$-valued functions $\psi=\{\psi_+,\psi_-\}$,
 which define trivializations of $\pi_2^*\CE$ over ${\tilde\CU}_\pm$, such that
$f_{+-}$ can be decomposed as
\eqn\eqforf{f_{+-}\ =\ \psi^{-1}_+\psi_-}
and
\eqn\eqforfi{\partial_{{\bar\l}_\pm}\psi_\pm\ =\ 0.}
Applying the vector fields \defofvf\ to \eqforf, we realize that
$$ \psi_+D^+_\a\psi_+^{-1}\ =\ 
   \psi_-D^+_\a\psi_-^{-1} 
   \qquad{\rm and}\qquad
   \psi_+D^i_+\psi_+^{-1}\ =\ 
   \psi_-D^i_+\psi_-^{-1}
$$
must be at most linear in $\l_+$. Therefore, we may introduce a (Lie algebra 
valued) one-form $\CA$ such that
\eqna\defofa
$$\eqalignno{D^+_\a\lrcorner\CA\ &\equiv\ 
             \CA^+_\a\ \equiv\  \l_+^\ad\CA_{\a\ad}\ = \ 
             \psi_\pm D^+_\a\psi_\pm^{-1},
                                   &   \defofa a\cr
             \partial_{{\bar\l}_+}\lrcorner\CA\ &\equiv\ \CA_{{\bar\l}_+}
             \ =\ 0,               &\defofa b\cr     
             D^i_+\lrcorner\CA\  &\equiv\ \CA^i_+ \equiv\ 
             \l_+^\ad\CA^i_\ad\ =\ \psi_\pm D^i_+\psi_\pm^{-1},
                                   &\defofa c\cr     
}$$ 
and hence,
\eqna\linsys
$$\eqalignno{(D^+_\a+\CA^+_\a)\psi_\pm\ &=\ 0, & \linsys a\cr
  \partial_{{\bar\l}_+}\psi_\pm         \ &=\ 0, & \linsys b\cr 
  (D^i_++\CA^i_+)\psi_\pm             \ &=\ 0. & \linsys c\cr
}
$$ 

The compatibility conditions for the linear system \linsys{} read as
\eqn\compcon{\eqalign{
             [\nabla_{\a\ad},\nabla_{\b\bd}]+[\nabla_{\a\bd},\nabla_{\b\ad}]
             \ &=\ 0,\qquad
             [\nabla^i_\ad,\nabla_{\b\bd}]+[\nabla^i_\bd,\nabla_{\b\ad}]
             \ =\ 0,\cr
             &\kern-.8cm
             \{\nabla^i_\ad,\nabla^j_\bd\}+\{\nabla^i_\bd,\nabla^j_\ad\}\ =\ 0,
             \cr
}}
where we have defined the covariant derivatives 
\eqn\defofcovder{\nabla_{\a\ad}\ \equiv\ \partial_{\a\ad}+\CA_{\a\ad}
                 \qquad{\rm and}\qquad
                 \nabla^i_\ad\ \equiv\ \partial^i_\ad+\CA^i_\ad.}
In particular, the field content of $\CN=4$ self-dual SYM theory consists of a 
self-dual gauge potential $\Ac~\!\!_{\a\ad}$, four positive chirality spinors 
$\cc\ \!\!^i_\a$, six scalars $\Wc~\!^{ij}=-\Wc~\!^{ji}$, four negative 
chirality spinors $\cc_{i\ad}$ and an anti-self-dual two-form $\Gc_{\ad\bd}$, 
all in the adjoint representation of $GL(n,\IC)$. The circle refers to the
lowest component in the superfield expansions of the corresponding superfields
$\CA_{\a\ad}$, $\chi^i_\a$, $W^{ij}$, $\chi_{i\ad}$ and $G_{\ad\bd}$, 
respectively. Applying the techniques as presented in reference \DevchandGV, 
i.e., imposing the transversal gauge,
\eqn\tg{\eta^\ad_i\CA^i_\ad\ =\ 0,}
in order to remove the superfluous gauge degrees of freedom associated with the
fermionic coordinates $\eta^\ad_i$, we expand $\CA_{\a\ad}$ and $\CA^i_\ad$ 
with respect to the odd coordinates according to
\eqn\componentexpi{\eqalign{
     \CA_{\a\ad}\ &=\ \Ac~\!\!_{\a\ad}+\epsilon_{\ad\bd}\cc\ 
                      \!\!^i_\a\eta^\bd_i+\cdots,\cr
     \CA^i_\ad  \ &=\ \epsilon_{\ad\bd}\Wc~\!^{ij}\eta^\bd_j+\sfrac{4}{3}
                      \epsilon^{ijkl}\epsilon_{\ad\bd}\cc_{k\gd}\eta^\gd_l
                      \eta^\bd_j\ -\cr
                  &~~~~~~~-\sfrac{6}{5}\epsilon^{ijkl}\epsilon_{\ad\bd}
                      (\Gc_{\gd\dd}\d^m_{l}+\epsilon_{\gd\dd}[\Wc~\!^{mn},
                      \Wc_{nl}])\eta^\gd_k\eta^\dd_m\eta^\bd_j+\cdots,\cr
}}
where $\Wc_{ij}\equiv{1\over2}\epsilon_{ijkl}\Wc~\!^{kl}$.
Therefore, \compcon\ is equivalent to
\eqn\fieldeqnfour{\eqalign{
  \fc_{\ad\bd}\ &=\ 0,\cr
  \epsilon^{\a\b}\cnab_{\a\ad}\cc\ \!\!^i_\b\ &=\ 0,\cr
  \epsilon^{\a\b}\epsilon^{\ad\bd}\cnab_{\a\ad}\cnab_{\b\bd}\Wc~\!^{ij}+
  \epsilon^{\a\b}\{\cc\ \!\!^i_\a,\cc\ \!\!^j_\b\}\ &=\ 0,\cr
   \epsilon^{\ad\bd}\cnab_{\a\ad}\cc_{i\bd}-{\sfrac{1}{2}}\epsilon_{ijkl}
  [\Wc~\!^{kl},\cc\ \!\!^j_\a]\ &=\ 0,\cr
  \epsilon^{\ad\bd}\cnab_{\a\ad}\Gc_{\bd\gd}+\{\cc\ \!\!^i_\a,\cc_{i\gd}\}+ 
  {\sfrac{1}{4}}\epsilon_{ijkl}[\cnab_{\a\gd}\Wc~\!^{ij},\Wc~\!^{kl}]\
  &=\ 0.\cr 
}}
In \fieldeqnfour, $\fc_{\ad\bd}$ denotes the anti-self-dual part of the 
curvature $[\cnab_{\a\ad},\cnab_{\b\bd}]=
\epsilon_{\a\b}\fc_{\ad\bd}+\epsilon_{\ad\bd}\fc_{\a\b}$. Note that the system
\fieldeqnfour\ takes the same form for the superfields
$\CA_{\a\ad}$, $\chi^i_\a$, $W^{ij}$, $\chi_{i\ad}$ and $G_{\ad\bd}$. We 
remark also that the field equations for $\CN<4$ are simply obtained by 
suitable truncations of \compcon\ and \fieldeqnfour, respectively.

Equations \defofa{} can be integrated to give the formulas
\eqn\pwtrans{\eqalign{
             \CA_{\a\ad}\ =\ {1\over 2\pi\i}\oint_c{\rm d}\l_+
             {\CA^+_\a\over\l_+\l_+^\ad} \qquad&{\rm and}\qquad
             \CA^i_\ad\ =\ {1\over 2\pi\i}\oint_c{\rm d}\l_+
             {\CA^i_+\over\l_+\l_+^\ad},
             \cr
}}
where the contour $c=\{\l_+\in\IC P^1\,|\,|\l_+|=1\}$ encircles $\l_+=0$.

Note that the equations \defofa{} imply that the gauge potentials $\CA_{\a\ad}$
and $\CA^i_\ad$ do not change when we perform transformations of the form
$$\psi_\pm\ \mapsto\ \psi_\pm h_\pm,$$
where $h=\{h_+,h_-\}$ is annihilated by the vector fields \defofvf.
Under such transformations the transition function $f_{+-}$ of $\pi^*_2\CE$
transforms into a transition function $h_+^{-1}f_{+-}h_-$ of a bundle which is 
said to be equivalent to $\pi^*_2\CE$,
$$f_{+-}\ \sim\ h_+^{-1}f_{+-}h_-.$$ 
In the following, the set of equivalence classes induced by this equivalence
relation, i.e., the moduli space of holomorphic vector bundles with the 
properties discussed above is denoted by $\CM_{\rm hol}(\CF^{5|2\CN})$.\foot{In
the appendix A, we give a precise mathematical definition of this space.} Note 
that we have   
$$\CM_{\rm hol}(\CP^{3|\CN})\ \cong\ \CM_{\rm hol}(\CF^{5|2\CN})
  \qquad{\rm via}\qquad f\ \mapsto\ \pi^*_2f,$$
by the definition of a pull-back. Moreover, gauge transformations of the gauge
potential $\CA$ are induced by transformations of the form 
$\psi_\pm\mapsto g^{-1}\psi_\pm$ for some smooth $GL(n,\IC)$-valued $g$.
Under such transformations the transition function $f_{+-}$ does not change. 

In summary, we have described a one-to-one correspondence -- known 
as the supertwistor correspondence -- between equivalence classes of 
holomorphic vector bundles over the supertwistor space $\CP^{3|\CN}$ which are 
topologically trivial and holomorphically trivial on the curves
$\IC P^1_{x,\eta}\hookrightarrow\CP^{3|\CN}$ and gauge equivalence classes 
of solutions to the $\CN$-extended self-dual SYM equations, i.e., we have 
the bijection
\eqn\otohbsdym{\CM_{\rm hol}(\CP^{3|\CN})\ \longleftrightarrow\
              \CM^\CN_{\rm SDYM},}
where $\CM^\CN_{\rm SDYM}$ denotes the moduli space of solutions to the
$\CN$-extended self-dual SYM equations. Equations \pwtrans\ give the explicit 
form of the Penrose-Ward transform.

%--------------------------------------------------------------------------
\subsec{Leznov gauge (light cone formalism)}

Let us now consider a certain gauge, which goes under the name Leznov gauge
\LeznovUP. One of the interesting issues of this (non-covariant) gauge is the 
fact that the self-duality equations \compcon\ can be reexpressed in terms of a
single Lie algebra valued chiral superfield, which we denote by $\Psi$. 
Another issue is its resemblance to the light cone gauge when the Minkowski 
signature is chosen.

To be explicit, assume the following expansion
\eqn\sac{\psi_+\ =\ 1+\l_+\Psi+\CO(\l_+^2)}
on ${\tilde\CU}_+$ and therefore on
the intersection ${\tilde\CU}_+\cap{\tilde\CU}_-$. In \sac, all 
$\l$-dependence has been made explicit. Note that generically $\psi_+$ is 
expanded according to $\psi_+=\psi^{(0)}_++\l_+\psi^{(1)}_++\cdots$. Then, the 
choice $\psi^{(0)}_+=1$ corresponds to fixing a particular gauge.\foot{Recall 
that gauge transformations are given by $\psi_+\mapsto g^{-1}\psi_+$. Thus, the
expansion \sac\ can be obtained from some general $\psi_+$ by performing
the gauge transformation  $\psi_+\mapsto (\psi^{(0)}_+)^{-1}\psi_+$.}
Upon substituting \sac\ 
into \linsys{}, we discover
\eqn\leznovgauge{\CA_{\a{\dot1}}\ =\ \partial_{\a{\dot2}}\Psi,\qquad
                 \CA_{\a{\dot2}}\ =\ 0\qquad{\rm and}\qquad
                 \CA^i_{\dot1}\ =\ \partial^i_{{\dot2}}\Psi,\qquad
                 \CA^i_{\dot2}\ =\ 0.}
The conditions \leznovgauge\ are called Leznov gauge. Plugging \leznovgauge\
into the self-duality equations \compcon, we obtain the following set of 
equations:\foot{Note that the essential equation of motion is just the first 
one, i.e., solving the first equation to all orders in $\eta^i_{\dot2}$ 
and substituting in the remaining two, one determines the superfield $\Psi$ to 
all orders in $\eta^i_{\dot1}$ and  $\eta^i_{\dot1}\eta^j_{\dot2}$, 
respectively; see also \SiegelZA.}
\eqn\compconii{\eqalign{
               \partial_{1{\dot1}}\partial_{2{\dot2}}\Psi-
               \partial_{2{\dot1}}\partial_{1{\dot2}}\Psi+
               [\partial_{1{\dot2}}\Psi,\partial_{2{\dot2}}\Psi]\ &=\ 0,\cr
               \partial^i_{{\dot1}}\partial_{\a{\dot2}}\Psi-
               \partial_{\a{\dot1}}\partial^i_{{\dot2}}\Psi+
               [\partial^i_{{\dot2}}\Psi,\partial_{\a{\dot2}}\Psi]\ &=\ 0,\cr
               \partial^i_{{\dot1}}\partial^j_{{\dot2}}\Psi+
               \partial^j_{{\dot1}}\partial^i_{{\dot2}}\Psi+
               \{\partial^i_{{\dot2}}\Psi,\partial^j_{{\dot2}}\Psi\}\ &=\ 0.
               \cr
}}
Now it is straightforward to show that the field content, for instance, in the
case of maximal $\CN=4$ supersymmetries in Leznov gauge is given by
\eqn\fieldcontentnfour{\eqalign{
     \CA_{\a{\dot1}}\ &=\ \partial_{\a{\dot2}}\Psi
     \quad\Rightarrow\quad
     f_{\a\b}\ =\ \partial_{\a{\dot2}}\partial_{\b{\dot2}}\Psi,\cr
     \chi^i_\a\ &=\ \partial^i_{{\dot2}}\partial_{\a{\dot2}}\Psi,\cr
     \chi_{i{\dot2}}\ &=\ {\sfrac{1}{2\cdot3!}}\epsilon_{ijkl}
       \partial^j_{{\dot2}}\partial^k_{{\dot2}}\partial^l_{{\dot2}}\Psi,\cr
     W^{ij}\ &=\ {\sfrac{1}{2}}\partial^i_{{\dot2}}\partial^j_{{\dot2}}
    \Psi,\cr
    G_{{\dot2}{\dot2}}\ &=\ {\sfrac{1}{2\cdot4!}}\epsilon_{ijkl}
      \partial^i_{{\dot2}}\partial^j_{{\dot2}}\partial^k_{{\dot2}}
       \partial^l_{{\dot2}}\Psi,\cr
}}
i.e., the superfield $\Psi$ plays the role of a potential. 

%--------------------------------------------------------------------------
\subsec{Reality conditions}

In the preceding discussion, we did only briefly mention the involutive 
automorphism $\tau$ which induces a real structure on the supertwistor space. 
Let us now be a bit more explicit. The Euclidean signature
is related to anti-linear transformations 
which act on the coordinates $x^{\a\ad}$ as
\eqn\actiontaui{\tau\left(\matrix{x^{1{\dot1}} & x^{1{\dot2}}\cr
                                  x^{2{\dot1}} & x^{2{\dot2}}\cr}
                    \right)\ =\ 
                    \left(\matrix{{\bar x}^{2{\dot2}} & 
                                  -{\bar x}^{2{\dot1}}\cr
                                  -{\bar x}^{1{\dot2}} & 
                                  {\bar x}^{1{\dot1}}\cr}\right),}
where the bar denotes complex conjugation. The real subspace $\IR^4$
of $\IC^4$ invariant under $\tau$ is defined by the equations
\eqn\actiontauii{x^{2{\dot2}}\ =\ {\bar x}^{1{\dot1}}\ \equiv\ x^4-\i x^3
                 \qquad{\rm and}\qquad
                 x^{2{\dot1}}\ =\ -{\bar x}^{1{\dot2}}\ \equiv\ -x^2+\i x^1,
}
and parametrized by real coordinates $x^\mu\in\IR$ with $\mu=1,\ldots,4$.
Moreover, $\tau$ acts on the $\l^\pm_\ad$ as
\eqn\actiontauiii{\tau(\l^+_\ad)\ =\ \left(\matrix{-{\bar\l}_+\cr1\cr}\right)
                  \qquad{\rm and}\qquad
                  \tau(\l^-_\ad)\ =\ \left(\matrix{-1\cr{\bar\l}_-\cr}\right).}
For later convenience, we define $\tau(\l^\pm_\ad)\equiv\hl^\pm_\ad$. 
The action of $\tau$ on the fermionic coordinates in the case of 
maximal $\CN=4$ supersymmetries is given by
\eqn\actiontauiv{\tau(\eta^\ad_i)\ =\ \epsilon^{\ad\bd} T_i^{\ j}
                 {\bar\eta}^\bd_j, 
                 \qquad{\rm with}\qquad
                 (T_i^{\ j})\ \equiv\ \left(\matrix{0  &1  & 0  & 0\cr
                                                    -1 & 0 & 0  & 0\cr
                                                    0  & 0 & 0  & 1\cr
                                                    0  & 0 & -1 & 0\cr}\right).
}
Therefore, we obtain the (symplectic) Majorana condition
\eqn\actiontauivi{\eta^\ad_i\ =\ \epsilon^{\ad\bd} T_i^{\ j}{\bar\eta}^\bd_j.} 
Note that reality of the fermionic coordinates on Euclidean space
can only be imposed if the number of supersymmetries is even
\refs{\KotrlaKY}; the $\CN=0$ and $\CN=2$ cases are obtained by suitable
truncations of the $\CN=4$ case \actiontauiv\ and \actiontauivi.
The action of $\tau$ on arbitrary functions is then immediate. 

Finally, skew-Hermitian self-dual SYM fields can be obtained by imposing the 
following condition
\eqn\actiontauivi{f_{+-}(x,\l_+,\eta)\ =\ [f_{+-}(\tau(x,\l_+,\eta))
                  ]^\dagger}
on the transition function. Here, the dagger denotes the extension of the 
complex conjugation to matrix-valued functions.

%--------------------------------------------------------------------------
%--------------------------------------------------------------------------
\newsec{Holomorphy and infinitesimal symmetries}

By now it is clear that we can relate -- by means of the Penrose-Ward transform
-- holomorphic vector bundles over the supertwistor space, which are 
holomorphically trivial along the curves
$\IC P^1_{x,\eta}\hookrightarrow\CP^{3|\CN}$ and characterized by the 
transition functions $f=\{f_{+-}\}$ representing 
$[f]\in\CM_{\rm hol}(\CP^{3|\CN})$, with solutions to the $\CN$-extended 
self-dual SYM equations. We can, however, associate with any open subset 
$\Omega\subset\CU_+\cap\CU_-\subset\CP^{3|\CN}$ an infinite number of such 
$[f]\in\CM_{\rm hol}(\CP^{3|\CN})$. Each class $[f]$ of 
equivalent holomorphic vector bundles $\CE\to\CP^{3|\CN}$ then
corresponds to a class $[\CA]\in\CM_{\rm SDYM}^\CN$ of gauge equivalent 
solutions to the $\CN$-extended self-dual SYM 
equations. Therefore, one might wonder about the possibility of the 
construction of a new solution from a given one. In this section, we are going 
to address this issue and describe so-called infinitesimal symmetries of the 
$\CN$-extended self-dual SYM equations. We thereby generalize results which 
have been discussed in the literature (cf., e.g., references 
\refs{\PohlmeyerYA-\MasonRF}) for the bosonic SDYM equations. 
 In order to 
discuss them, we will need to consider infinitesimal deformations of the 
transition functions of holomorphic vector bundles
$$\CE\ \to\ \CP^{3|\CN}$$
having the above properties. By virtue of the Penrose-Ward transform, we relate
them to infinitesimal perturbations $\d\CA$ of the gauge potential $\CA$. Note 
that nontrivial infinitesimal deformations of the transition functions, denoted
by $\d f_{+-}$, correspond to vector fields on the moduli space 
$\CM_{\rm hol}(\CP^{3|\CN})$. Nontrivial infinitesimal perturbations $\d\CA$ 
determine vector fields on the moduli space $\CM_{\rm SDYM}^\CN$.\foot{More 
precisely, one should rather say on the solution space than on the moduli 
space. The latter one is obtained from the former one by factorizing with 
respect to the group of gauge transformations. Similar arguments hold for 
$(\d f_{+-})f_{+-}^{-1}$. 
However, here and in the following we just ignore this subtlety.} Therefore, 
the linearized Penrose-Ward transform gives us an isomorphism
\eqn\iso{T_{[f]}\CM_{\rm hol}(\CP^{3|\CN})\ \cong\ 
         T_{[\CA]}\CM_{\rm SDYM}^\CN,}
between the tangent spaces. We refer again to appendix A for precise 
mathematical definitions.

%--------------------------------------------------------------------------
\subsec{Infinitesimal perturbations}

In the following, we study small perturbations of the transition function 
$f_{+-}$ of a holomorphic vector bundle $\CE\to\CP^{3|\CN}$ and the pull-back 
bundle $\pi_2^*\CE\to\CF^{5|2\CN}$. Any infinitesimal perturbation of $f_{+-}$ 
is allowed, as small enough perturbations of the bundle $\CE$ will -- by a 
version of the Kodaira theorem -- preserve its trivializability properties on 
the curves $\IC P^1_{x,\eta}\hookrightarrow\CP^{3|\CN}$. Take now 
$\CF^{5|2\CN}$ and consider an infinitesimal transformation $\d$ of the 
transition function $f_{+-}$ of the form
\eqn\infperttrans{\delta\,:\,f_{+-}\ \mapsto\ \d f_{+-}\ 
                  \equiv\ \sum_a\ve_a\d_a f_{+-},}
where $\ve_a$ are the infinitesimal parameters of the transformation. 
Generically, they are either Gra{\ss}mann even (parity $p_a=0$) or Gra{\ss}mann
odd (parity $p_a=1$). Note that the total parity $p_\d$ of $\d$ is zero. 
Occasionally, we will use the Latin letters $a,b,c,\ldots$ to denote
symbolically both, bosonic as well as fermionic indices. We may write
\eqn\deformfa{f_{+-}+\d f_{+-}\ =\ (\psi_++\d\psi_+)^{-1}(\psi_-+\d\psi_-).}
Expanding the right hand side of this equation up to first order in the 
perturbation, we obtain
\eqn\deformf{\d f_{+-}\ =\ f_{+-}\psi_-^{-1}\d\psi_-
                           -\psi_+^{-1}\d\psi_+f_{+-}.}
Let us define the following ${\frak g}{\frak l}(n,\IC)$-valued function
\eqn\defofphi{\varphi_{+-}\ \equiv\ \psi_+(\d f_{+-})\psi_-^{-1}.}
The substitution of \deformf\ into \defofphi\ yields
\eqn\eqphii{\varphi_{+-}\ =\ \phi_+-\phi_-,}
where 
\eqn\eqforhpmi{\d\psi_\pm\ =\ -\phi_\pm\psi_\pm.}
The ${\frak g}{\frak l}(n,\IC)$-valued functions $\phi_+$ and $\phi_-$
can be extended to holomorphic functions in $\l_+$ and $\l_-$ on the patches 
${\tilde\CU}_+$ and ${\tilde\CU}_-$, respectively. We remark that finding such
$\phi_\pm$ means to solve the infinitesimal variant of the Riemann-Hilbert 
problem. Obviously, the splitting \eqphii, \eqforhpmi\ and hence solutions to 
the Riemann-Hilbert problem are not unique, as we certainly have the freedom to
consider a new ${\tilde\phi}_\pm$ shifted by functions $\g_\pm$, 
${\tilde\phi}_\pm=\phi_\pm+\gamma_\pm$, such that $\gamma_+=\gamma_-$ on the 
intersection ${\tilde\CU}_+\cap{\tilde\CU}_-$, that is, there exists some $\g$ 
with $\g_\pm=\g|_{{\tilde\CU}_\pm}$.

Equation \defofa{} yields
\eqna\deforeqa
$$\eqalignno{\d\CA^+_\a\ &=\ \d\psi_\pm D^+_\a\psi_\pm^{-1}+
                             \psi_\pm D^+_\a\d\psi_\pm^{-1},
                         &\deforeqa a\cr
             \d\CA^i_+ \ &=\ \d\psi_\pm D^i_+\psi_\pm^{-1}+
                             \psi_\pm D^i_+\d\psi_\pm^{-1}.
                         &\deforeqa b\cr                 
}
$$ 
Substituting \eqforhpmi\ into these equations, we arrive at
\eqn\deforeqai{\d\CA^+_\a\ =\ \nabla^+_\a\phi_\pm
               \qquad{\rm and}\qquad
               \d\CA^i_+\ =\ \nabla^i_+\phi_\pm.}
Here, we have introduced the definitions
\eqn\defofnabi{\nabla^+_\a\ \equiv\  D^+_\a+\CA^+_\a
               \qquad{\rm and}\qquad
               \nabla^i_+\ \equiv\ D^i_++\CA^i_+.}
Note that \defofnabi\ acts adjointly in \deforeqai. Therefore, \deforeqai\
together with \eqphii\ imply that
\eqn\deforeqaii{\nabla^+_\a\varphi_{+-}\ =\ 0\qquad{\rm and}\qquad
               \nabla^i_+\varphi_{+-}\ =\ 0.}
Moreover, the equation \linsys{b} remains untouched, since
$\d\psi_\pm$ and $\phi_\pm$ are annihilated by $\partial_{{\bar\l}_\pm}$.

One may easily check that for the choice 
$\phi_\pm=\psi_\pm\chi_\pm\psi_\pm^{-1}$, where the matrix-valued functions
$\chi_\pm$ depend holomorphically on the twistor coordinates 
$(x^{\a\ad}\l_\ad^\pm,\l^\pm_\ad,\eta^\ad_i\l^\pm_\ad)$,
we have $\d\CA_{\a\ad}=0$ and $\d\CA^i_\ad=0$, respectively. Hence, such
$\phi_\pm$ define trivial perturbations. On the other hand, infinitesimal gauge
transformations of the form
$$\d\CA_{\a\ad}\ =\ \nabla_{\a\ad}\o
  \qquad{\rm and}\qquad
  \d\CA^i_\ad\ =\ \nabla^i_\ad\o,$$ 
where $\o$ is some smooth ${\frak g}{\frak l}(n,\IC)$-valued function
imply (for irreducible gauge potentials) that 
$$\phi_\pm\ =\ \o.$$ 
Therefore, such $\phi_\pm$ do not depend on $\l_+$. In particular, we have 
$\varphi_{+-}=0$ and hence $\d f_{+-}=0$. 

Finally, we obtain the formulas
\eqn\infpwtrans{\eqalign{
             \d\CA_{\a\ad}\ =\ {1\over 2\pi\i}\oint_c{\rm d}\l_+
             {\nabla^+_\a\phi_\pm\over\l_+\l_+^\ad}  \qquad&{\rm and}\qquad
             \d\CA^i_\ad  \ =\ {1\over 2\pi\i}\oint_c{\rm d}\l_+
             {\nabla^i_+\phi_\pm\over\l_+\l_+^\ad},\cr
}}
where the contour is again taken as $c=\{\l_+\in\IC P^1\,|\,|\l_+|=1\}$.\foot{
We see that generically the outcome of the transformation $\d\CA$ is a
highly nonlocal expression depending on $\CA$, i.e., we may write 
$\d\CA=F[\CA,\partial\CA,\ldots]$, where $F$ is a functional whose explicit
form is determined by $\phi_\pm$.}

Summarizing, to any infinitesimal deformation, 
$f_{+-}\mapsto f_{+-}+\d f_{+-}$, of the transition function $f_{+-}$ we have
associated an 
infinitesimal symmetry transformation $\CA\mapsto\CA+\d\CA$. Clearly, by 
construction, the components of $\d\CA$ satisfy the linearized version of 
the $\CN$-extended self-dual SYM equations.

%--------------------------------------------------------------------------
\subsec{An example: Hidden super Kac-Moody symmetries}

After this somewhat general discussion of symmetries, let us give a first 
example of symmetries. Let ${\frak g}$ be a some (matrix) Lie 
superalgebra\foot{The only assumption we want to made at this stage is that the
corresponding Lie supergroup consists of matrices that are of $c$-type. See
\DeWittCY\ for details.} with structure constants $f_{ab}^{\ \ c}$ and (matrix)
generators $X_a$, which satisfy
\eqn\slaKMY{[X_a,X_b\}\ =\ f_{ab}^{\ \ c}X_c,}
where $[\cdot,\cdot\}$ denotes the supercommutator
\eqn\defofsupercom{[A,B\}\ \equiv\ AB-(-)^{p_Ap_B}BA.}
Then, we consider the following perturbation of the transition function:
\eqn\perturbationKMY{\d_a^mf_{+-}\ \equiv\ \l_+^m[X_a,f_{+-}],\qquad
                     {\rm with}\qquad m\in\IZ.}
One may readily verify that the corresponding algebra is given by
\eqn\KMalg{[\d^m_a,\d^n_b\}\ =\ f_{ab}^{\ \ c}\d^{m+n}_c}
when acting on $f_{+-}$, i.e., we obtain the centerless super Kac-Moody algebra
${\frak g}\otimes\IC[[\l,\l^{-1}]]$. Note that in case one whishes to 
preserve the reality conditions, one must slightly change the above
deformation. For instance, in case one chooses for ${\frak g}$
the gauge algebra ${\frak s}{\frak u}(n)$, the deformation is
given according to $$\d_a^mf_{+-}\ =\ (\l_+^m+(-\l_-)^m)[X_a,f_{+-}].$$ 
We shall be continuing with our above deformation
\perturbationKMY.

Next we need to find -- by means of the discussion of the previous subsection 
-- the action of $\d^m_a$ on the components $\CA_{\a\ad}$ and $\CA^i_\ad$ of 
the gauge potential $\CA$. Consider the function $\varphi^m_{+-a}$ as defined 
in \defofphi. We obtain
$$\eqalign{\varphi^m_{+-a}\
           &=\ \phi^m_{+a}-\phi^m_{-a}\cr 
           &=\ \psi_+\l_+^m[X_a,f_{+-}]\psi_-^{-1}\ =\ 
               \l_+^m\psi_+[X_a,\psi_+^{-1}\psi_-]\psi_-^{-1}\cr
           &=\ -\l_+^m[X_a,\psi_+]\psi_+^{-1}+\l_+^m[X_a,\psi_-]\psi_-^{-1}\cr
           &=\ \l_+^m\phi^0_{+a}-\l_+^m\phi^0_{-a},
}$$
where in the last step we have introduced the functions $\phi^0_{\pm a}$ 
which are solutions of the Riemann-Hilbert problem for $m=0$, 
\eqn\phisolzeroKMY{\phi^0_{\pm a}\ =\ -[X_a,\psi_\pm]\psi_\pm^{-1}.} As the 
$\phi^0_{\pm a}$ are holomorphic and nonsingular in $\l_\pm$ on their 
respective domains, we expand them in powers of $\l_+$ on the intersection 
${\tilde\CU}_+\cap{\tilde\CU}_-$ according to
\eqn\phiexpKMY{\phi_{\pm a}^0\ =\ \sum_{n=0}^\infty\l_+^{\pm n}
               \phi^{0(n)}_{\pm a},}
where the coefficients $\phi^{0(n)}_{\pm a}$ are conserved nonlocal charges
\PohlmeyerYA.
Furthermore, equations \infpwtrans\ imply
\eqn\perturbationAKMYpi{\eqalign{
     \d^m_a\CA_{\a{\dot1}}\ &=\ \nabla_{\a{\dot1}}\phi^{m(0)}_{+a}-
                                \nabla_{\a{\dot2}}\phi^{m(1)}_{+a}      
                           \ =\ \nabla_{\a{\dot1}}\phi^{m(0)}_{-a},\cr
     \d^m_a\CA_{\a{\dot2}}\ &=\ \nabla_{\a{\dot2}}\phi^{m(0)}_{+a}      
                           \ =\ -\nabla_{\a{\dot1}}\phi^{m(1)}_{-a}
                                +\nabla_{\a{\dot2}}\phi^{m(0)}_{-a},\cr
}}
and similarly for $\CA^i_\ad$.

Assume for a moment that $m\geq 0$. Then $\varphi^m_{+-a}$ can be split into 
$\phi^m_{\pm a}$ with
\eqn\phisolKMY{\eqalign{
               \phi^m_{+a}\ &=\  \sum_{n=0}^\infty\l_+^n\phi_{+a}^{m(n)}
               \ =\ \sum_{n=0}^\infty\l_+^{m+n}\phi^{0(n)}_{+a}-
               \sum_{n=0}^{m-1}\l_+^{m-n}\phi^{0(n)}_{-a},\cr
               \phi^m_{-a}\ &=\ \sum_{n=0}^\infty\l_+^{-n}\phi_{-a}^{m(n)}
               \ =\ \sum_{n=0}^\infty\l_+^{-n}\phi^{0(m+n)}_{-a}.\cr
}}
Note that for $m>0$ only $\phi^m_{-a}$ contains zero modes of 
$\partial_{\l_+}$. We know from our previous discussion, however, that 
solutions to the Riemann-Hilbert problem are not unique. So, for instance,
we could have added to $\phi^m_{+a}$ any differentiable function which does not
depend on $\l_+$. But at the same time we had to add the same function to 
$\phi^m_{-a}$, as well. However, such shifts eventually result in gauge 
transformations of the gauge potential $\CA$.\foot{See also the discussion at 
the end of subsection 3.1.} As we are not interested in such trivial 
symmetries, 
the solution \phisolKMY\ turns out to be the appropriate choice for our further
discussion. Expanding the functions $\phi^m_{\pm a}$ in powers of $\l_+$, we
obtain the following coefficients
\eqn\phiexpKMY{\phi_{+a}^{m(n)}\ =\ \cases{
                       \d_{m,0}\phi_{+a}^{0(0)}&for $n=0$\cr
                       \phi^{0(n-m)}_{+a}-\phi^{0(m-n)}_{-a} &for $n>0$}
              \qquad{\rm and}\qquad
               \phi_{-a}^{m(n)}\ =\ \phi_{-a}^{0(m+n)},
               }
Note that $\phi^{0(-k)}_{\pm a}=0$ for $k>0$.  Finally, equations 
\perturbationAKMYpi\ together with \phiexpKMY\ yield the desired transformation
rules for the components of the gauge potential for $m\geq0$
\eqn\perturbationAKMYi{\eqalign{
             \d^m_a\CA_{\a{\dot1}}\ =\ \nabla_{\a{\dot1}}\phi^{0(m)}_{-a}
             \quad&{\rm and}\quad
             \d^m_a\CA_{\a{\dot2}}\ =\ -\nabla_{\a{\dot1}}\phi^{0(m+1)}_{-a}+
             \nabla_{\a{\dot2}}\phi^{0(m)}_{-a},\cr
             \d^m_a\CA^i_{\dot1}\ =\ (-)^{p_a} \nabla^i_{\dot1}\phi^{0(m)}_{-a}
             \quad&{\rm and}\quad
             \d^m_a\CA^i_{\dot2}\ =\ -(-)^{p_a}\nabla^i_{\dot1}
             \phi^{0(m+1)}_{-a}+(-)^{p_a}\nabla^i_{\dot2}\phi^{0(m)}_{-a}.
             \cr
}}
We stress that not all of the transformations \perturbationAKMYi\ are 
nontrivial. Namely, we have 
$$\d^m_a\CA_{\a{\dot2}}\ =\ 0\ =\ \d^m_a\CA^i_{\dot2}$$ for $m>0$, as 
can easily be seen from \perturbationAKMYpi\ by remembering that 
the coefficient $\phi_{+a}^{m(0)}$ is identically zero for all $m>0$.
This observation motivates us to assume that $\phi_{+a}^{0(0)}$ is zero, as 
well, implying that $\CA_{\a{\dot2}}$ and $\CA^i_{\dot2}$ are invariant
under the action of $\d^m_a$ for all $m\geq0$.\foot{This means that all the 
zero modes of $\partial_{\l_+}$ of $\varphi_{+-a}^m=\phi^m_{+a}-\phi_{-a}^m$ 
are contained in $\phi_{-a}^m$. Alternatively, we could have changed the 
transformation laws to
$${\tilde\d}^m_a\CA_{\a\ad}\ \equiv\ \d^m_a\CA_{\a\ad}-
   \nabla_{\a\ad}\phi^{m(0)}_{+a} $$
and similarly for $\CA^i_\ad$ to get similar results; see below.} 
Equation \phisolzeroKMY\ therefore 
yields that $\psi^{(0)}_+$ is proportional to the identity, i.e., without loss 
of generality we may assume that the expansion of $\psi_+$ is given by \sac.
Thus, the simplification $\phi_{+a}^{0(0)}=0$ leads us to the 
Leznov gauge, where the components  $\CA_{\a{\dot2}}$ and 
$\CA^i_{\dot2}$ are gauge transformed to zero.

The next step is to compute the action of the commutator of two successive
infinite-\break simal 
transformations, $[\d_1,\d_2]$, on the components of the gauge
potential. In particular, we have
$$[\d_1,\d_2]\ =\ (-)^{p_ap_b}\epsilon_m^a\rho_n^b[\d^m_a,\d^n_b\},$$
where $\epsilon_m^a$ and $\rho_n^b$ are the infinitesimal parameters of the 
transformations $\d_1$ and $\d_2$, respectively. Explicitly, we may write 
\eqn\SConA{[\d_1,\d_2]\CA_{\a\ad}\ =\ 
           \d_1(\CA_{\a\ad}+\d_2\CA_{\a\ad})-\d_1\CA_{\a\ad}-
           \d_2(\CA_{\a\ad}+\d_1\CA_{\a\ad})+\d_2\CA_{\a\ad}}
and similarly for $\CA^i_\ad$. Then a straightforward calculation shows that
\eqn\SConAI{[\d^m_a,\d^n_b\}\CA_{\a{\dot1}}\ =\ 
            \nabla_{\a{\dot1}}\Sigma^{mn}_{ab}\qquad{\rm and}\qquad
            [\d^m_a,\d^n_b\}\CA^i_{{\dot1}}\ =\ (-)^{p_a+p_b}
            \nabla^i_{{\dot1}}\Sigma^{mn}_{ab},}
where
\eqn\SConAII{\Sigma^{mn}_{ab}\ \equiv\ 
            -\d^m_a\phi^{0(n)}_{-b}+(-)^{p_ap_b}\d^n_b\phi^{0(m)}_{-a}-
            [\phi^{0(m)}_{-a},\phi^{0(n)}_{-b}\}.}
In this derivation we have used the coefficients \phiexpKMY. The expressions
$\d^m_a\phi^{0(n)}_{-b}$ are determined via the contour integral
\eqn\SConAIII{\d^m_a\phi^{0(n)}_{-b}\ =\ \oint_c{{\rm d}\l_+\over 2\pi\i}\,
             \l_+^{n-1}\,\d^m_a\phi^0_{-b},}
where the contour is taken as $c=\{\l_+\in\IC P^1\,|\,|\l_+|=1\}$.
Remember that $\phi_{-a}^0$ is given by 
$\phi_{-a}^0=-[X_a,\psi_-]\psi_-^{-1}$. Hence, 
\eqna\SConAIV
$$\eqalignno{\d^m_a\phi^0_{-b}
             \ &=\ -(-)^{p_ap_b}\left([X_b,\d^m_a\psi_-\}
             \psi^{-1}_-+\phi_{-b}^0(\d^m_a\psi_-)\psi_-^{-1}\right)&{}\cr
             &=\ -(-)^{p_ap_b}\left([X_b,(\d^m_a\psi_-)\psi^{-1}_-\}\ +
               \right.&{}\cr
             &\left.\kern2.5cm
             + \ (-)^{p_ap_b}(\d^m_a\psi_-)\psi^{-1}_-[X_b,\psi_-]\psi^{-1}_-
             +\phi_{-b}^0(\d^m_a\psi_-)\psi_-^{-1}\right)&{}\cr
             &=\ (-)^{p_ap_b}[X_b+\phi^0_{-b},\phi^m_{-a}\}. & \SConAIV{}\cr
}$$
Equation \SConAIII\ implies
\eqn\SConAV{\d^m_a\phi^{0(n)}_{-b}\ =\ (-)^{p_ap_b}\left(
            [X_b,\phi^{0(m+n)}_{-a}\}+\sum_{k=0}^n\ [\phi^{0(k)}_{-b},
            \phi^{0(m+n-k)}_{-a}\}\right),}
where we have used the expansions \phiexpKMY. Furthermore, we find
$$\eqalign{[X_b,\phi^{0(m+n)}_{-a}\}\ &=\ - \oint_c{{\rm d}\l_+\over 2\pi\i}\,
    \l_+^{m+n-1}\,[X_b,[X_a,\psi_-]\psi_-^{-1}\}\cr
    &=\  \oint_c{{\rm d}\l_+\over 2\pi\i}\,
    \l_+^{m+n-1}\,(-)^{p_ap_b}[X_a,\psi_-]\psi_-^{-1}[X_b,\psi_-]\psi_-^{-1}\ -
    \cr
    &\kern1.5cm-\ \oint_c{{\rm d}\l_+\over 2\pi\i}\,\l_+^{m+n-1}\,
    [X_b,[X_a,\psi_-]\}\psi_-^{-1}\cr
    &=\ (-)^{p_ap_b}\sum_{k=0}^{m+n}\ \phi^{0(k)}_{-a}\phi^{0(m+n-k)}_{-b}
            -\ \oint_c{{\rm d}\l_+\over 2\pi\i}\,\l_+^{m+n-1}\,
    [X_b,[X_a,\psi_-]\}\psi_-^{-1}.\cr
}$$
Upon using the super Jacobi identity for the triple $(X_a,X_b,\psi_-)$, we find
$$\eqalign{
 \d^m_a\phi^{0(n)}_{-b}-(-)^{p_ap_b}\d^n_b\phi^{0(m)}_{-a}\ 
 &=\ -f_{ab}^{\ \ c}\phi_{-c}^{0(m+n)}+\sum_{k=0}^{m+n}\ [\phi^{0(k)}_{-a},
 \phi^{0(m+n-k)}_{-b}\}\ -\cr
 &\kern-3cm-\ \sum_{k=0}^n\ [\phi^{0(m+n-k)}_{-a}, \phi^{0(k)}_{-b}\}-
 \sum_{k=0}^m\ [\phi^{0(k)}_{-a},\phi^{0(m+n-k)}_{-b}\},\cr
}$$
which simplifies further to
\eqn\SConAVI{\d^m_a\phi^{0(n)}_{-b}-(-)^{p_ap_b}\d^n_b\phi^{0(m)}_{-a}\ =\ 
     -f_{ab}^{\ \ c}\phi_{-c}^{0(m+n)}-[\phi_{-a}^{0(m)},\phi_{-b}^{0(n)}\}.}
Altogether, the function $\Sigma^{mn}_{ab}$ as defined in \SConAII\ turns out 
to be
\eqn\SConAVII{\Sigma^{mn}_{ab}\ =\ f_{ab}^{\ \ c}\phi^{0(m+n)}_{-c}.}
Therefore, the supercommutator \SConAI\ is given by
\eqn\SConAVIII{[\d^m_a,\d^n_b\}\ =\ 
             f_{ab}^{\ \ c}\d_c^{m+n},}
when acting on $\CA_{\a\ad}$ and $\CA^i_\ad$, respectively. The superalgebra 
\SConAVIII\ defines the analytic half of the centerless super Kac-Moody algebra
${\frak g}\otimes\IC[[\l,\l^{-1}]]$.

So far, we have restricted to the case when $m\geq 0$. However, the line of
argumentation is quite similar for negative $m$. In this case one may assume 
without loss of generality that $\phi_{-a}^{0(0)}$ is zero, which obviously 
implies that the other half of the components of the gauge potential are gauge 
transformed to zero, i.e.,  $\CA_{\a{\dot1}}$ and $\CA^i_{\dot1}$. Putting both
cases together one eventually obtains the full affine extension of 
${\frak g}$, that is, the super Kac-Moody algebra 
${\frak g}\otimes\IC[[\l,\l^{-1}]]$.\foot{For a similar discussion in the
context of $2D$ sigma models, see, e.g., \SchwarzTD.}

%--------------------------------------------------------------------------
\subsec{Superconformal symmetries}

In our preceding discussion, we explained how one can describe super Kac-Moody 
symmetries associated with a given (matrix) Lie superalgebra ${\frak g}$,
and which act on the solution space of the $\CN$-extended self-dual SYM 
equations. In the 
remainder of this section we are going to push those ideas a little further by 
introducing affine extensions of the superconformal algebra. In doing so,
we obtain an infinite number of conserved nonlocal charges, which are 
associated with the superconformal algebra.
In the following, we shall assume that we are given 
a real structure $\tau$ as discussed in subsection 2.3. That means, we deal 
with the single fibration \doublefibrationi.

It is well known that (classically) the $\CN$-extended self-dual SYM equations 
are invariant under superconformal transformations. The superconformal group 
for $\CN\neq 4$ is locally isomorphic to a real form of the super matrix group 
$SU(4|\CN)$.
In the case of maximal $\CN=4$ supersymmetries, the supergroup $SU(4|4)$ is not
semi-simple and the superconformal group is considered to be a real form of the
semi-simple part $PSU(4|4)\subset SU(4|4)$.
The generators of the superconformal group  are the translation generators 
$P_{\a\ad}$, $Q_{i\a}$ and $Q^i_\ad$, the dilatation generator $D$, the 
generators of special conformal transformations $K_{\a\ad}$, $K^{i\a}$ and
$K^\ad_i$, the rotation generators $J_{\a\b}$ and $J_{\ad\bd}$, the generators 
$T_i^j$ of the internal symmetry and the generator of the axial symmetry 
$A$. The latter one is absent in the case of maximal $\CN=4$ supersymmetries. 
These generators can be realized in terms of the following vector fields on the
anti-chiral superspace $\IR^{4|2\CN}$:
\eqn\scgen{\eqalign{
           P_{\a\ad} \ &=\ \partial_{\a\ad},\qquad
           Q_{i\a}   \  =\ \eta^\ad_i\partial_{\a\ad},\qquad
           Q^i_\ad   \  =\ \partial_\ad^i,\cr
           D         \ &=\ -x^{\a\ad}\partial_{\a\ad}-{\sfrac{1}{2}}
                           \eta^\ad_i\partial^i_\ad,\cr
           K^{\a\ad} \ &=\ x^{\a\bd}(x^{\b\ad}\partial_{\b\bd}+
                           \eta^\ad_i\partial^i_\bd),\cr
           K^{i\a}   \ &=\ -x^{\a\ad}\partial^i_\ad,\qquad
           K^\ad_i   \  =\ \eta^\bd_i(x^{\a\ad}\partial_{\a\bd}+
                           \eta^\ad_j\partial^j_\bd),\cr
           J_{\a\b}  \ &=\ {\sfrac{1}{2}}x^{\g\ad}\epsilon_{\g(\a}
                           \partial_{\b)\ad},\qquad
           J_{\ad\bd}\  =\ {\sfrac{1}{2}}\left(x^{\a\gd}\epsilon_{\gd(\ad}
                           \partial_{\a\bd)}+\eta^\gd_i\epsilon_{\gd(\ad}
                           \partial^i_{\bd)}\right),\cr
           T_i^j\ &=\ \eta^\ad_i\partial^j_\ad-{\sfrac{1}{\CN}}\,\d_i^j
                           \eta^\ad_k\partial_\ad^k,\qquad
           A         \  =\ {\sfrac{1}{2}}\eta^\ad_i\partial^i_\ad,\cr
}}
where parentheses mean normalized symmetrization of the enclosed indices.

Infinitesimal transformations of the components $\CA_{\a\ad}$ and $\CA^i_\ad$
of the gauge potential $\CA$ under the action of the superconformal group are 
given by
\eqn\infliea{\d_{N_a}\CA_{\a\ad}\ =\ \CL_{N_a}\CA_{\a\ad}\qquad{\rm and}\qquad
             \d_{N_a}\CA_\ad^i  \ =\ \CL_{N_a}\CA_\ad^i,}
where $N_a=N_a^{\a\ad}\partial_{\a\ad}+N^\ad_{a\,i}\partial^i_\ad$ is any 
generator of the superconformal group, and $\CL_{N_a}$ is the Lie 
superderivative along the vector field $N_a$. Explicitly, equations \infliea\ 
read as
\eqn\inflieai{\eqalign{
    \CL_{N_a}\CA_{\a\ad}\ &=\ N_a\CA_{\a\ad}+\CA_{\b\bd}\partial_{\a\ad}
                              N_a^{\b\bd}+(-)^{p_a+1}\CA^i_\bd
                              \partial_{\a\ad}N^\bd_{a\,i},\cr
    \CL_{N_a}\CA_\ad^i  \ &=\ N_a\CA^i_\ad+(-)^{p_a}\CA_{\b\bd}
                              \partial^i_\ad N_a^{\b\bd}+\CA^j_\bd
                              \partial^i_\ad N^\bd_{a\,j}.\cr
}}
It is not too difficult to show that for any generator $N_a$ as given in
\scgen, the transformations \infliea\ together with \inflieai\ give a symmetry
of the $\CN$-extended self-dual SYM equations \compcon\ and \fieldeqnfour,
respectively.

So far, we have given the action of the superconformal group on the 
components of the gauge potential $\CA$. The linear system \linsys{}, whose
compatibility conditions are the $\CN$-extended self-dual SYM equations is, 
however, defined on the supertwistor space $\CP^{3|\CN}$. Therefore, 
the question is how to define a proper action of the superconformal group on 
the supertwistor space such that it preserves the linear system \linsys{}. 

Remember that the (super)twistor space describes constant complex structures
on the (super)space $\IR^{4|2\CN}$. Thus, the action of the 
superconformal group on $\CP^{3|\CN}$ must be chosen such that it does not
change a fixed constant complex structure. Consider the body $\IR^{4|0}$ of the
anti-chiral superspace $\IR^{4|2\CN}$. Constant complex structures on 
$\IR^{4|0}$ are parametrized by the two-sphere $S^2\cong SO(4)/U(2)$. 
The two-sphere $S^2$ can be viewed as the complex projective line $\IC P^1$ 
parametrized by the coordinates $\l_\pm$. Then, the complex structure 
$\CJ=(\CJ_{\a\ad}^{~~~\b\bd})$ on $\IR^{4|0}$, compatible with \actiontauii\ 
and \actiontauiii, can be written as
\eqn\complstronr{\CJ_{\a\ad}^{~~~\b\bd}\ =\ -\i\g_\pm\d^\b_\a
                 (\l^\pm_\ad\hl_\pm^\bd+\l_\pm^\bd\hl^\pm_\ad),}
where $\l^\pm_\ad$ and  $\hl^\pm_\ad$ are as introduced in section 2 and 
$\g_\pm\equiv(1+\l_\pm{\bar\l}_\pm)^{-1}$. Using the explicit form 
\complstronr\ of the complex structure, one readily verifies that 
$\CJ_{\a\ad}^{~~~\g\gd}\CJ_{\g\gd}^{~~~\b\bd}=-\d^\b_\a\d^\bd_\ad$.\foot{We see
that in the present case the corresponding K\"ahler two-form $\o$ is 
anti-self-dual, i.e., its components are of the form 
$\o_{\a\ad\b\bd}=\epsilon_{\a\b}\CJ_{\ad\bd}$. If we had chosen the 
($\CN$-extended) anti-self-dual (super) YM equations from the very
beginning, the K\"ahler form  would have been self-dual.}
On the two-sphere $S^2$ which parametrizes the different complex 
structures of $\IR^{4|0}$, we introduce the standard complex structure 
${\frak J}$ which, for instance, on the $U_+$ patch is given by 
${\frak J}_{\l_+}^{~~\l_+}=\i=-{\frak J}_{{\bar\l}_+}^{~~{\bar\l}_+}$. Thus,
the complex structure on the body $\IR^{4|0}\times\IC P^1$ of the supertwistor
space $\CP^{3|\CN}$ can be taken as $J=(\CJ,{\frak J})$.

After introducing a complex structure on the bosonic part 
$\IR^{4|0}\times\IC P^1$ of $\IR^{4|2\CN}\times\IC P^1$, 
we need to extend the above discussion to the full supertwistor space. In order
to define a complex structure on $\IR^{4|2\CN}\times\IC P^1$, recall that only 
an even amount of supersymmetries is possible, i.e., $\CN=0,2$ or $4$. Our 
particular choice of the 
symplectic Majorana condition \actiontauiv\ allows us to introduce a complex 
structure on the fermionic part $\IR^{0|2\CN}$ similar to \complstronr, namely
\eqn\complstronri{{\bf J}_{i~\bd}^{\ad~j}\ =\ -\i\g_\pm\d^j_i
                  (\l^\pm_\bd\hl_\pm^\ad+\l_\pm^\ad\hl^\pm_\bd).}
Therefore, $J=(\CJ,{\bf J},{\frak J})$ will be the proper choice of the complex
structure\foot{Clearly, this choice of the complex structure does not exhaust  
the space of all admissible complex structures. However, in the present case it
is enough to restrict ourselves to this class of complex structures. For 
further discussions, see appendix B.} on the supertwistor
space $\CP^{3|\CN}\cong\IR^{4|2\CN}\times\IC P^1$. 

We can now answer the initial question, namely the generators $N_a$ 
given by \scgen\ of the superconformal group should be lifted to vector fields 
$\tN_a=\tN_a^{\a\ad}\partial_{\a\ad}+\tN^\ad_{a\,i}\partial^i_\ad+
\tN_a^{\l_\pm}\partial_{\l_\pm}+\tN_a^{{\bar\l}_\pm}\partial_{{\bar\l}_\pm}$ 
on the supertwistor space such that the Lie superderivative of the complex 
structure $J$ along the lifted vector fields $\tN_a$ vanishes, i.e.,
\eqn\cononcompstr{\CL_{\tN_a} J\ =\ 0.} 
Letting $\CJ_\ad^{~\bd}={\g_\pm\over 2\i}(\l^\pm_\ad\hl_\pm^\bd+
\l_\pm^\bd\hl^\pm_\ad)$, we can write \cononcompstr\ explicitly as
\eqn\cononcompstri{\eqalign{
     2\,\tN_a\CJ_\ad^{~\bd}+\CJ_\gd^{~\ad}\partial_{\a\bd}
     \tN_a^{\a\gd}-\CJ_\bd^{~\gd}\partial_{\a\gd}\tN_a^{\a\ad}\ &=\ 0,\cr
     \CJ_\dd^{~\gd}\partial_{\a\ad}\tN^\dd_{a\,i}-\CJ_\ad^{~\bd}
     \partial_{\a\bd}\tN^\gd_{a\,i}\ &=\ 0,\cr
     \CJ_\ad^{~\bd}\partial^i_\gd\tN_a^{\a\ad}-\CJ_\gd^{~\dd}\partial^i_\dd
     \tN^{\a\bd}_a\ &=\ 0,\cr     
     \d^i_j\,\tN_a\CJ_\ad^{~\bd}-(-)^{p_a}\CJ_\ad^{~\gd}\partial^j_\gd
     \tN^\bd_{a\,i}
     +(-)^{p_a}\CJ_\gd^{~\bd}\partial^j_\bd\tN^\gd_{a\,i}\ &=\ 0,\cr
}}
whereas the equations involving ${\frak J}$ tell us that the components
$\tN^{\l_\pm}_a$ and $\tN^{{\bar\l}_\pm}_a$ are holomorphic in $\l_\pm$ and
${\bar\l_\pm}$, respectively. The final expressions for the generators  \scgen\
lifted to the supertwistor space $\CP^{3|\CN}$ satisfying \cononcompstri\ are
\eqn\scgenlifted{\eqalign{
           \tP_{\a\ad}\ &=\ P_{\a\ad},\qquad
           \tQ_{i\a}\ =\  Q_{i\a},
           \qquad
           \tQ^i_\ad\ =\ Q^i_\ad,\cr
           \tD\ &=\ D,\cr
           \tK^{\a\ad}\ &=\ K^{\a\ad}+x^{\a\bd}Z_\bd^{~\ad},\qquad
           \tK^{i\a}\ =\ K^{i\a},\qquad
           \tK^\ad_i\ =\ K^\ad_i+\eta^\bd_i Z_\bd^{~\ad},\cr
           \tJ_{\a\b}\ &=\ J_{\a\b},\qquad
           \tJ_{\ad\bd}\ =\ J_{\ad\bd}-{\sfrac{1}{2}}Z_{\ad\bd},\cr
           \tT_i^j\ &=\ T_i^j,\qquad
           \tA\ =\ A,\cr
}}
where 
\eqn\scgenliftedi{Z_{\ad\bd}\ \equiv\ \l^\pm_\ad\l^\pm_\bd\partial_{\l_\pm}
                  +\hl^\pm_\ad\hl^\pm_\bd\partial_{{\bar\l}_\pm}.}
Of course, these generators fullfil the same algebra.

Now we can define the infinitesimal transformation of the $GL(n,\IC)$-valued
functions $\psi_\pm$ participating in \linsys{} under the action of the 
superconformal group
\eqn\infliepsi{\d_{\tN_a}\psi_\pm\ =\ \CL_{\tN_a}\psi_\pm\ =\ \tN_a\psi_\pm,}
where $\tN_a$ is any of the generators given in \scgenlifted. It is a 
straightforward exercise to verify explicitly that the linear system 
\linsys{} is invariant under the transformations \infliea\ and \infliepsi.

\bigskip\noindent\nobreak{\it An example of an infinite-dimensional hidden
symmetry algebra}

To jump ahead of our story a bit, consider the Abelian subalgebra
of the superconformal algebra which is spanned by the (super)translation
generators $\tP_{\a\ad}$ and $\tQ^i_\ad$. On the supertwistor space we
may use the coordinates $z^\a_\pm=x^{\a\ad}\l^\pm_\ad$, $\l^\pm_\ad$ and 
$\eta^\pm_i=\eta^\ad_i\l^\pm_\ad$. Expressing the generators
 $\tP_{\a\ad}$ and $\tQ^i_\ad$ in terms of these coordinates, we obtain
\eqn\indactp{{\widetilde P}_{\a\ad}\ =\ 
              \l^\pm_\ad{\partial\over\partial z^\a_\pm}
             \qquad{\rm and}\qquad
             {\widetilde Q}^i_{\ad}\ =\ 
             \l^\pm_\ad{\partial\over\partial \eta^\pm_i},}
when acting on holomorphic functions of $(z^\a_\pm,\l^\pm_\ad,\eta_i^\pm)$
on $\CP^{3|\CN}$.
Now, consider the vector fields ${\widetilde P}_{\a\ad_1\cdots\ad_{2s_\a}}^\pm$
and 
${\widetilde Q}^i_{\pm\ad_1\cdots\ad_{2f_i}}$ which are given by
\eqn\defofps{{\widetilde P}^\pm_{\a\ad_1\cdots\ad_{2b_\a }}\ \equiv\ 
               \l^\pm_{\ad_1}\cdots\l^\pm_{\ad_{2b_\a }}
             {\partial\over\partial z^\a_\pm}
             \qquad{\rm and}\qquad
             {\widetilde Q}^i_{\pm\ad_1\cdots\ad_{2f_i}}\ \equiv\ 
   \l^\pm_{\ad_1}\cdots\l^\pm_{\ad_{2f_i}}{\partial\over\partial\eta^\pm_i},}
where we have introduced
\eqn\defoflests{(\l^+_{\ad_n})\ \equiv \ \left(\matrix{1\cr \l_+}\right)
                \quad{\rm and}\quad 
                 (\l^-_{\ad_n})\  \equiv \ \left(\matrix{\l_-\cr 1}\right),
                \quad{\rm for}\quad
                n=\cases{1,\ldots,2b_\a& \cr
                         1,\ldots,2f_i&}\kern-.4cm.}
In these formulas, the parameters $b_\a,f_i$ are elements of 
$\{{n\over2}\,|\,n\in\IN\cup\{\infty\}\}$ whereas ``$b$'' refers to bosonic
and ``$f$'' to fermionic. Clearly, the vector fields
$\tP^\pm_{\a\ad_1\cdots\ad_{2b_\a}}$ and 
$\tQ^i_{\pm\ad_1\cdots\ad_{2f_i}}$ are totally symmetric under an 
exchange of their dotted indices. 
Using \defofps, we can define the infinitesimal transformations
\eqn\defoftrafosi{f_{+-}\ \mapsto\ \d f_{+-},
                   \qquad{\rm with}\qquad
                  \d\in\{\tP^\pm_{\a\ad_1\cdots\ad_{2b_\a }}, 
                   \tQ^i_{\pm\ad_1\cdots\ad_{2f_i}}\}}
of the transition function $f_{+-}$. 
Now we can plug $\d f_{+-}$ into \defofphi\ in order to obtain the functions 
$\phi_\pm$ which in turn give us the infinitesimal perturbations 
$\d_{\a\ad_1\cdots\ad_{2b_\a }}\CA$ and $\d^i_{\ad_1\cdots\ad_{2f_i}}\CA$ 
of the gauge potential $\CA$. We postpone the explicit construction
of the infinitesimal transformations of the gauge potential to the next 
subsection. 

Note that with any of the transformations
$$\d_{\a\ad_1\cdots\ad_{2b_\a }}f_{+-},\qquad 
  \d^i_{\ad_1\cdots\ad_{2f_i}}f_{+-},\qquad 
  \d_{\a\ad_1\cdots\ad_{2b_\a }}\CA,\qquad{\rm and}\qquad
  \d^i_{\ad_1\cdots\ad_{2f_i}}\CA$$ 
one may associate dynamical systems on the respective moduli space and try to 
solve the obtained differential equations. Interestingly, integral curves of 
dynamical systems on $\CM_{\rm hol}(\CP^{3|\CN})$ can be described 
explicitly.\foot{In the next section, we will also describe a corresponding 
dynamical system on the moduli space $\CM_{\rm SDYM}^\CN$. Note again that
strictly speaking we should say on the solution space.} Namely, consider the 
following system of differential equations
\eqn\dynsystrans{\eqalign{
                 {\partial\over\partial t^{\a\ad_1\cdots\ad_{2b_\a }}}f_{+-}\ 
                 &=\ \l^\pm_{\ad_1}\cdots\l^\pm_{\ad_{2b_\a }}
                 {\partial\over\partial z^\a_\pm}f_{+-},\cr
                 {\partial\over\partial\xi^{\ad_1\cdots\ad_{2f_i}}_i}f_{+-}\ 
                 &=\ \l^\pm_{\ad_1}\cdots\l^\pm_{\ad_{2f_i}}
                 {\partial\over\partial \eta^\pm_i}f_{+-},
}}
where $t^{\a\ad_1\cdots\ad_{2b_\a }}$ and $\xi^{\ad_1\cdots\ad_{2f_i}}_i$ are 
parameters. These equations can easily be
solved. The solution to \dynsystrans\ reads as
\eqn\dynsystransi{f_{+-}\ =\ f_{+-}(z^\a_\pm+ t^{\a\ad_1\cdots\ad_{2b_\a }}
                  \l^\pm_{\ad_1}\cdots\l^\pm_{\ad_{2b_\a }},\l^\pm_\ad,
                  \eta^i_\pm+\xi^{\ad_1\cdots\ad_{2f_i}}_i
                  \l^\pm_{\ad_1}\cdots\l^\pm_{\ad_{2f_i}}).}
Note that any point of $\IR^{4|2\CN}$ can be obtained by a shift of the
origin and hence we may put, without loss of generality, $z^\a_\pm$ and 
$\eta^i_\pm$ to zero. Therefore, \dynsystransi\ simplifies to
\eqn\dynsystransii{f_{+-}\ =\ f_{+-}(t^{\a\ad_1\cdots\ad_{2b_\a }}
                  \l^\pm_{\ad_1}\cdots\l^\pm_{\ad_{2b_\a }},\l^\pm_\ad,
                  \xi^{\ad_1\cdots\ad_{2f_i}}_i
                  \l^\pm_{\ad_1}\cdots\l^\pm_{\ad_{2f_i}}),}
where now 
$$(t^{\a{\dot1}\cdots{\dot1}},t^{\a{\dot2}{\dot 1}\cdots{\dot1}},
\xi^{{\dot1}\cdots{\dot1}}_i,\xi^{{\dot2}{\dot 1}\cdots{\dot1}}_i)$$ 
are interpreted as coordinates on the anti-chiral superspace $\IR^{4|2\CN}$
whereas the others are additional moduli. 

Finally, we remark that for finite sums in \dynsystransii,  
$b_\a,f_i<\infty$, the polynomials
\eqn\defofpol{z^\a_\pm\ =\ t^{\a\ad_1\cdots\ad_{2b_\a }}
              \l^\pm_{\ad_1}\cdots\l^\pm_{\ad_{2b_\a }}\qquad{\rm and}\qquad
              \eta_i^\pm\ =\  \xi^{\ad_1\cdots\ad_{2f_i}}_i
              \l^\pm_{\ad_1}\cdots\l^\pm_{\ad_{2f_i}}}
can be regarded as holomorphic sections of the bundle
$$\CO(2b_1)\oplus\CO(2b_2)\oplus\bigoplus_{i=1}^\CN
  \Pi\CO(2f_i)\ \to\ \IC P^1.$$
We shall call this space -- for reasons which become clear below -- 
{\it enhanced supertwistor space} and denote it by 
$\CP^{3|\CN}[2b_1,2b_2|2f_1,\ldots,2f_\CN]$. The transition functions 
\dynsystransii\ can then be interpreted as the transition functions of
holomorphic vector bundles $\CE\to\CP^{3|\CN}[2b_1,2b_2|2f_1,\ldots,2f_\CN]$.

%--------------------------------------------------------------------------
\subsec{Affine extensions of the superconformal algebra}

Similar to our discussion presented in subsection 3.2, we will now focus on
the construction of affine extensions of the superconformal algebra. Note that 
as $\CN=3\ (4)$ SYM theory in four dimensions is related via the Penrose-Ward 
transform\foot{See, e.g., reference \PopovRB\ for details.} to hCS theory 
on the superambitwistor space, our discussion translates, as already 
indicated, directly to the twistor construction of hidden symmetry algebras for
the full $\CN=4$ SYM theory. 

In the previous subsection, we have explained how the generators of the 
superconformal group need to be lifted to the supertwistor space, such that
the complex structure is not deformed. Hence, we would like to find an affine 
extension of the superconformal algebra which does not change the complex 
structure. To do this, we remember that the components $\tN_a^{\l_\pm}$ and 
$\tN_a^{{\bar\l}_\pm}$ of the generators $\tN_a$ are holomorphic functions in 
$\l_\pm$ and ${\bar\l}_\pm$, respectively. Therefore, a natural choice of an 
extension is given by
\eqn\affineEI{\tN_a^m\ \equiv\ \l_+^m\tN_a^{\a\ad}\partial_{\a\ad}+
        \l_+^m\tN^\ad_{a\,i}\partial^i_\ad+\l_+^m\tN_a^{\l_+}\partial_{\l_+}+
        {\bar\l}_+^m\tN_a^{{\bar\l}_+}\partial_{{\bar\l}_+},
        \quad{\rm for}\quad m\in\IZ}
on $\CU_+\cap\CU_-\subset\CP^{3|\CN}$. One may readily check that the Lie 
superderivative of the complex structure on $\CP^{3|\CN}$ along any $\tN_a^m$ 
vanishes, thus preserving it. The Lie superbracket of two such vector fields 
computes to
\eqn\affineEII{[\tN_a^m,\tN_b^n\}\ =\ \left(f_{ab}^{\ \ c}+n\,g_a\,\d_b^c-
               (-)^{p_ap_b}m\,g_b\,\d^c_a\right)\tN_c^{m+n}
               +K_{ab}^{mn}\,\partial_{{\bar\l}_+},}
where the $f_{ab}^{\ \ c}$ are the structure constants of the superconformal
algebra (see appendix C for details) and 
\eqn\affineEIII{g_a\ \equiv\ \l_+^{-1}\tN^{\l_+}_a\qquad{\rm and}\qquad
                {\bar g}_a\ \equiv\ {\bar\l}_+^{-1}\tN^{{\bar\l}_+}_a}
as well as
\eqn\affineEIV{\eqalign{
     K_{ab}^{mn}\ &\equiv\ \left(n({\bar\l}_+^{m+n}{\bar g}_a-
     \l^{m+n}_+g_a)\tN_b^{{\bar\l}_+}-(-)^{p_ap_b}m({\bar\l}_+^{m+n}{\bar g}_b-
     \l^{m+n}_+g_b)\tN_a^{{\bar\l}_+}\ +\right.\cr
     &\kern1cm \left.+\ (\l_+^m{\bar\l}_+^n-{\bar\l}_+^{m+n} )\tN_a
     (\tN_b^{{\bar\l}_+})-(-)^{p_ap_b}(\l_+^n{\bar\l}_+^m-{\bar\l}_+^{m+n} )
     \tN_b(\tN_a^{{\bar\l}_+})\right).\cr
}}
Therefore, we define the following perturbation of the transition function
$f_{+-}$
\eqn\affineEV{\d^m_af_{+-}\ \equiv\ \tN^m_a f_{+-}}
and hence upon action on $f_{+-}$, \affineEII\ reduces to
\eqn\affineEVI{[\d_a^m,\d_b^n\}\ =\ \left(f_{ab}^{\ \ c}+n\,g_a\,\d_b^c-
               (-)^{p_ap_b}m\,g_b\,\d^c_a\right)\d_c^{m+n}.}
In particular, if one considers the maximal subalgebra ${\frak h}$ of the
superconformal algebra which consists only of those generators $\tN_a$ that
do not contain terms proportional to $Z_{\ad\bd}$ given by \scgenliftedi,
the algebra \affineEVI\ simplifies further to
\eqn\affineEVII{[\d_a^m,\d_b^n\}\ =\ h_{ab}^{\ \ c}\d_c^{m+n},}
where $h_{ab}^{\ \ c}$ are the structure constants of ${\frak h}$. Therefore,
we obtain the centerless super Kac-Moody algebra 
${\frak h}\otimes\IC[[\l,\l^{-1}]]$. Moreover, we have
\eqn\affineEVIII{[\d^m_a,\d^n_a\}\ =\ -g_a(m-n)\,\d^{m+n}_a,}
which resembles for $a=({\dot1}{\dot2})$, i.e., 
$\tN_{{\dot1}{\dot2}}=\tJ_{{\dot1}{\dot2}}$, a centerless Virasoro algebra.
In general, the algebra \affineEVI\ can be seen as a super Kac-Moody-Virasoro 
type algebra.\foot{At this point we stress that for the generators 
$\tK^{\a\ad}$ and $\tK^\ad_i$ of special conformal transformations, the 
corresponding functions $g_a$ also depend on $x^{\a\ad}$ and $\eta^\ad_i$ 
(besides $\l_\pm$). This will eventually lead to the
exclusion of the corresponding generators $\d_a^m$ for $m>0$ from the 
representation of the symmetry algebra on the solution space of the 
$\CN$-extended self-dual SYM equations; see below.}

The next step is to construct the corresponding action of the $\d^m_a$ on the
components of the gauge potential and to compute their supercommutator. 
First, we restrict to the case when $m\geq0$. We have already given the 
transformation laws of the components of the gauge potential in subsection 3.2,
i.e., 
we can take the transformations \perturbationAKMYi. This time, however, the
$\phi^0_{\pm a}$ are given by
\eqn\affineEIX{\phi^0_{\pm a}\ =\ -(\tN_a\psi_\pm)\psi_\pm^{-1}.}
Again, the coefficients $\phi^{m(0)}_{+a}$ are identically zero when $m$ is 
strictly positive. In order to proceed as in subsection 3.2, it is tempting to
assume that $\phi^{0(0)}_{+a}$ vanishes, as well, 
such that only $\CA_{\a{\dot1}}$ 
and $\CA_{\dot1}^i$ get transformed. However, this is not possible as $\d^0_a$ 
represents by construction superconformal transformations\foot{Explicitly, we 
have $\d^0_a\CA_{\a\ad}=\CL_{N_a}\CA_{\a\ad}$ and similarly for $\CA^i_\ad$;
cf. subsection 3.3.}, and the assumption $\phi^{0(0)}_{+a}=0$ certainly 
implies that $\d^0_a\CA_{\a{\dot2}}=0=\d^0_a\CA_{\dot2}^i$ in contradiction to
the former statement. The best we can do is to consider instead the 
transformations
\eqn\affineEX{\eqalign{
    \td^m_a\CA_{\a{\dot1}}\ &\equiv\ \d^m_a\CA_{\a{\dot1}}-
    \nabla_{\a{\dot1}}\phi^{m(0)}_{+a}\  =\ -\nabla_{\a{\dot2}}\phi^{m(1)}_{+a}
    \ =\ \nabla_{\a{\dot1}}(\phi^{m(0)}_{-a}-\phi^{m(0)}_{+a}),\cr
    \td^m_a\CA_{\a{\dot2}}\ &\equiv\ \d^m_a\CA_{\a{\dot2}}-
    \nabla_{\a{\dot2}}\phi^{m(0)}_{+a}\ =\ 0\cr
}}
and similarly for $\CA^i_\ad$. Clearly, for $m>0$ we have $\td^m_a=\d^m_a$.
For $m=0$, the transformations \affineEX\ can be interpreted as a 
superconformal transformation accompanied by a gauge transformation mediated by
the function $\phi^{0(0)}_{+a}$. We now follow subsection 3.2 and assume 
without loss of generality
that the power series expansion of $\psi_+$ is given according to \sac. Thus, 
the components $\CA_{\a{\dot2}}$ and $\CA^i_{\dot2}$ are put to zero and we 
again work in Leznov gauge. Therefore, due to \affineEIX\ the coefficient 
$\phi^{0(0)}_{+a}$ is identically zero only for the generators of the 
superconformal group for which $\tN_a=N_a$. For those generators we also have
$$\CL_{N_a}\CA_{\a{\dot2}}\Big|_{(\CA_{\a{\dot2}},\CA_{\dot2}^i)=(0,0)}
  \ =\ 0\ =\ 
  \CL_{N_a}\CA_{\dot2}^i\Big|_{(\CA_{\a{\dot2}},\CA_{\dot2}^i)=(0,0)}.$$ 
We stress that by construction the transformations \affineEX, when written 
in Leznov gauge, do preserve this gauge.\foot{Note that the only nonvanishing
Lie superderivatives (in Leznov gauge) are
$$\eqalign{
    \CL_{K^{\a{\dot2}}}\CA_{\b{\dot2}}\ =\ \CA_{\b{\dot1}}x^{\a{\dot1}},\qquad
    \CL_{K^{\dot2}_i}\CA_{\a{\dot2}}\ &=\ \CA_{\a{\dot1}}\eta^{\dot1}_i,\qquad
    \CL_{J_{{\dot1}{\dot1}}}\CA_{\b{\dot2}}\ =\ \sfrac{1}{2}\CA_{\a{\dot1}},\cr
    \CL_{K^{\a{\dot2}}}\CA^i_{\dot2}\ =\ \CA_{\dot1}^ix^{\a{\dot1}},\qquad
    \CL_{K^{\dot2}_i}\CA_{\dot2}^j\ &=\ -\CA_{\dot1}^j\eta^{\dot1}_i,\qquad
    \CL_{J_{{\dot1}{\dot1}}}\CA^i_{\dot2}\ =\ \sfrac{1}{2}\CA_{\dot1}^i.\cr
}$$
Computing  the expressions 
$\partial_{\a{\dot2}}\phi_{+a}^{0(0)}$ and 
$\partial^i_{\dot2}\phi_{+a}^{0(0)}$ for those generators, one realizes that
$\td^m_a\CA_{\a{\dot2}}=0=\td^m_a\CA^i_{\dot2}$ is indeed true.}

Next, we need to compute the commutator $[\td_1,\td_2]$ of two successive 
transformations. 
We do this computation in two steps: first, we consider the case where $m,n>0$
and second the case where $m>0$ and $n=0$.\foot{The case $m=n=0$ is obvious.} 

Assume that $m,n>0$. In this case, we can use the equations \SConAI\ together 
with \SConAII. We simply need to replace $\d^m_a$ by $\td^m_a$. Remember that
for $m>0$ we have
$\td^m_a\CA_{\a{\dot1}}=\nabla_{\a{\dot1}}\phi_{-a}^{0(m)}$ and
$\td^m_a\CA^i_{\dot1}=(-)^{p_a}\nabla^i_{\dot1}\phi_{-a}^{0(m)}$. 
Therefore, similar to \SConAIV{} the expression $\td^m_a\phi^0_{-b}$ is given 
by
\eqn\affineEXI{\td^m_a\phi^0_{-b}\ =\ (-)^{p_ap_b}
               [\tN_b+\phi^0_{-b},\phi^m_{-a}\}\ =\ (-)^{p_ap_b}
               \left(\tN_b\phi^m_{-a}+[\phi^0_{-b},\phi^m_{-a}\}\right),}
which directly follows from $\phi^0_{-b}=-(\tN_b\psi_-)\psi_-^{-1}$, and hence
\eqna\affineEXII
$$\eqalignno{\td^m_a\phi^{0(n)}_{-b}\ &=\ 
           (-)^{p_ap_b}\oint_c{{\rm d}\l_+\over 2\pi\i}\,\l_+^{n-1}\,
           \left(\tN_b\phi^m_{-a}+[\phi^0_{-b},\phi^m_{-a}\}\right)&{}\cr
           &=\ (-)^{p_ap_b}\left(N_b\phi^{0(m+n)}_{-a}-\sum_{k=-1}^1\ (n+k)
           \,g_b^{(k)}\phi_{-a}^{0(m+n+k)}\ +\right.&{}\cr
            &\kern3cm\left.+\ \sum_{k=0}^n\ [\phi^{0(k)}_{-b},
            \phi^{0(m+n-k)}_{-a}\}\right),& \affineEXII{}\cr
}$$
where the coefficients $g_a^{(k)}$ are those of the function $g_a$ defined in
\affineEIII\ when expanded in powers of $\l_+$. Moreover, we have
$$\eqalign{N_b\phi_{-a}^{0(m+n)}
          \ &=\ -\oint_c{{\rm d}\l_+\over 2\pi\i}\,\l_+^{m+n-1}\,
          N_b((\tN_a\psi_-)\psi_-^{-1})\cr
          \ &=\ -\oint_c{{\rm d}\l_+\over 2\pi\i}\,\l_+^{m+n-1}\,\left(
          \tN_b((\tN_a\psi_-)\psi_-^{-1})-
          \tN_b^{\l_+}\partial_{\l_+}((\tN_a\psi_-)\psi_-^{-1})\right)\cr
          \ &=\ -\oint_c{{\rm d}\l_+\over 2\pi\i}\,\l_+^{m+n-1}\,
          (\tN_b\tN_a\psi_-)\psi_-^{-1}+(-)^{p_ap_b}\sum_{k=0}^{m+n}\ 
          \phi^{0(k)}_{-a}\phi^{0(m+n-k)}_{-b}\ +\cr
          &\kern3cm+\ \sum_{k=-1}^1\ (m+n+k)\,g_b^{(k)}\phi_{-a}^{0(m+n+k)}.\cr
}$$
Putting everything together, we obtain
\eqn\affineEXIII{\eqalign{
    \td^m_a\phi^{0(n)}_{-b}-(-)^{p_ap_b}\td^n_b\phi^{0(m)}_{-a}\
    &=\ -f_{ab}^{\ \ c}\phi_{-c}^{0(m+n)}-[\phi_{-a}^{0(m)},\phi_{-b}^{0(n)}\}
    \ -\cr
    &\kern-1cm-\ \sum_{k=-1}^1\ \left(n\,g_a^{(k)}\d_b^c
     -(-)^{p_ap_b}m\,g_b^{(k)}\d_a^c\right)\phi^{0(m+n+k)}_{-c}.\cr
}}
Therefore, by virtue of the transformation laws \affineEX, we conclude
that the algebra does only close, when all the $g_a^{(k)}$ are constants. 
From \scgenlifted\ we realize that we need to exclude the generators 
$\tK^{\a\ad}$ and $\tK_i^\ad$. In the following, let ${\frak h}$ be the 
maximal subalgebra of the superconformal algebra which does not contain
$\tK^{\a\ad}$ and $\tK_i^\ad$.\foot{See also appendix C.} Hence, equation 
\affineEXIII\ implies
\eqn\affineEXIV{[\td^m_a,\td^n_b\}\ =\ h_{ab}^{\ \ c}\td_c^{m+n}
             +\sum_{k=-1}^1\ \left(n\,g_a^{(k)}\d_b^c
             -(-)^{p_ap_b}m\,g_b^{(k)}\d_a^c\right)\td^{m+n+k}_c,}
for $m,n>0$. In \affineEXIV, the $h_{ab}^{\ \ c}$ are the structure constants
of ${\frak h}$.

It remains to compute $[\td^m_a,\td^0_b\}$ for $m>0$. A straightforward 
calculation shows that the function $\Sigma^{m0}_{ab}$ entering the formulas 
\SConAI\ is given by
\eqn\affineEXV{\eqalign{\Sigma^{m0}_{ab}\ &=\  
            -\td^m_a(\phi^{0(0)}_{-b}-\phi^{0(0)}_{+b})
            +(-)^{p_ap_b}\td^0_b\phi^{0(m)}_{-a}-
            [\phi^{0(m)}_{-a},\phi^{0(0)}_{-b}-\phi^{0(0)}_{+b}\}\cr
            &=\ -\td^m_a\phi^{0(0)}_{-b}
            +(-)^{p_ap_b}\td^0_b\phi^{0(m)}_{-a}-
            [\phi^{0(m)}_{-a},\phi^{0(0)}_{-b}\}
            +[\phi^{0(m)}_{-a},\phi^{0(0)}_{+b}\}+\td^m_a\phi^{0(0)}_{+b}.
}}
Recall that for $m=0$ the transformation $\td^0_b$ consists of a superconformal
transformation accompanied by a gauge transformation mediated by the function
$\phi_{+b}^{0(0)}$, that is, $\td^0_b=\d^0_b+\d^g_b$. Therefore, from
varying $\phi_{-a}^0=-(\tN_a\psi_-)\psi_-^{-1}$, we obtain\foot{Generally, we 
could have written $\td^n_b\phi^0_{-a}=(-)^{p_ap_b}
[\tN_a+\phi^0_{-a},\phi^n_{-b}-\phi^{n(0)}_{+b}\}$ from the very beginning, 
since for $n>0$ we have $\phi^{n(0)}_{+b}=0$.}
\eqn\affineEXVI{\td^0_b\phi^0_{-a}\ =\ (-)^{p_ap_b}
               [\tN_a+\phi^0_{-a},\phi^0_{-b}-\phi^{0(0)}_{+b}\}.}
Thus, the sum over the first four terms of \affineEXV\ gives
\eqna\affineEXVII
$$\eqalignno{
            \td^m_a\phi^{0(0)}_{-b}-(-)^{p_ap_b}\td^0_b\phi^{0(m)}_{-a}+
            [\phi^{0(m)}_{-a},\phi^{0(0)}_{-b}-\phi^{0(0)}_{+b}\}\ &=&{}\cr 
            &\kern-7cm=\ -f_{ab}^{\ \ c}\phi_{-c}^{0(m)}+(-)^{p_ap_b}
            \sum_{k=-1}^1\ m\,g_b^{(k)}\phi^{0(m+k)}_{-a}-(-)^{p_ap_b}
            g_b^{(-1)}\phi^{0(m-1)}_{-a}.&\affineEXVII{}\cr
}$$
Furthermore, in a similar manner we find
\eqna\affineEXVIII
$$\eqalignno{\td^m_a\phi^{0(0)}_{+b}\ &=\ 
           (-)^{p_ap_b}\oint_c{{\rm d}\l_+\over 2\pi\i}\,\l_+^{-1}\,
           \left(\tN_b\phi^m_{+a}+[\phi^0_{+b},\phi^m_{+a}\}\right)&{}\cr
           &=\ (-)^{p_ap_b}g_b^{(-1)}\phi_{+a}^{m(1)}\ =\ g_b^{(-1)}
           (-)^{p_ap_b}\left(
           \phi_{+a}^{0(1-m)}-\phi_{-a}^{0(m-1)}\right),& \affineEXVIII{}\cr
}$$
where in the last step we have used \phiexpKMY. Therefore, we end up with
\eqn\affineEXVIX{\Sigma^{m0}_{ab}
                  \ =\ f_{ab}^{\ \ c}\phi_{-c}^{0(m)}-
                  (-)^{p_ap_b}\sum_{k=-1}^1\ m\,g_b^{(k)}
                  \phi^{0(m+k)}_{-a}+
                  (-)^{p_ap_b}g_b^{(-1)}\phi^{0(1-m)}_{+a}
}
for $m>0$.
Assuming again that the coefficients $g_a^{(k)}$ are independent of $x^{\a\ad}$
and $\eta^\ad_i$, we arrive at
\eqn\affineEXX{[\td^m_a,\td^0_b\}\CA_{\a{\dot1}}\ =\ h_{ab}^{\ \ c}\td_c^m
    \CA_{\a{\dot1}}-(-)^{p_ap_b}\sum_{k=-1}^1\ m\,g_b^{(k)}\td^{m+k}_a
    \CA_{\a{\dot1}}}
and similarly for $\CA^i_{\dot1}$. 

In summary, we have obtained the algebra
\eqn\affineEXXXI{[\td^m_a,\td^n_b\}\ =\ h_{ab}^{\ \ c}\td_c^{m+n}
             +\sum_{k=-1}^1\ \left(n\,g_a^{(k)}\d_b^c
             -(-)^{p_ap_b}m\,g_b^{(k)}\d_a^c\right)\td^{m+n+k}_c}
for $m,n\geq0$. The computations for negative $m,n$ go along similar lines and
one eventually finds (as in subsection 3.2) the algebra \affineEXXXI\
for all $m,n\in\IZ$.

\break\bigskip\noindent{\it Some remarks}

At first sight it seems a little surprising that we had to
exclude the generators of special conformal transformations $\tK^{\a\ad}$ and
$\tK^\ad_i$ and hence to restrict to the subalgebra ${\frak h}$. However, 
remember that we work on the open subset 
$\CP^{3|\CN}=\IC P^{3|\CN}\setminus\IC P^{1|\CN}$ of $\IC P^{3|\CN}$. The 
latter space is the compactification of $\CP^{3|\CN}$ which corresponds via
the supertwistor correspondence to compactified superspacetime, i.e., 
one relates holomorphic vector bundles over $\IC P^{3|\CN}$ via the
Penrose-Ward transform to the $\CN$-extended self-dual SYM theory on 
compactified spacetime.\foot{For Euclidean signature one has $S^4$, for
Minkowski $S^3\times S^1$ and for Kleinian $S^2\times S^2$, respectively.}
Recall that $\CP^{3|\CN}$ is covered by two coordinate 
patches, while the space $\IC P^{3|\CN}$ by four. Special conformal 
transformations relate different coordinate patches and hence by removing the
subspace $\IC P^{1|\CN}$ one does not have as much freedom as one has for 
$\IC P^{3|\CN}$. Moreover, special conformal transformations transform $\l_\pm$
into a function which depends on $x^{\a\ad}$, $\eta^\ad_i$ and $\l_\pm$. 
Thus, they do not preserve the fibration $\CP^{3|\CN}\to\IC P^1$.
Therefore, in order to extend the full superconformal symmetry
algebra  one should rather
work on the compactified supertwistor space
$\IC P^{3|\CN}$. Then, however, the linear system \linsys{} needs
to be changed to incorporate the Levi-Civita superconnection which is
induced from compactified superspacetime.

%--------------------------------------------------------------------------
%--------------------------------------------------------------------------
\newsec{Holomorphy and self-dual super Yang-Mills hierarchies}

In the previous section, we have introduced infinitesimal symmetries which were
related with translational symmetries on  
$\IR^{4|2\CN}$. Moreover, we discussed the dynamical system \dynsystrans. 
By solving this system, we discovered the space
\eqn\defofenhtwistor{\CP^{3|\CN}[2b_1,2b_2|2f_1,\ldots,2f_\CN]\ \equiv\    
         \CO(2b_1)\oplus \CO(2b_2)\oplus\bigoplus_{i=1}^\CN\Pi\CO(2f_i)
         \ \to\ \IC P^1,}
which we called the enhanced supertwistor space. Note that this space
can be viewed as an open subset of the weighted projective space
$$W\IC P^{3|\CN}[2b_1,2b_2,1,1|2f_1,\ldots,2f_\CN]$$ by removing the subspace 
$W\IC P^{1|\CN}[2b_1,2b_2|2f_1,\ldots,2f_\CN]$.\foot{For related discussions on
weighted projective spaces in the context of the supertwistor correspondence 
see references \refs{\WittenNN,\PopovNK,\AhnXS}.} The goal of this section is 
to explore the space $\CP^{3|\CN}[2b_1,2b_2|2f_1,\ldots,2f_\CN]$ in more detail
and furthermore to discuss holomorphic vector bundles over it. We will 
eventually obtain the {\it $\CN$-extended self-dual SYM hierarchy}.
We remark that for $b_\a=f_i={1\over2}$, the subsequent discussion reduces, of
course, to the one presented in section 2. If no confusion arises, we will
write, for brevity, $\CP^{3|\CN}_{b,f}$ instead of 
$\CP^{3|\CN}[2b_1,2b_2|2f_1,\ldots,2f_\CN]$.

%--------------------------------------------------------------------------
\subsec{Enhanced supertwistor space}

In the following, we again work in the complex setup. As before, one can impose
proper reality conditions afterwards (cf. subsection 2.3).

Obviously, since the enhanced supertwistor space $\CP^{3|\CN}_{b,f}$ is
fibered over the Riemann sphere\foot{That is, it is a vector bundle over
$\IC P^1$.}, it can be covered by two coordinate patches 
which we denote -- as before -- by ${\frak U}=\{\CU_+,\CU_-\}$. Moreover, 
we introduce the following coordinates on $\CU_\pm$:
\eqn\coordsonets{(z_\pm^\a,\l_\pm,\eta_i^\pm),\quad{\rm with}\quad
                 z_+^\a\ =\ \l_+^{2b_\a}\,z_-^\a,
                 \quad\l_+\ =\ \l_+^2\l_-
                 \quad{\rm and}\quad
                 \eta_i^+\ =\ \l_+^{2f_i}\,\eta_i^-.}
According to the preceding discussion, we are interested in holomorphic 
sections of the bundle \defofenhtwistor. Contrary to the supertwistor space,
these are rational curves of degree $(2b_1,2b_2;2f_1,\ldots,2f_\CN)$, 
$\IC P^1_{x,\eta}\hookrightarrow\CP^{3|\CN}_{b,f}$, 
which are given by the expressions
\eqn\ratcurvestwos{z_\pm^\a\ =\ x^{\a\ad_1\cdots\ad_{2b_\a}}
                  \l^\pm_{\ad_1}\cdots\l^\pm_{\ad_{2b_\a}}
                  \quad{\rm and}\quad
                  \eta^\pm_i\ =\ \eta_i^{\ad_1\cdots\ad_{2f_i}}
                   \l^\pm_{\ad_1}\cdots\l^\pm_{\ad_{2f_i}}
                   \quad{\rm on}\quad
                  \CU_\pm,}
and parametrized by the supermoduli 
$$(x,\eta)\ =\ (x^{\a\ad_1\cdots\ad_{2b_\a}},\eta_i^{\ad_1\cdots\ad_{2f_i}})\
              \in\ \IC^{2(b_1+b_2+1)|2(f_1+\cdots+f_\CN)+\CN}.$$
Therefore, we have enhanced the moduli space 
$\IC^{4|2\CN}$ of rational curves of degree one\foot{More precisely, we should 
write of degree $(1,1;1,\ldots,1)$.} living in the supertwistor space to the 
moduli space $\IC^{2(b_1+b_2+1)|2(f_1+\cdots+f_\CN)+\CN}$ of rational degree 
$(2b_1,2b_2;2f_1,\ldots,2f_\CN)$ curves sitting inside the enhanced 
supertwistor space. We remark that for the choices for which at most one of the
dotted indices is equal to two, the coordinates 
$(x^{\a\ad_1\cdots\ad_{2b_\a}}, \eta_i^{\ad_1\cdots\ad_{2f_i}})$ might be 
interpreted -- up to some unimportant prefactors entering through 
permutations of their dotted indices -- as coordinates on the
anti-chiral superspace $\IC^{4|2\CN}$.\foot{Cf. subsection 3.2.} 
Therefore, the complexified (super)spacetime is naturally embedded in the space
$\IC^{2(b_1+b_2+1)|2(f_1+\cdots+f_\CN)+\CN}$ and we may write
$$ \IC^{2(b_1+b_2+1)|2(f_1+\cdots+f_\CN)+\CN}\ \cong\ \IC^{4|2\CN}\oplus 
   \IC^{2(b_1+b_2-1)|2(f_1+\cdots+f_\CN)-\CN}.
$$
Altogether, the equations \ratcurvestwos\ allow us to introduce  
the double fibration\foot{Note that in the real setup this double fibration
does not reduce to a single fibration like \doublefibrationi.}
\eqn\doublefibrationcsdh{\CP^{3|\CN}_{b,f} 
                         ~~~\underleftarrow{\pi_2}~~~
                        \CF^{2(b_1+b_2)+3|2(f_1+\cdots+f_\CN)+\CN}
                         ~~~\underrightarrow{\pi_1}~~~
                        \IC^{2(b_1+b_2+1)|2(f_1+\cdots+f_\CN)+\CN},}
where now the correspondence space is given by the direct product
$$\CF^{2(b_1+b_2)+3|2(f_1+\cdots+f_\CN)+\CN}\ =\ 
  \IC^{2(b_1+b_2+1)|2(f_1+\cdots+f_\CN)+\CN}\times\IC P^1.$$
Furthermore, the natural choice of coordinates on 
$\CF^{2(b_1+b_2)+3|2(f_1+\cdots+f_\CN)+\CN}$ reads as
\eqn\defofcoordsonets{(x^{\a\ad_1\cdots\ad_{2b_\a}},\l^\pm_\ad,
                      \eta_i^{\ad_1\cdots\ad_{2f_i}})}
and hence the action of projections $\pi_1$ and $\pi_2$ in the fibration
\doublefibrationcsdh\ is 
\eqn\defofpisdh{\eqalign{
             &\pi_1\,:\,(x^{\a\ad_1\cdots\ad_{2b_\a}},\l^\pm_\ad, 
                       \eta_i^{\ad_1\cdots\ad_{2f_i}})\to
             (x^{\a\ad_1\cdots\ad_{2b_\a}}, \eta_i^{\ad_1\cdots\ad_{2f_i}}),\cr
              &\pi_2\,:\,(x^{\a\ad_1\cdots\ad_{2b_\a}},\l^\pm_\ad, 
                \eta_i^{\ad_1\cdots\ad_{2f_i}})\to
             (x^{\a\ad_1\cdots\ad_{2b_\a}}\l^\pm_{\ad_1}
                      \cdots\l^\pm_{\ad_{2b_\a}},\l^\pm_\ad, 
               \eta_i^{\ad_1\cdots\ad_{2f_i}}\l^\pm_{\ad_1}
                      \cdots\l^\pm_{\ad_{2f_i}}),\cr
}}
where
\eqn\defoflests{(\l^+_{\ad_n})\ \equiv \ \left(\matrix{1\cr \l_+}\right)
                \quad{\rm and}\quad 
                 (\l^-_{\ad_n})\  \equiv \ \left(\matrix{\l_-\cr 1}\right),
                \quad{\rm for}\quad
                n=\cases{1,\ldots,2b_\a&\cr
                         1,\ldots,2f_i&\cr}\kern-.4cm,}
as before.

In summary, we learn that a point in 
$\IC^{2(b_1+b_2+1)|2(f_1+\cdots+f_\CN)+\CN}$ corresponds to a projective line 
$\IC P^1_{x,\eta}$ in $\CP^{3|\CN}_{b,f}$ given by solutions to \ratcurvestwos\
for fixed $(x^{\a\ad_1\cdots\ad_{2b_\a}},\eta_i^{\ad_1\cdots\ad_{2f_i}})$.
Conversely, a point in $\CP^{3|\CN}_{b,f}$ corresponds to an affine 
subspace (strictly speaking an affine $\beta$-subspace) in 
$\IC^{2(b_1+b_2+1)|2(f_1+\cdots+f_\CN)+\CN}$ of dimension 
$(2(b_1+b_2)|2(f_1+\cdots+f_\CN))$ which is defined by solutions to 
\ratcurvestwos\ for fixed $(z^\a_\pm,\l_\pm,\eta^\pm_i)$.

%--------------------------------------------------------------------------
\subsec{Self-dual super Yang-Mills hierarchies}

In subsection 4.1, we have introduced the enhanced supertwistor space 
$\CP^{3|\CN}_{b,f}$
and the double fibration \doublefibrationcsdh. Now it is natural to consider
holomorphic vector bundles over $\CP^{3|\CN}_{b,f}$ and to ask for the 
output of the Penrose-Ward transform.
Eventually, we will obtain the promised $\CN$-extended self-dual SYM 
hierarchies.

Let $\CE$ be a holomorphic vector bundle over  $\CP^{3|\CN}_{b,f}$ and
$\pi_2^*\CE$ the pull-back of $\CE$ to the correspondence space 
$\CF^{2(b_1+b_2)+3|2(f_1+\cdots+f_\CN)+\CN}$. The covering of the latter is 
denoted by ${\tilde{\frak U}}=\{{\tilde\CU}_+,{\tilde\CU}_-\}$.
These bundles are defined by transition functions\foot{Here, we again use 
the same letter $f$ for both bundles.} $f=\{f_{+-}\}$ which are annihilated
by the vector fields
\eqn\defofvfsdh{\eqalign{
                D^\pm_{\a\ad_1\cdots\ad_{2b_\a-1}}\ &=\ 
                \l^\ad_\pm\partial_{\a\ad\ad_1\cdots\ad_{2b_\a-1}},\qquad
                D^\pm_3\ =\ \partial_{{\bar\l}_\pm},\cr
                & D^i_{\pm\ad_1\cdots\ad_{2f_i-1}}
                 \ =\ \l^\ad_\pm\partial^i_{\ad\ad_1\cdots\ad_{2f_i-1}}.
}} 
In these expressions, we have introduced the abbreviations
$$ \partial_{\a\ad_1\cdots\ad_{2b_\a}}\ \equiv\ {\partial\over\partial
x^{\a\ad_1\cdots\ad_{2b_\a}}}\qquad{\rm and}\qquad
\partial^i_{\ad_1\cdots\ad_{2f_i}}\ \equiv\ 
{\partial\over\partial\eta^{\ad_1\cdots\ad_{2f_i}}_i}.$$
We note that the vector fields $D^\pm_{\a\ad_1\cdots\ad_{2b_\a-1}}$ and 
$D^i_{\pm\ad_1\cdots\ad_{2f_i-1}}$
in \defofvfsdh\ form a basis of the tangent spaces
of the $(2(b_1+b_2)|2(f_1+\cdots+f_\CN))$-dimensional leaves of the fibration 
$\pi_2\,:\,\CF^{2(b_1+b_2)+3|2(f_1+\cdots+f_\CN)+\CN}\to\CP^{3|\CN}_{b,f}$.

The requirement of topological triviality of the bundle 
$\CE\to\CP_{b,f}^{3|\CN}$
allows us to split the transition function $f_{+-}$ according to
\eqn\eqforfbig{f_{+-}\ =\ \psi^{-1}_+\psi_-,}
whereas holomorphic triviality of $\CE\to\CP_{b,f}^{3|\CN}$ along the 
rational curves 
$\IC P^1_{x,\eta}\hookrightarrow\CP_{b,f}^{3|\CN}$ ensures that there exist
$\psi_\pm$ such that
$\partial_{{\bar\l}_\pm}\psi_\pm=0$. Proceeding as in section 2, we write
\eqna\defofasdh
$$\eqalignno{\CA^+_{\a\ad_1\cdots\ad_{2b_\a-1}}\  &\equiv\ 
                \l^\ad_+\CA_{\a\ad\ad_1\cdots\ad_{2b_\a-1}} \ = \ 
                 \psi_\pm D^+_{\a\ad_1\cdots\ad_{2b_\a-1}}
                    \psi_\pm^{-1}, &   \defofasdh a\cr
             \CA_{{\bar\l}_+}\ &=\ 0,&   \defofasdh b\cr     
             \CA^i_{+\ad_1\cdots\ad_{2f_i-1}}\ &\equiv\ 
             \l_+^\ad\CA^i_{\ad\ad_1\cdots\ad_{2f_i-1}}\ =\  
               \psi_\pm D^i_{+\ad_1\cdots\ad_{2f_i-1}}\psi_\pm^{-1},
             &\defofasdh c\cr
}$$ 
and therefore
\eqna\linsyssdh
$$\eqalignno{(D^+_{\a\ad_1\cdots\ad_{2b_\a-1}}+
              \CA^+_{\a\ad_1\cdots\ad_{2b_\a-1}})\psi_\pm\ &=\ 0,& 
             \linsyssdh a\cr
             \partial_{{\bar\l}_+}\psi_\pm\ &=\ 0, & \linsyssdh b\cr
            ( D^i_{+\ad_1\cdots\ad_{2f_i-1}}+
               \CA^i_{+\ad_1\cdots\ad_{2f_i-1}})\psi_\pm\ &=\ 0. 
             &\linsyssdh c\cr
}$$ 
We note that for $b_\a=f_i={1\over2}$ this system reduces, of
course, to the old one given by \linsys{}. Moreover, we have the following
symmetry properties
\eqn\soioa{\CA_{\a\ad\ad_1\cdots\ad_{2b_\a-1}}\ =\ 
           \CA_{\a\ad(\ad_1\cdots\ad_{2b_\a-1})}\quad{\rm and}\quad
           \CA^i_{\ad\ad_1\cdots\ad_{2f_i-1}}\ =\ 
           \CA^i_{\ad(\ad_1\cdots\ad_{2f_i-1})}.}
The compatibility conditions for \linsyssdh{} read as
\eqn\compconsdh{\eqalign{
             [\nabla_{\a\ad\ad_1\cdots\ad_{2b_\a-1}},
             \nabla_{\b\bd\bd_1\cdots\bd_{2b_\b-1}}]+
             [\nabla_{\a\bd\ad_1\cdots\ad_{2b_\a-1}},
             \nabla_{\b\ad\bd_1\cdots\bd_{2b_\b-1}}]\ &=\ 0,\cr
             [\nabla^i_{\ad\ad_1\cdots\ad_{2f_i-1}},
             \nabla_{\b\bd\bd_1\cdots\bd_{2b_\b-1}}]+
             [\nabla^i_{\bd\ad_1\cdots\ad_{2f_i-1}},
             \nabla_{\b\ad\bd_1\cdots\bd_{2b_\b-1}}]\ &=\ 0,\cr
             \{\nabla^i_{\ad\ad_1\cdots\ad_{2f_i-1}},
             \nabla^j_{\bd\bd_1\cdots\bd_{2f_j-1}}\}+
             \{\nabla^i_{\bd\ad_1\cdots\ad_{2f_i-1}},
             \nabla^j_{\ad\bd_1\cdots\bd_{2f_j-1}}\}\ &=\ 0.\cr
}}
Here, we have defined
\eqn\defofcovdersdh{\eqalign{
                 \nabla_{\a\ad\ad_1\cdots\ad_{2b_\a-1}}\ &\equiv\ 
                 \partial_{\a\ad\ad_1\cdots\ad_{2b_\a-1}}+
                 \CA_{\a\ad\ad_1\cdots\ad_{2b_\a-1}},\cr
                 \nabla^i_{\ad\ad_1\cdots\ad_{2f_i-1}}\ &\equiv\ 
                 \partial^i_{\ad\ad_1\cdots\ad_{2f_i-1}}+
                 \CA^i_{\ad\ad_1\cdots\ad_{2f_i-1}}.\cr
}}
We remark that the components 
$$\CA_{\a\ad{\dot1}\cdots{\dot1}}\qquad{\rm and}\qquad
  \CA^i_{\ad{\dot1}\cdots{\dot1}}$$ 
coincide with the components $\CA_{\a\ad}$ and $\CA^i_\ad$ of the gauge 
potential on $\IC^{4|\CN}$ we have introduced in \defofa{}. In the sequel, we
shall refer to \compconsdh\ as the {\it truncated $\CN$-extended self-dual SYM
hierarchy}. The {\it full hierarchy} are then obtained by taking the 
limit $b_\a,f_i\to\infty$. In this asymptotic regime, we define (symbolically)
the space
\eqn\dofoftsfh{\CP^{3|\CN}_\infty\ \equiv\ \lim_{b_\a,f_i\to\infty}
                \CP^{3|\CN}_{b,f},}
which we call the {\it fully enhanced supertwistor space}. 

From the equations \defofasdh{} it follows that
\eqn\pwtranssdh{\eqalign{
             \CA_{\a\ad\ad_1\cdots\ad_{2b_\a-1}}\ &=\ 
             {1\over 2\pi\i}\oint_c{\rm d}\l_+
             {\CA^+_{\a\ad_1\cdots\ad_{2b_\a-1}}\over\l_+\l_+^\ad},\cr
             \CA^i_{\ad\ad_1\cdots\ad_{2f_i-1}}\ &=\ 
             {1\over 2\pi\i} \oint_c{\rm d}\l_+
             {\CA^i_{+\ad_1\cdots\ad_{2f_i-1}}\over\l_+\l_+^\ad} ,
             \cr
}}
where $c=\{\l_+\in\IC P^1\,|\,|\l_+|=1\}$. As in the previous discussion, the 
equations \pwtranssdh\ make the Penrose-Ward transform explicit.

In summary, we have extended the one-to-one correspondence between equivalence 
classes of holomorphic vector bundles over the supertwistor space and gauge 
equivalence classes of solutions to the $\CN$-extended self-dual SYM equations 
to the level of the hierarchies, i.e., now we have a one-to-one correspondence
between equivalence classes holomorphic vector bundles over the enhanced 
supertwistor space and gauge equivalence classes of solutions to the truncated
$\CN$-extended self-dual SYM hierarchy. Of course, by way of construction, 
the old correspondence is just a subset of this extension. In summary, we
have the bijection
\eqn\otohbsdh{\CM_{\rm hol}(\CP^{3|\CN}_{b,f})\ \longleftrightarrow\
              \CM^\CN_{\rm SDYMH}(b,f),}
where $\CM^\CN_{\rm SDYMH}(b,f)$ denotes the moduli space of solutions to the 
respective truncated $\CN$-extended self-dual SYM hierarchy.

%--------------------------------------------------------------------------
\subsec{Superfield equations of motion}

So far, we have written down the truncated self-dual SYM 
hierarchies \compconsdh\ quite abstractly as the compatibility conditions of 
the linear system \linsyssdh{}. By recalling the discussion of the 
$\CN$-extended super SDYM equations presented at the end of subsection 2.1, the
next step in our discussion is to look for the equations of motion on 
superfield level equivalent to \compconsdh. To do this, we need to identify the
(super)field content. At first sight, we expect to find -- as {\it fundamental}
field content (in a covariant formulation)
-- the field content of the $\CN$-extended self-dual SYM theory 
plus a tower of additional fields which depends on the parameters $b_\a$ and 
$f_i$. However, as we shortly realize, this will not be entirely true. Instead 
we shall find that for $f_i>{1\over2}$ certain combinations of the  
$\CA^i_{\ad\ad_1\cdots\ad_{2f_i-1}}$ play the role of potentials for a lot 
of the naively expected fields, such that those combinations should be regarded
as fundamental fields. 

For the rest of this section, we shall for simplicity consider the case 
where $b_1=b_2\equiv b$ and $f_1=\cdots=f_\CN\equiv f$. For a moment, let us 
also introduce a shorthand index notation
\eqn\shorthandI{\CA_{\a\ad\ad_1\cdots\ad_{2b-1}}\ \equiv\
                \CA_{\a\ad\Ad}\qquad{\rm and}\qquad
                \CA_{\ad\ad_1\cdots\ad_{2f-1}}^i\ \equiv\
                \CA^I_\ad,}
which simplifies the subsequent formulas.

First, we point out that the equations \compconsdh\ can concisely be rewritten
as
\eqn\compconsdhI{\eqalign{
             [\nabla_{\a\ad\Ad},\nabla_{\b\bd\Bd}]+
             [\nabla_{\a\bd\Ad},\nabla_{\b\ad\Bd}]
             \ &=\ 0,\qquad
             [\nabla^I_\ad,\nabla_{\b\bd\Bd}]+[\nabla^I_\bd,\nabla_{\b\ad\Bd}]
             \ =\ 0,\cr
        &\kern-.8cm
           \{\nabla^I_\ad,\nabla^J_\bd\}+\{\nabla^I_\bd,\nabla^J_\ad\}\ =\ 0,
             \cr
}}
which translates to the following superfield definitions 
\eqna\defofsfI
$$\eqalignno{[\nabla_{\a\ad\Ad},\nabla_{\b\bd\Bd}]\ &\equiv\ \epsilon_{\ad\bd}
             f_{\a\Ad\b\Bd},& \defofsfI a\cr
             [\nabla^I_\ad,\nabla_{\b\bd\Bd}]\ &\equiv\ \epsilon_{\ad\bd}
             \chi^I_{\b\Bd},& \defofsfI b\cr
             \{\nabla^I_\ad,\nabla^J_\bd\}\ &\equiv\ 2\epsilon_{\ad\bd}
             W^{IJ}.& \defofsfI c\cr
}$$
Note that quite generally we have
\eqn\defofcI{
   \eqalign{\CF_{\a\ad\Ad\b\bd\Bd}\ &=\ [\nabla_{\a\ad\Ad},\nabla_{\b\bd\Bd}]
   \ =\ \sfrac{1}{2}(\CF_{\a\ad\Ad\b\bd\Bd}-\CF_{\b\bd\Bd\a\ad\Ad})\cr
    &=\ \epsilon_{\ad\bd}f_{\a\Ad\b\Bd}+\epsilon_{\a\b}f_{\ad\Ad\bd\Bd}+
             \CF_{\a\ad[\Ad\b\bd\Bd]},}}
where
\eqn\defofcII{\eqalign{f_{\a\Ad\b\Bd}\ &\equiv\ 
           -\sfrac{1}{2}\epsilon^{\ad\bd}\CF_{\a\ad\Ad\b\bd\Bd},\cr
           f_{\ad\Ad\bd\Bd}\ &\equiv\ 
           \sfrac{1}{2}\epsilon^{\a\b}\CF_{\a\ad\Ad\b\bd\Bd},\cr
            \CF_{\a\ad[\Ad\b\bd\Bd]}\ &\equiv\ 
           \sfrac{1}{2}(\CF_{\a\ad\Ad\b\bd\Bd}-\CF_{\a\ad\Bd\b\bd\Ad}).}}
Equation \defofcI\ can be simplified further to
\eqn\defofcIII{\CF_{\a\ad\Ad\b\bd\Bd}\ =\ 
               \epsilon_{\ad\bd}f_{\a\Ad\b\Bd}+
               \epsilon_{\a\b}f_{\ad(\Ad\bd\Bd)}+\CF_{(\a\ad[\Ad\b)\bd\Bd]}.}
Therefore, equation \defofsfI{a} implies that
\eqn\superfieldeqhI{f_{\ad(\Ad\bd\Bd)}\ =\ 0\qquad{\rm and}\qquad
                    \CF_{(\a\ad[\Ad\b)\bd\Bd]}\ =\ 0,}
which are the first two of the superfield equations of motion. We point out 
that for the choice $\Ad=\Bd=({\dot1}\cdots{\dot1})$ the set
\superfieldeqhI\ represents nothing but the ordinary SDYM equations 
(cf. the first equation of \fieldeqnfour). 

Next, we consider the Bianchi identity for the triple 
$(\nabla_{\a\ad\Ad},\nabla^I_\bd,\nabla_{\g\gd\Cd})$. We find
\eqn\superfieldeqhII{\nabla^I_\ad f_{\a\Ad\b\Bd}\ =\ 
                     \nabla_{\a\ad\Ad}\chi^I_{\b\Bd}.}
From this equation we deduce another two field equations, namely
\eqn\superfieldeqhIII{\epsilon^{\a\b}\nabla_{\a\ad(\Ad}\chi^I_{\b\Bd)}\ =\ 0
                     \qquad{\rm and}\qquad
                     \nabla_{(\a\ad[\Ad}\chi^I_{\b)\Bd]}\ =\ 0.}
Now the Bianchi identity for $(\nabla_{\a\ad\Ad},\nabla^I_\bd,\nabla^J_\gd)$ 
says that
\eqn\superfieldeqhIV{\nabla_{\a\ad\Ad}W^{IJ}\ =\ \sfrac{1}{2}\nabla^I_\ad
                     \chi^J_{\a\Ad}.}
Applying $\nabla_{\b\bd\Bd}$ to \superfieldeqhIV, we obtain upon 
(anti)symmetrization the following two equations of motion 
\eqn\superfieldeqhV{\eqalign{
     \epsilon^{\a\b}\epsilon^{\ad\bd}\nabla_{\a\ad(\Ad}\nabla_{\b\bd\Bd)}W^{IJ}
     +\epsilon^{\a\b}\{\chi^I_{\a(\Ad},\chi^J_{\b\Bd)}\}\ &=\ 0,\cr
     \epsilon^{\ad\bd}\nabla_{(\a\ad[\Ad}\nabla_{\b)\bd\Bd]}W^{IJ}+
     \{\chi^I_{(\a[\Ad},\chi^J_{\b)\Bd]}\}\ &=\ 0.
}}
The Bianchi identity for the combination 
$(\nabla^I_\ad,\nabla^J_\bd,\nabla^K_\gd)$ shows that $\nabla^I_\ad W^{JK}$ 
determines a superfield which is totally antisymmetric in the indices
$IJK$, i.e.,
\eqn\superfieldeqhVI{\nabla^I_\ad W^{JK}\ \equiv\ \chi^{IJK}_\ad.}
Upon acting on both sides by $\nabla_{\a\ad\Ad}$ and contracting the dotted
indices, we obtain the field equation for $\chi^{IJK}_\ad$
\eqn\superfieldeqhVII{\epsilon^{\ad\bd}\nabla_{\a\ad\Ad}\chi^{IJK}_\bd-
                      3[W^{[IJ},\chi^{K]}_{\a\Ad}]\ =\ 0.}
The application of $\nabla^I_\ad$ to $\chi^{JKL}_\bd$ and symmetrization in 
$\ad$ and $\bd$ leads by virtue of \superfieldeqhVI\ to a new superfield which
is totally antisymmetric in $IJKL$,
\eqn\superfieldeqhVIII{\nabla^I_{(\ad}\chi^{JKL}_{\bd)}\ \equiv\ 
                       G_{\ad\bd}^{IJKL}.}
Furthermore, some algebraic manipulations show that
\eqn\superfieldeqhIX{\nabla^I_\ad\chi^{JKL}_\bd\ =\ 
       \nabla^I_{(\ad}\chi^{JKL}_{\bd)}+\nabla^I_{[\ad}\chi^{JKL}_{\bd]}
       \ =\ G_{\ad\bd}^{IJKL}+3\epsilon_{\ad\bd}[W^{I[J},W^{KL]}],}
where equation \superfieldeqhVI\ and the definition \superfieldeqhVIII\ have
been used. From this equation, the equation of motion for the superfield
$G_{\ad\bd}^{IJKL}$ can readily be derived. We obtain
\eqn\superfieldeqhX{\epsilon^{\ad\gd}\nabla_{\a\ad\Ad}G_{\bd\gd}^{IJKL}+
     4\{\chi^{[I}_{\a\Ad},\chi^{JKL]}_\bd\}+6[W^{[JK},\nabla_{\a\bd\Ad}W^{LI]}]
     \ =\ 0.}
As \superfieldeqhVI\ implies the existence of the superfield
$G_{\ad\bd}^{IJKL}$, the definition \superfieldeqhVIII\ determines a new
superfield $\psi^{IJKLM}_{\ad\bd\gd}$ being totally antisymmetric in
$IJKLM$ and totally symmetric in $\ad\bd\gd$, i.e.,
\eqn\superfieldeqhXI{\nabla^I_{(\ad}G^{JKLM}_{\bd\gd)}\ \equiv\
                     \psi^{IJKLM}_{\ad\bd\gd}.}
It is easily shown that
\eqn\superfieldeqhXII{\nabla^I_{\ad}G^{JKLM}_{\bd\gd}\ =\
                      \nabla^I_{(\ad}G^{JKLM}_{\bd\gd)}-\sfrac{2}{3}
         \epsilon_{\ad(\bd}\epsilon^{\dd\ed}\nabla^I_\dd G^{JKLM}_{\ed\gd)}.}
After some tedious algebra, we obtain from \superfieldeqhXII\ the formula
\eqn\superfieldeqhXIII{\nabla^I_{\ad}G^{JKLM}_{\bd\gd}\ =\ 
    \psi^{IJKLM}_{\ad\bd\gd}+\sfrac{4}{3}\epsilon_{\ad(\bd}
    \left(4[W^{I[J},\chi^{KLM]}_{\gd)}]+3[\chi^{I[JK}_{\gd)},W^{LM]}]\right),}
where the definition \superfieldeqhXI\ has been substituted. This
equation in turn implies the equation of motion for $\psi^{IJKLM}_{\ad\bd\gd}$,
\eqn\superfieldeqhXIV{\eqalign{\epsilon^{\ad\dd}\nabla_{\a\ad\Ad}
     \psi^{IJKLM}_{\bd\gd\dd}&+5[\chi^{[I}_{\a\Ad},G^{JKLM]}_{\bd\gd}]\ +\cr 
     &+\sfrac{40}{3}[\nabla_{\a(\bd\Ad}W^{[IJ},\chi^{KLM]}_{\gd)}]-
     \sfrac{20}{3}[W^{[IJ},\nabla_{\a(\bd\Ad}\chi^{KLM]}_{\gd)}]\ =\ 0,
}}
which follows after a somewhat lengthy calculation. 

Now one can continue this procedure of defining superfields via the action of
$\nabla^I_\ad$ and of finding the corresponding equations of motion. 
Generically, the number of fields one obtains in this way is determined by the
parameter $f$, i.e., the most one can get is
$$\psi^{I_1\cdots I_{2\CN f}}_{\ad_1\cdots\ad_{2\CN f-2}},$$
which is, as before, totally antisymmetric in $I_1\cdots I_{2\CN f}$ and 
totally symmetric in $\ad_1\cdots\ad_{2\CN f-2}$. 

Let us collect the superfield equations of motion for the $\CN$-extended 
self-dual SYM hierarchy:
\eqn\superfieldeqhXV{\eqalign{
  f_{\ad(\Ad\bd\Bd)}\ =\ 0 \qquad{\rm and}\qquad
  \CF_{(\a\ad[\Ad\b)\bd\Bd]}\ &=\ 0,\cr
  \epsilon^{\a\b}\nabla_{\a\ad(\Ad}\chi^I_{\b\Bd)}\ =\ 0 \qquad{\rm and}\qquad
  \nabla_{(\a\ad[\Ad}\chi^I_{\b)\Bd]}\ &=\ 0,\cr
  \epsilon^{\a\b}\epsilon^{\ad\bd}\nabla_{\a\ad(\Ad}\nabla_{\b\bd\Bd)}W^{IJ}
  +\epsilon^{\a\b}\{\chi^I_{\a(\Ad},\chi^J_{\b\Bd)}\}\ &=\ 0,\cr
  \epsilon^{\ad\bd}\nabla_{(\a\ad[\Ad}\nabla_{\b)\bd\Bd]}W^{IJ}+
  \{\chi^I_{(\a[\Ad},\chi^J_{\b)\Bd]}\}\ &=\ 0,\cr
  \epsilon^{\ad\bd}\nabla_{\a\ad\Ad}\chi^{IJK}_\bd-
  3[W^{[IJ},\chi^{K]}_{\a\Ad}]\ &=\ 0,\cr
  \epsilon^{\ad\gd}\nabla_{\a\ad\Ad}G_{\bd\gd}^{IJKL}+
  4\{\chi^{[I}_{\a\Ad},\chi^{JKL]}_\bd\}+6[W^{[JK},\nabla_{\a\bd\Ad}W^{LI]}]
  \ &=\ 0,\cr
  &\kern-2cm\vdots\cr
  \epsilon^{\ad\bd}\nabla_{\a\ad\Ad}
  \psi^{I_1\cdots I_{2\CN f}}_{\bd\ad_1\cdots\ad_{2\CN f-3}}+ 
  J^{I_1\cdots I_{2\CN f}}_{\a\Ad\ad_1\cdots\ad_{2\CN f-3}}\ &=\ 0,\cr
}}
where the currents
$J^{I_1\cdots I_{2\CN f}}_{\a\Ad\ad_1\cdots\ad_{2\CN f-3}}$ are determined in 
an obvious manner. Clearly, the system \superfieldeqhXV\ contains as a subset
the $\CN$-extended self-dual SYM equations. In particular, for the choice 
$b=f={1\over2}$ it reduces to \fieldeqnfour. Altogether, we have obtained the 
field content of the $\CN$-extended self-dual SYM theory plus a number of 
additional fields together with their superfield equations of motion. 

However, as we have already indicated, this is not the end of the story. The 
system \superfieldeqhXV, though describing the truncated hierarchy, contains a 
lot of 
redundant information. So, it should not be regarded as the {\it fundamental} 
system displaying the truncated hierarchy. Namely, the use of
the shorthand index notation \shorthandI\ does not entirely reflect all of the
possible index symmetry properties of the appearing superfields. In 
order to incorporate all possibilities, we instead need to write out the 
explicit form of $I,J,K,\ldots$.

As before, let us impose the transversal gauge condition
\eqn\tgH{
    \eta^{\ad\ad_1\cdots\ad_{2f-1}}_i\CA^i_{\ad\ad_1\cdots\ad_{2f-1}}\ =\ 0,}
which again reduces super gauge transformations to ordinary ones. Note that
in \tgH\ only $\CA^i_{(\ad\ad_1\cdots\ad_{2f-1})}$ contributes, since 
the fermionic coordinates are totally symmetric under an exchange of their 
dotted indices. The condition \tgH\ then allows to define the recursion 
operator $\CD$ according to
\eqn\defofrec{\CD\ \equiv\ \eta^{\ad\ad_1\cdots\ad_{2f-1}}_i
              \nabla^i_{\ad\ad_1\cdots\ad_{2f-1}}\ =\ 
              \eta^{\ad\ad_1\cdots\ad_{2f-1}}_i
              \partial^i_{\ad\ad_1\cdots\ad_{2f-1}},}
i.e., $\CD$ is a positive definite homogeneity operator. Equation \defofsfI{c} 
yields 
\eqn\recursionI{\eqalign{
                 (1+\CD)\CA^i_{(\ad\ad_1\cdots\ad_{2f-1})}\ &=\ 
                 -2\epsilon_{\bd(\ad}\eta^{\bd\bd_1\cdots\bd_{2f-1}}_j
                 W^{ij}_{\ad_1\cdots\ad_{2f-1})\,\bd_1\cdots\bd_{2f-1}},\cr
                 \CD\CA^i_{[\ad,\bd]\ad_1\cdots\ad_{2f-2}}\ &=\ 
                 \epsilon_{\ad\bd}\eta^{\gd\bd_1\cdots\bd_{2f-1}}_j
                 W^{ij}_{\gd\ad_1\cdots\ad_{2f-2}\,\bd_1\cdots\bd_{2f-1}},
}}
which states that $\CA^i_{(\ad\ad_1\cdots\ad_{2f-1})}$ does not
have a zeroth order component in the $\eta$-expansion while 
$\CA^i_{[\ad,\bd]\ad_1\cdots\ad_{2f-2}}$ does. Therefore, we obtain as a
fundamental superfield 
\eqn\fundamentfieldsI{\phi^i_{\ad_1\cdots\ad_{2f-2}}\ \equiv\ 
                      \CA^i_{[{\dot1},{\dot2}]\ad_1\cdots\ad_{2f-2}},
                      \qquad{\rm for}\qquad f>\sfrac{1}{2}.}
Note that as $\phi^i_{\ad_1\cdots\ad_{2f-2}}=\phi^i_{(\ad_1\cdots\ad_{2f-2})}$
it defines for each $i$ a spin $f-1$ superfield of odd Gra{\ss}mann 
parity. 

Equation \defofsfI{b} reads explicitly as
$$[\nabla^i_{\ad\ad_1\cdots\ad_{2f-1}},\nabla_{\b\bd\bd_1\cdots\bd_{2b-1}}]
  \ =\ 
 \epsilon_{\ad\bd}\chi^i_{\ad_1\cdots\ad_{2f-1}\,\b\,\bd_1\cdots\bd_{2b-1}}.$$
The contraction with $\epsilon^{\ad\ad_1}$ shows that
\eqn\prepotentialI{\chi^i_{\ad_1\cdots\ad_{2f-1}\,\b\,\bd_1\cdots\bd_{2b-1}}
  \ =\ 2\nabla_{\b(\ad_1\bd_1\cdots\bd_{2b-1}}\phi^i_{\ad_2\cdots\ad_{2f-1})},}
where the symmetrization is only ment between the $\ad_1\cdots\ad_{2f-1}$.
Therefore, the superfield 
$\chi^i_{\ad_1\cdots\ad_{2f-1}\,\b\,\bd_1\cdots\bd_{2b-1}}$ cannot be regarded
as a fundamental field -- the superfield \fundamentfieldsI\ plays the role
of a potential for the former. 

Next, we discuss the superfield 
$W^{ij}_{\ad_1\cdots\ad_{2f-1}\,\bd_1\cdots\bd_{2f-1}}$. To show which
combinations of it are really fundamental, we need some preliminaries. Consider
the index set
$$ \ad_1\cdots\ad_n\bd_1\cdots\bd_n, $$
which is separately totally symmetric in $\ad_1\cdots\ad_n$ and 
$\bd_1\cdots\bd_n$, respectively. Then we have the useful formula
\eqn\usefulformula{\eqalign{
                   \ad_1\cdots\ad_n\bd_1\cdots\bd_n\ &=\ 
                   (\ad_1\cdots\ad_n\bd_1\cdots\bd_n)\ +\ 
                   \sum\ {\rm all\ possible\ contractions}\cr
                    &=\ (\ad_1\cdots\ad_n\bd_1\cdots\bd_n)\ +\cr
                    &\kern2cm +\ A_1\sum_{i,j}
                 \ad_1\cdots\!\!\contra{62}{\ \ad_i\cdots\ad_n\bd_1\cdots\bd_j}
                  \cdots\bd_n\ +\ \cdots,\cr
}}
where the parantheses denote, as before, symmetrization of the enclosed 
indices and ``contraction'' means antisymmetrization in the respective index 
pair. The $A_i$ for $i=1,\ldots,n$ are combinatorial coefficients, whose
explicit form is not needed in the sequel. The proof of \usefulformula\ is 
quite similar to the one of the Wick theorem and we thus leave it to the 
interested reader. Equation \defofsfI{c} is explicitly given by
$$ \{\nabla^i_{\ad\ad_1\cdots\ad_{2f-1}},\nabla^j_{\bd\bd_1\cdots\bd_{2f-1}}\}
\ =\ 2\epsilon_{\ad\bd}W^{ij}_{\ad_1\cdots\ad_{2f-1}\,\bd_1\cdots\bd_{2f-1}}.$$
After contraction with $\epsilon^{\ad\ad_1}$ we obtain
\eqn\fundamentfieldsII{W^{ij}_{\ad_1\cdots\ad_{2f-1}\,\bd_1\cdots\bd_{2f-1}}
  \ =\ -\nabla^j_{\ad_1\bd_1\cdots\bd_{2f-1}}\phi^i_{\ad_2\cdots\ad_{2f-1}},}
where the definition \fundamentfieldsI\ has been inserted. Contracting this
equation with $\epsilon^{\ad_1\bd_1}$, we get
\eqn\fundamentfieldsIII{\epsilon^{\ad_1\bd_1}
         W^{ij}_{\ad_1\cdots\ad_{2f-1}\,\bd_1\cdots\bd_{2f-1}}\ =\ 
         -2\{\phi^i_{\ad_2\cdots\ad_{2f-1}},\phi^j_{\bd_2\cdots\bd_{2f-1}}\}.}
Thus, we conclude that  
$W^{ij}_{\ad_1\cdots\ad_{2f-2}[\ad_{2f-1}\bd_1]\bd_2\cdots\bd_{2f-1}}$ is
a composite field and hence not a fundamental one. Using formula
\usefulformula, we may schematically write
\eqn\fundamentfieldsIV{W^{ij}_{\ad_1\cdots\ad_{2f-1}\,\bd_1\cdots\bd_{2f-1}}
       \ =\ W^{ij}_{(\ad_1\cdots\ad_{2f-1}\,\bd_1\cdots\bd_{2f-1})}
            +\sum\ {\rm all\ possible\ contractions}.}
The contraction terms in \fundamentfieldsIV, however, are all composite 
expressions due to \fundamentfieldsIII. Therefore, only the superfield
\eqn\fundamentfieldsV{W^{ij}_{(\ad_1\cdots\ad_{2f-1}\,\bd_1\cdots\bd_{2f-1})}
              \ =\ W^{[ij]}_{(\ad_1\cdots\ad_{2f-1}\,\bd_1\cdots\bd_{2f-1})}}
is fundamental. For each combination $[ij]$ it represents a Gra{\ss}mann even
superfield with spin $2f-1$.

Now we need to consider the superfield defined in \superfieldeqhVI,
$$ \chi^{ijk}_{\ad\,\ad_1\cdots\ad_{2f-1}\,\bd_1\cdots\bd_{2f-1}\,
   \gd_1\cdots\gd_{2f-1}}\ =\ \nabla^i_{\ad\ad_1\cdots\ad_{2f-1}}
   W^{ij}_{\bd_1\cdots\bd_{2f-1}\,\gd_1\cdots\gd_{2f-1}}. $$
By extending the formula \usefulformula\ to the index triple 
$$\ad_1\cdots\ad_{2f-1}\,\bd_1\cdots\bd_{2f-1}\,\gd_1\cdots\gd_{2f-1}$$
and by utilizing the symmetry properties of 
$\chi^{ijk}_{\ad\,\ad_1\cdots\ad_{2f-1}\,\bd_1\cdots\bd_{2f-1}\,\gd_1\cdots
\gd_{2f-1}}$, one can show, by virtue of the above arguments, that only the
combination
\eqn\fundamentfieldsV{\chi^{ijk}_{(\ad\,\ad_1\cdots\ad_{2f-1}\,
   \bd_1\cdots\bd_{2f-1}\,\gd_1\cdots\gd_{2f-1})}\ =\ 
  \chi^{[ijk]}_{(\ad\,\ad_1\cdots\ad_{2f-1}\,
   \bd_1\cdots\bd_{2f-1}\,\gd_1\cdots\gd_{2f-1})}}
of $\chi^{ijk}_{\ad\,\ad_1\cdots\ad_{2f-1}\,\bd_1\cdots\bd_{2f-1}\,
\gd_1\cdots\gd_{2f-1}}$ remains as a fundamental superfield. It defines
for each $[ijk]$ a Gra{\ss}mann odd superfield with spin $3f-1$.

Repeating this procedure, we deduce from the definition \superfieldeqhVIII\ 
that 
\eqn\fundamentfieldsVI{\eqalign{
  G^{ijkl}_{(\ad\bd\,\ad_1\cdots\ad_{2f-1}\,
   \bd_1\cdots\bd_{2f-1}\,\gd_1\cdots\gd_{2f-1}\,\dd_1\cdots\dd_{2f-1})}&\cr
   &\kern-1.5cm=\ 
  G^{[ijkl]}_{(\ad\bd\,\ad_1\cdots\ad_{2f-1}\,
   \bd_1\cdots\bd_{2f-1}\,\gd_1\cdots\gd_{2f-1}\,\dd_1\cdots\dd_{2f-1})}\cr
}}
is fundamental and it represents one\foot{Note that we assume $\CN\leq4$.}
spin $4f-1$ superfield which is Gra{\ss}mann even. All higher order fields,
such as \superfieldeqhXI, yield no further fundamental fields due to the
antisymmetrization of $ijklm$, etc. In summary, the fundamental field 
content of the truncated self-dual SYM hierarchies is given by
\eqn\fundamentfieldsVII{\eqalign{
           \kern1cm\CA_{\a\ad\ad_1\cdots\ad_{2b-1}},\quad
                        \phi^i_{\ad_1\cdots\ad_{2f-2}},&\quad
            W^{[ij]}_{(\ad_1\cdots\ad_{2f-1}\,\bd_1\cdots\bd_{2f-1})},\cr
       &\kern-5cm\chi^{[ijk]}_{(\ad\,\ad_1\cdots\ad_{2f-1}\,
   \bd_1\cdots\bd_{2f-1}\,\gd_1\cdots\gd_{2f-1})},\quad
         G^{[ijkl]}_{(\ad\bd\,\ad_1\cdots\ad_{2f-1}\,
   \bd_1\cdots\bd_{2f-1}\,\gd_1\cdots\gd_{2f-1}\,\dd_1\cdots\dd_{2f-1})},\cr
}}
where we assume that $f>{1\over2}$.\foot{For $f={1\over2}$, the field 
$\phi^i_{\ad_1\cdots\ad_{2f-2}}$ must be replaced by $\chi^i_\a$.} All other 
naively expected fields, which for instance appear in \superfieldeqhXV, are 
composite expressions of the above fields.

It remains to find the superfield equations of motion for the fields 
\fundamentfieldsVII. This, however, is easily done since we have already 
derived \superfieldeqhXV. By following the lines which led to \superfieldeqhXV\
and by taking into account the definition \fundamentfieldsI, the system 
\superfieldeqhXV\ reduces for $f>{1\over2}$ to
\eqn\superfieldeqhXVII{\eqalign{
 f_{\ad(\ad_1\cdots\ad_{2b-1}\bd\bd_1\cdots\bd_{2b-1})}\ =\ 0 
 \qquad{\rm and}\qquad
 \CF_{(\a\ad[\ad_1\cdots\ad_{2b-1}\b)\bd\bd_1\cdots\bd_{2b-1}]}\ &=\ 0,\cr
 \epsilon^{\a\b}\epsilon^{\ad\bd}\nabla_{\a\ad(\ad_1\cdots\ad_{2b-1}}
 \nabla_{\b\bd\bd_1\cdots\bd_{2b-1})}\phi^i_{\gd_1\cdots\gd_{2f-2}}\ &=\ 0,\cr
 \nabla_{(\a\ad[\ad_1\cdots\ad_{2b-1}}\nabla_{\b)(\gd_1\bd_1\cdots\bd_{2b-1}]}
 \phi^i_{\gd_2\cdots\gd_{2f-1})}\ &=\ 0,\cr
 &\kern-10cm\epsilon^{\ad\bd_1}\nabla_{\a\ad\ad_1\cdots\ad_{2b-1}}
 W^{[ij]}_{(\bd_1\cdots\bd_{4f-2})}\ -\cr 
 -\ 2\{\phi^{[i}_{(\bd_2\cdots\bd_{2f-1}},\nabla_{\a\bd_{2f}\ad_1\cdots
 \ad_{2b-1}}\phi^{j]}_{\bd_{2f+1}\cdots\gd_{4f-2})}\}\ &=\ 0,\cr
 &\kern-10cm\epsilon^{\ad\bd}\nabla_{\a\ad\ad_1\cdots\bd_{2b-1}}
 \chi^{[ijk]}_{(\bd\bd_1\cdots\bd_{6f-3})}\ -\cr
 -\ 6[W^{[ij}_{(\bd_1\cdots\bd_{4f-2}},\nabla_{\a\bd_{4f-1}\ad_1\cdots
  \ad_{2b-1}}\phi^{k]}_{\bd_{4f}\cdots\bd_{6f-3})}]\ &=\ 0,\cr
 &\kern-10cm\epsilon^{\ad\gd}\nabla_{\a\ad\ad_1\cdots\ad_{2b-1}}
 G_{(\bd\gd\bd_1\cdots\bd_{8f-4})}^{[ijkl]}\ +\cr
 &\kern-9.5cm\ -8\{\nabla_{\a(\bd_1\ad_1\cdots\ad_{2b-1}}\phi^{[i}_{\bd_2\cdots
 \bd_{2f-1}},\chi^{ijk]}_{\bd\bd_{2f}\cdots\bd_{8f-4})}\}\ -\cr
 -\ 6[W^{[ij}_{(\bd_1\cdots\bd_{4f-2}},
 \nabla_{\a\bd\ad_1\cdots\ad_{2b-1}}W^{kl]}_{\bd_{4f-1}\cdots\bd_{8f-4})}
 ]\ &=\ 0,\cr
}}
which are the superfield equations of motion for the truncated
$\CN$-extended self-dual SYM hierarchy.

%--------------------------------------------------------------------------
\subsec{Equivalence of the field equations and the compatibility conditions}

Above we have derived the superfield equations of motion. What remains is to 
show how the superfields \fundamentfieldsVII\ are expressed in terms of their 
zeroth order components
\eqn\zerothorderfundamentfieldsI{
    \Ac_{\a\ad\ad_1\cdots\ad_{2b-1}},\quad
    \pc\ \!\!^i_{\ad_1\cdots\ad_{2f-2}},\quad
    \Wc\ \!\!^{[ij]}_{(\ad_1\cdots\ad_{4f-2})},\quad
    \cc\ \!\!^{[ijk]}_{(\ad_1\cdots\ad_{6f-2})},\quad
    \Gc\ \!\!^{[ijkl]}_{(\ad_1\cdots\ad_{8f-2})},
}
and furthermore that the field equations on $\IC^4$ (or $\IR^4$
after reality conditions have been imposed)\foot{Note that in total
we have $4b+2$ complex
bosonic coordinates. Here, we interpret the spacetime $\IC^4$ as a subset of 
$\IC^{4b+2}\cong\IC^4\oplus\IC^{4b-2}$ (see also the discussion in subsection 
4.1), i.e., the remaining $4b-2$ bosonic coordinates are regarded as additional
moduli of the fields on $\IC^4$.}, i.e., those 
equations that are obtained from the set \superfieldeqhXVII\ by projecting
onto the zeroth order components \zerothorderfundamentfieldsI\ of the 
superfields \fundamentfieldsVII, imply the compatibility conditions 
\compconsdhI. We will, however, be not too explicit in showing this
equivalence, since the argumentation goes along similar lines as those 
given for the $\CN$-extended self-dual SYM equations (cf. \DevchandGV). Here, 
we just sketch the idea.

In order to write down 
the superfield expansions, remember that we have imposed the gauge
\tgH\ which led to the recursion operator $\CD$ according to \defofrec.
Using the formulas \defofsfI{}, \superfieldeqhIV, \superfieldeqhVI,
\superfieldeqhVIII, \superfieldeqhXI\ and \prepotentialI, we obtain the 
following recursion relations:
\eqna\recursionrelations
$$\eqalignno{
     \kern-1cm(1+\CD)\CA^i_{(\ad_1\cdots\ad_{2f})}\ &=\ 
     -2\epsilon_{\bd(\ad_1}\eta^{\bd\bd_1\cdots\bd_{2f-1}}_j
     W^{ij}_{\ad_2\cdots\ad_{2f})\,\bd_1\cdots\bd_{2f-1}},
     &\recursionrelations a\cr
      \kern-1cm\CD\CA_{\a\ad\ad_1\cdots\ad_{2b-1}}\ &=\ -2\epsilon_{\ad\bd}
     \eta_j^{\bd\bd_1\cdots\bd_{2f-1}}\nabla_{\a\bd_1\ad_1\cdots\ad_{2b-1}}
     \phi^i_{\bd_2\cdots\bd_{2f-1}},&\recursionrelations b\cr
      \kern-1cm\CD\phi^i_{\ad_1\cdots\ad_{2f-2}}\ &=\ 
     \eta^{\gd\bd_1\cdots\bd_{2f-1}}_j
     W^{ij}_{\gd\ad_1\cdots\ad_{2f-2}\,\bd_1\cdots\bd_{2f-1}},
     &\recursionrelations c\cr
      \kern-1cm\CD W^{[ij]}_{(\ad_1\cdots\ad_{4f-2})}\ &=\
    \eta^{\bd\bd_1\cdots\bd_{2f-1}}_k\chi^{[ij]k}_{\bd(\ad_1\cdots\ad_{4f-2})\,
    \bd_1\cdots\bd_{2f-1}},&\recursionrelations d\cr
     \kern-1cm\CD\chi^{[ijk]}_{(\ad_1\cdots\ad_{6f-2})}\ &=\ 
    \eta^{\bd\bd_1\cdots\bd_{2f-1}}_lG^{[ijk]l}_{\bd(\ad_1\cdots\ad_{6f-2})\,
    \bd_1\cdots\bd_{2f-1}}\ +&{}\cr
    &\kern.5cm+\ 3\epsilon_{\bd(\ad_1}\eta^{\bd\bd_1\cdots\bd_{2f-1}}_l
    [W^{l[i}_{\bd_1\cdots\bd_{2f-1}\,\ad_2\cdots\ad_{2f}},
    W^{jk]}_{\ad_{2f+1}\cdots\ad_{6f-2})}],&\recursionrelations e\cr
    \kern-1.5cm\CD G^{[ijkl]}_{(\ad_1\cdots\ad_{8f-2})}\ &=\ 
    \eta^{\bd\bd_1\cdots\bd_{2f-1}}_m\psi^{[ijkl]m}_{\bd
     (\ad_1\cdots\ad_{8f-2})\,\bd_1\cdots\bd_{2f-1}}\ + &{}\cr
    &\kern.5cm + \ \sfrac{4}{3}\epsilon_{\bd(\ad_1}
    \eta^{\bd\bd_1\cdots\bd_{2f-1}}_m
    \left(4[W^{m[i}_{\bd_1\cdots\bd_{2f-1}\,\ad_2\cdots\ad_{2f}},
    \chi^{jkl]}_{\ad_{2f+1}\cdots\ad_{8f-2})}]\ +
    \right.&{}\cr
    &\kern1cm\left.+\ 3[\chi^{m[ij}_{\ad_2\,\bd_1\cdots\bd_{2f-1}\,\ad_3\cdots
    \ad_{4f-4}},W^{kl]}_{\ad_{4f-4}\cdots\ad_{8f-2})}]\right).
    &\recursionrelations f\cr
}$$
An explanation of these formulas is in order. The right hand sides of the
equations \recursionrelations{} depend not only on the fundamental fields
\fundamentfieldsVII\ but also on composite expressions of those fields: 
For instance, consider the recursion relation \recursionrelations{c} of the 
field $\phi^i_{\ad_1\cdots\ad_{2f-2}}$. The right hand side of
\recursionrelations{c} depends on the superfield 
$W^{ij}_{\ad_1\cdots\ad_{2f-1}\,\bd_1\cdots\bd_{2f-1}}$. However, as we
learned in \fundamentfieldsIV, it can be rewritten as the fundamental field
$W^{[ij]}_{(\ad_1\cdots\ad_{2f-1}\,\bd_1\cdots\bd_{2f-1})}$ plus contraction
terms which are of the form \fundamentfieldsIII. Similar arguments hold
for the other recursion relations. Therefore, the right hand 
sides of \recursionrelations{} can solely be written in terms of the
fundamental fields. However, as these formulas in terms of the fundamental 
fields will look pretty messy, we refrain from writing them down but always
have in mind their explicit expansions. 
Note that in \recursionrelations{f} the field
$\psi^{[ijkl]m}_{\bd(\ad_1\cdots\ad_{8f-2})\,\bd_1\cdots\bd_{2f-1}}$ consists
only of composite expressions of the fields \fundamentfieldsIII. Using
the equations \recursionrelations{}, one can now straightforwardly determine
the superfield expansions by a successive application of the recursion
operator $\CD$, since 
if one knows the expansions to $n$-th order in the 
fermionic coordinates, the recursions \recursionrelations{} yield them to 
$(n+1)$-th order, because of the positivity of $\CD$. 
But again, this procedure will lead to both unenlightning and
complicated looking expressions, so we do not present them here.
  
Finally, the recursion operator can be used to show the equivalence between
the field equations and the constraint equations \compcon. This can be done
inductively, i.e., one first assumes that the equations \superfieldeqhXVII\
hold to $n$-th order in the fermionic coordinates, then one applies $k+\CD$ to
\superfieldeqhXVII, where $k\in\IN_0$ is some properly chosen integer, and 
shows that they also hold to $(n+1)$-th order. 
To give an easy example, consider the curvature equations 
\eqn\someexamplecurveq{
 f_{\ad(\ad_1\cdots\ad_{2b-1}\bd\bd_1\cdots\bd_{2b-1})}\ =\ 0 
 \qquad{\rm and}\qquad
 \CF_{(\a\ad[\ad_1\cdots\ad_{2b-1}\b)\bd\bd_1\cdots\bd_{2b-1}]}\ =\ 0}
and assume that they hold to $n$-th order. For the first equation one obtains
\eqna\exrec
$$\eqalignno{
  \CD f_{\ad(\ad_1\cdots\ad_{2b-1}\bd\bd_1\cdots\bd_{2b-1})}\ &=\ &{}\cr
  &\kern-3cm=\ \sfrac{1}{2}\epsilon^{\a\b}\CD
  [\nabla_{\a\ad(\ad_1\cdots\ad_{2b-1}},
  \nabla_{\b\bd\bd_1\cdots\bd_{2b-1})}]&{}\cr
  &\kern-3cm=\ \sfrac{1}{2}\epsilon^{\a\b}\left(
  \nabla_{\a\ad(\ad_1\cdots\ad_{2b-1}}\CD
      \CA_{\b\bd\bd_1\cdots\bd_{2b-1})}-
      \nabla_{\b\bd(\bd_1\cdots\bd_{2b-1}}\CD
 \CA_{\a\ad\ad_1\cdots\ad_{2b-1})}\right)&{}\cr
 &\kern-3cm=\ -\epsilon^{\a\b}\left(
  \epsilon_{\bd\gd}\eta^{\gd\gd_1\cdots\gd_{2f-1}}_i
  \nabla_{\a\ad(\ad_1\cdots\ad_{2b-1}}\nabla_{\b\gd_1\bd_1\cdots\bd_{2b-1})}
  \phi^i_{\gd_2\cdots\gd_{2f-1}}\ -\right.&{}\cr
  &\kern-1cm\left.-\ \epsilon_{\ad\gd}\eta^{\gd\gd_1\cdots\gd_{2f-1}}_i
  \nabla_{\b\bd(\bd_1\cdots\bd_{2b-1}}\nabla_{\a\gd_1\ad_1\cdots\ad_{2b-1})}
  \phi^i_{\gd_2\cdots\gd_{2f-1}}\right)&{}\cr
  &\kern-3cm=\ 0,&\exrec a\cr
}$$
and for the second one 
$$\eqalignno{
  \CD\CF_{(\a\ad[\ad_1\cdots\ad_{2b-1}\b)\bd\bd_1\cdots\bd_{2b-1}]}\ &=&{}\cr
  &\kern-3cm=\ \CD[\nabla_{(\a\ad[\ad_1\cdots\ad_{2b-1}},
  \nabla_{\b)\bd\bd_1\cdots\bd_{2b-1}]}]&{}\cr
  &\kern-3cm=\ \nabla_{(\a\ad[\ad_1\cdots\ad_{2b-1}}\CD
      \CA_{\b)\bd\bd_1\cdots\bd_{2b-1}]}-
      \nabla_{(\b\bd[\bd_1\cdots\bd_{2b-1}}\CD
      \CA_{\a)\ad\ad_1\cdots\ad_{2b-1}]}&{}\cr
  &\kern-3cm=\ -2\epsilon_{\bd\gd}\eta^{\gd\gd_1\cdots\gd_{2f-1}}_i
  \nabla_{(\a\ad[\ad_1\cdots\ad_{2b-1}}\nabla_{\b)\gd_1\bd_1\cdots\bd_{2b-1}]}
  \phi^i_{\gd_2\cdots\gd_{2f-1}}\ +&{}\cr
  &\kern-2cm+\ 2\epsilon_{\ad\gd}\eta^{\gd\gd_1\cdots\gd_{2f-1}}_i
  \nabla_{(\b\bd[\bd_1\cdots\bd_{2b-1}}\nabla_{\a)\gd_1\ad_1\cdots\ad_{2b-1}]}
  \phi^i_{\gd_2\cdots\gd_{2f-1}}&{}\cr
  &\kern-3cm=\ 0,&\exrec b\cr
}$$
where we have used \recursionrelations{b} together with the superfield
equations of $\phi^i_{\ad_1\cdots\ad_{2f-2}}$ given in \superfieldeqhXVII\ to
$n$-th order, which shows that the equations \someexamplecurveq\ are indeed 
satisfied to $(n+1)$-th order. In the derivation \exrec{a}, we have
used the identity
\eqna\exreci
$$\eqalignno{\epsilon^{\a\b}\nabla_{\a\ad(\ad_1\cdots\ad_{2b-1}}
     \nabla_{\b\bd\bd_1\cdots\bd_{2b-1})}\ &=\ &{}\cr
     &\kern-3cm=\ \sfrac{1}{2}\epsilon^{\a\b}\left(
     \nabla_{\a\ad(\ad_1\cdots\ad_{2b-1}}\nabla_{\b\bd\bd_1\cdots\bd_{2b-1})}
    +\nabla_{\a\ad(\ad_1\cdots\ad_{2b-1}}\nabla_{\b\bd\bd_1\cdots\bd_{2b-1})}
    \right)&{}\cr
    &\kern-3cm=\ \sfrac{1}{2}\epsilon^{\a\b}\left(
     -\nabla_{\a\bd(\ad_1\cdots\ad_{2b-1}}\nabla_{\b\ad\bd_1\cdots\bd_{2b-1})}
    +\nabla_{\a\ad(\ad_1\cdots\ad_{2b-1}}\nabla_{\b\bd\bd_1\cdots\bd_{2b-1})}
    \right)&{}\cr
     &\kern-3cm=\ \sfrac{1}{2}\epsilon_{\ad\bd}\epsilon^{\a\b}\epsilon^{\gd\dd}
      \nabla_{\a\gd(\ad_1\cdots\ad_{2b-1}}\nabla_{\b\dd\bd_1\cdots\bd_{2b-1})}
     &\exreci{}\cr
}$$
upon inserting the first equation of \someexamplecurveq\  (again to 
$n$-th order). Similar calculations apply for the remaining equations.

%--------------------------------------------------------------------------
\subsec{Leznov gauge (light cone formalism)}

Again, we may introduce a Lie algebra valued potential $\Psi$ leading
to a simpler form of the truncated hierarchies \compconsdh. Moreover,
the components of the gauge potential are then given by suitable derivatives
of $\Psi$. 

The gauge fixing condition in this case is as before
\eqn\sacsdh{\psi_+\ =\ 1+\l_+\Psi+\CO(\l_+^2).}
One readily verifies that this parametrization leads to
\eqn\leznovgaugesdh{\eqalign{
                    \CA_{\a{\dot1}\ad_1\cdots\ad_{2b-1}}\ =\ 
                    \partial_{\a{\dot2}\ad_1\cdots\ad_{2b-1}}\Psi
                    \qquad&{\rm and}\qquad
                    \CA_{\a{\dot2}\ad_1\cdots\ad_{2b-1}}\ =\ 0,\cr
                    \CA^i_{{\dot1}\ad_1\cdots\ad_{2f-1}}\ =\ 
                    \partial^i_{{\dot2}\ad_1\cdots\ad_{2f-1}}\Psi
                    \qquad&{\rm and}\qquad
                    \CA^i_{{\dot2}\ad_1\cdots\ad_{2f-1}}\ =\ 0.
}}
Therefore, the truncated hierarchy \compconsdh\ turns into the following 
system:
\eqn\compconsdhii{\eqalign{
               \partial_{\a{\dot1}\ad_1\cdots\ad_{2b-1}}
               \partial_{\b{\dot2}\bd_1\cdots\bd_{2b-1}}\Psi&-
               \partial_{\b{\dot1}\bd_1\cdots\bd_{2b-1}}
               \partial_{\a{\dot2}\ad_1\cdots\ad_{2b-1}}\Psi\ +\cr
               &\kern1cm+\ [\partial_{\a{\dot2}\ad_1\cdots\ad_{2b-1}}\Psi,
               \partial_{\b{\dot2}\bd_1\cdots\bd_{2b-1}}\Psi]\ =\ 0,\cr
               \partial^i_{{\dot1}\ad_1\cdots\ad_{2f-1}}
               \partial_{\b{\dot2}\bd_1\cdots\bd_{2b-1}}\Psi&-
               \partial_{\b{\dot1}\bd_1\cdots\bd_{2b-1}}
               \partial^i_{{\dot2}\ad_1\cdots\ad_{2f-1}}\Psi\ +\cr
               &\kern1cm+\ [\partial^i_{{\dot2}\ad_1\cdots\ad_{2f-1}}\Psi,
               \partial_{\b{\dot2}\bd_1\cdots\bd_{2b-1}}\Psi]\ =\ 0,\cr
               \partial^i_{{\dot1}\ad_1\cdots\ad_{2f-1}}
               \partial^j_{{\dot2}\bd_1\cdots\bd_{2f-1}}\Psi&+
               \partial^j_{{\dot1}\bd_1\cdots\bd_{2f-1}}
               \partial^i_{{\dot2}\ad_1\cdots\ad_{2f-1}}\Psi\ +\cr
               &\kern1cm+\ \{\partial^i_{{\dot2}\ad_1\cdots\ad_{2f-1}}\Psi,
               \partial^j_{{\dot2}\bd_1\cdots\bd_{2f-1}}\Psi\}\ =\ 0.
               \cr
}}
Clearly, when all of the free $\ad$ and $\bd$ indices in \compconsdhii\ are 
chosen to be one, we recover the $\CN$-extended self-dual SYM equations in 
Leznov gauge \compconii.

Choosing, for instance, all of the $\ad$ indices equal to one and keeping the 
$\bd$ indices arbitrary, we can interpret the equations \compconsdhii\
as equations on {\it symmetries}
\eqn\symmetries{\d_{\a\ad_1\cdots\ad_{2b}}\Psi\ \equiv\ 
               \partial_{\a\ad_1\cdots\ad_{2b}}\Psi
               \qquad{\rm and}\qquad
               \d_{\ad_1\cdots\ad_{2f}}^i\Psi\ \equiv\ 
               \partial^i_{\ad_1\cdots\ad_{2f}}\Psi,}
i.e.,
\eqn\compconsdhiii{\eqalign{
               \partial_{\a{\dot1}\cdots{\dot1}}
               \d_{\b{\dot2}\bd_1\cdots\bd_{2b-1}}\Psi-
               \partial_{\a{\dot2}{\dot1}\cdots{\dot1}}
               \d_{\b{\dot1}\bd_1\cdots\bd_{2b-1}}\Psi\ +
               [\partial_{\a{\dot2}{\dot1}\cdots{\dot1}}\Psi,
               \d_{\b{\dot2}\bd_1\cdots\bd_{2b-1}}\Psi]\ &=\ 0,\cr
               \partial^i_{{\dot1}\cdots{\dot1}}
               \d_{\b{\dot2}\bd_1\cdots\bd_{2b-1}}\Psi-
               \partial^i_{{\dot2}{\dot1}\cdots{\dot1}}
               \d_{\b{\dot1}\bd_1\cdots\bd_{2b-1}}\Psi\ +
               [\partial^i_{{\dot2}{\dot1}\cdots{\dot1}}\Psi,
               \d_{\b{\dot2}\bd_1\cdots\bd_{2b-1}}\Psi]\ &=\ 0,\cr
               \partial^i_{{\dot1}\cdots{\dot1}}
               \d^j_{{\dot2}\bd_1\cdots\bd_{2f-1}}\Psi-
               \partial^i_{{\dot2}{\dot1}\cdots{\dot1}} 
               \d^j_{{\dot1}\bd_1\cdots\bd_{2f-1}}\Psi\ +
               \{\partial^i_{{\dot2}{\dot1}\cdots{\dot1}}\Psi,
               \d^j_{{\dot2}\bd_1\cdots\bd_{2f-1}}\Psi\}\ &=\ 0.
               \cr
}}
In other words, the differential equations \symmetries\ describe graded
Abelian flows on the space of solutions to \compconii. These flows
are integral curves for the dynamical system \symmetries. This system is the
pendant -- albeit in a particular gauge -- to the system \dynsystrans. 

Finally, the field content in Leznov gauge is given by
\eqn\fieldcontentLeH{\eqalign{
     \CA_{\a{\dot1}\ad_1\cdots\ad_{2b-1}}\ &=\ 
    \partial_{\a{\dot2}\ad_1\cdots\ad_{2b-1}}\Psi,\cr
     \phi^i_{\ad_1\cdots\ad_{2f-2}}\ &=\ \sfrac{1}{2}
         \partial^i_{{\dot1}{\dot2}\ad_1\cdots\ad_{2f-2}}\Psi,\cr
     W^{[ij]}_{(\ad_1\cdots\ad_{4f-2})}
    \ &=\ {\sfrac{1}{2}}\partial^{[i}_{{\dot2}(\ad_1\cdots\ad_{2f-1}}
      \partial^{j]}_{{\dot2}\ad_{2f}\cdots\ad_{4f-2})}\Psi,\cr
     \chi^{[ijk]}_{({\dot2}\ad_1\cdots\ad_{6f-3})}\ &=\ {\sfrac{1}{2}}
       \partial^{[i}_{{\dot2}(\ad_1\cdots\ad_{2f-1}}
       \partial^j_{{\dot2}\ad_{2f}\cdots\bd_{4f-2}}
       \partial^{k]}_{{\dot2}\ad_{4f-1}\cdots\gd_{6f-3})}\Psi,\cr
 G^{[ijkl]}_{({\dot2}{\dot2}\ad_1\cdots\ad_{8f-4})}
   \ &=\ {\sfrac{1}{2}}
       \partial^{[i}_{{\dot2}(\ad_1\cdots\ad_{2f-1}}
       \partial^j_{{\dot2}\ad_{2f}\cdots\ad_{4f-2}}
       \partial^k_{{\dot2}\ad_{4f-1}\cdots\ad_{6f-3}}
       \partial^{l]}_{{\dot2}\ad_{6f-2}\cdots\ad_{8f-4})}\Psi.\cr
}}
Interestingly, the superfield expansion of the potential $\Psi$ of the 
hierarchy does not involve the nonlocal operator $\partial_{\a{\dot2}}^{-1}$ 
which is due to equation \fundamentfieldsI.\foot{For the definition of this 
operator using the Mandelstam-Leibbrandt prescription see, e.g., 
\MandelstamCB.} This is contrary to the self-dual SYM case (cf. equations 
\fieldcontentnfour).

%--------------------------------------------------------------------------
%--------------------------------------------------------------------------
\newsec{Open topological $B$-model on the enhanced supertwistor space}

The above chosen trivializiations of the holomorphic vector bundles 
$\CE\to\CP^{3|\CN}$ and $\CE\to\CP^{3|\CN}_{b,f}$ were convenient for the 
discussion of the self-dual SYM equations and the self-dual SYM hierarchies, 
respectively. It is well known, however, that there is a variety of other
possible trivializations. In particular, it has been argued in \WittenNN\ and 
explained in detail in \PopovRB\ that the $\CN$-extended self-dual SYM 
equations are gauge equivalent to hCS theory \WittenFB\ on the supertwistor 
space $\CP^{3|\CN}$ by choosing certain non-holomorphic trivializations. In 
this section, we are going to extend this discussion and show that the moduli 
space $\CM_{\rm hCS}^\CN(b,f)$ of solutions to the equations of motion of hCS 
theory on the enhanced supertwistor space $\CP^{3|\CN}_{b,f}$ can bijectively
be mapped onto the moduli space $\CM^\CN_{\rm SYMH}(b,f)$ of solutions to the
$\CN$-extended self-dual SYM hierarchy, i.e., we extend \otohbsdh\ 
according to
\eqn\otohbsdhhcs{\CM_{\rm hol}(\CP^{3|\CN}_{b,f})\longleftrightarrow\ 
                \CM_{\rm SYMH}^\CN(b,f)
                \ \longleftrightarrow\ \CM_{\rm hCS}^\CN(b,f).}
Furthermore, we then show that the open topological $B$-model describes the
(truncated) hierarchies.

%--------------------------------------------------------------------------
\subsec{Holomorphic Chern-Simons theory on the enhanced supertwistor space}

Consider a holomorphic rank $n$ vector bundle 
$\CE\to\CP^{3|\CN}_{b,f}$. Recall that on $\CP^{3|\CN}_{b,f}$ we have 
introduced the coordinates \coordsonets. Thus, the transition function
$f=\{f_{+-}\}$ of $\CE$ is annihilated by the vector fields
\eqn\somevfhcs{{\bar V}^\pm_1\ =\ {\partial\over\partial{\bar z}^2_\pm},\qquad
  {\bar V}^\pm_2\ =\ {\partial\over\partial{\bar z}^1_\pm},\qquad
  \partial_{{\bar\l}_\pm}\ =\ {\partial\over\partial{\bar\l}_\pm}  
  \qquad{\rm and}\qquad
  {\bar\partial}^i_\pm\ =\  {\partial\over\partial{\bar\eta}_i^\pm},}
which form a basis of vector fields of type $(0,1)$ on the (complexified) 
tangent bundle of the enhanced supertwistor space $\CP^{3|\CN}_{b,f}$.

Remember that the exterior derivative ${\rm d}$ on $\CP^{3|\CN}_{b,f}$ 
can always be decomposed into holomorphic and anti-holomorphic parts, i.e., 
${\rm d}=\partial+{\bar\partial}$. Using \somevfhcs, we obtain for 
the anti-holomorphic part
$${\bar\partial}\ =\ {\bar\Theta}^\a_\pm\,{\bar V}^\pm_\a+{\rm d}{\bar\l}_\pm\,
  \partial_{{\bar\l}_\pm}+{\rm d}{\bar\eta}_i^\pm\,{\bar\partial}^i_\pm,$$
where the $(0,1)$-forms are defined by 
$${\bar V}^\pm_\b\lrcorner{\bar\Theta}^\a_\pm\ =\ \d^\a_\b,\qquad
  \partial_{{\bar\l}_\pm}\lrcorner{\rm d}{\bar\l}_\pm\ =\ 1
  \qquad{\rm and}\qquad
  {\bar\partial}^i_\pm\lrcorner{\rm d}{\bar\eta}_j^\pm\ =\ \d^i_j$$

Let us now assume that the holomorphic vector bundle $\CE$ is topologically
trivial which implies that there exist some smooth matrix-valued functions 
${\hat\psi}=\{{\hat\psi}_+,{\hat\psi}_-\}$ such that the transition function 
$f_{+-}$ of $\CE$ is given by 
\eqn\transhcs{f_{+-}={\hat\psi}_+^{-1}{\hat\psi}_-.}
Therefore, we deduce from the condition ${\bar\partial}f_{+-}=0$
\eqn\linsyshcspre{({\bar\partial}+\CA^{0,1}){\hat\psi}_\pm\ =\ 0
                  \qquad\Longleftrightarrow\qquad
                  \CA^{0,1}\ =\ {\hat\psi}_\pm{\bar\partial}
                  {\hat\psi}_\pm^{-1},}
what can concisely be rewritten as
\eqn\linsyshcsprei{{\bar\partial}_\CA{\hat\psi}_\pm\ =\ 0,}
where we have defined ${\bar\partial}_\CA\equiv{\bar\partial}+\CA^{0,1}$. 
Letting\foot{In the sequel, we also assume
that $\CA^{0,1}$ does not contain anti-holomorphic fermionic components and
depends holomorphically on $\eta^\pm_i$ \refs{\WittenNN,\PopovRB,\PopovNK}.}
$$\CA^{0,1}\ =\ {\bar\Theta}^\a_\pm\CA_\a^\pm+
                {\rm d}{\bar\l}_\pm\CA_{{\bar\l}_\pm}+
                {\rm d}{\bar\eta}_i^\pm\CA^i_\pm,$$
we can rewrite the system \linsyshcsprei\ in components according to
\eqna\linsyshcs
$$\eqalignno{({\bar V}^+_\a+\CA^+_\a){\hat\psi}_\pm\ &=\ 0, & \linsyshcs a\cr
(\partial_{{\bar\l}_+}+\CA_{{\bar\l}_+}){\hat\psi}_\pm\ &=\ 0, &\linsyshcs b\cr
  ({\bar\partial}^i_++\CA^i_+){\hat\psi}_\pm\ &=\ 0. &\linsyshcs c\cr
}$$ 
Note that since the $(0,1)$-forms ${\bar\Theta}^\a_\pm$, ${\rm d}{\bar\l}_\pm$ 
and ${\rm d}{\bar\eta}_i^\pm$ are basis sections of the bundles
$\overline{\CO}(-2b_\a)$, $\overline{\CO}(-2)$ and 
$\Pi\overline{\CO}(2f_i)$, the components 
$\CA_\a^\pm$, $\CA_{{\bar\l}_\pm}$ and $\CA^i_\pm$ must be sections of 
$\overline{\CO}(-2b_\a)$, $\overline{\CO}(-2)$ and 
$\Pi\overline{\CO}(2f_i)$, respectively, such that $\CA^{0,1}$ is
$\IC$-valued.\foot{Here, we again consider generic $b_\a$s and $f_i$s.} 

Finally, the compatibility conditions of \linsyshcsprei\ are the equations of 
motion of hCS theory,  
\eqn\eomhcs{{\bar\partial}_{\CA}^2\ =\ 0\qquad\Longleftrightarrow\qquad
            {\bar\partial}\CA^{0,1}+\CA^{0,1}\wedge\CA^{0,1}\ =\ 0,}
on the enhanced supertwistor space.

%--------------------------------------------------------------------------
\subsec{Bijection between moduli spaces}

Above we have described hCS theory on the enhanced supertwistor space.
In the following, we shall explain how the moduli space of solutions to the
equations of motion of hCS theory is related to the moduli space of the
$\CN$-extended self-dual SYM hierarchy of the previous section.

Consider again the topologically trivial holomorphic vector bundle 
$\CE\to\CP^{3|\CN}_{b,f}$ of the preceding subsection, which is 
characterized by the transition function $f=\{f_{+-}\}$, and furthermore the
double fibration
$$\CP^{3|\CN}_{b,f}~~~\underleftarrow{\pi_2}~~~
  \CF^{2(b_1+b_2)+3|2(f_1+\cdots+f_\CN)+\CN}~~~\underrightarrow{\pi_1}~~~
  \IC^{2(b_1+b_2+1)|2(f_1+\cdots+f_\CN)+\CN}.
$$
Let us now pull back $\CE$ with the help of the projection $\pi_2$ to the 
correspondence space $\CF^{2(b_1+b_2)+3|2(f_1+\cdots+f_\CN)+\CN}$. By 
definition of the pull-back bundle, the pulled back transition functions
are constant along the leaves 
$$ \CF^{2(b_1+b_2)+3|2(f_1+\cdots+f_\CN)+\CN}/\CP^{3|\CN}_{b,f}\ \cong\
   \IC^{2(b_1+b_2)|2(f_1+\cdots+f_\CN)} $$
of the fibration
$\CF^{2(b_1+b_2)+3|2(f_1+\cdots+f_\CN)+\CN}\to\CP^{3|\CN}_{b,f}$. 
In other words, they are subject to the conditions
\eqn\eqonpbtrans{
    D^\pm_{\a\ad_1\cdots\ad_{2b_\a-1}}f_{+-}\ =\ 0\ =\ 
    D^i_{\pm\ad_1\cdots\ad_{2f_i-1}}f_{+-},}
where the vector fields have been defined in \defofvfsdh\ and,
as before, we use the same letter $f_{+-}$ for the pulled back transition 
function. Thus, we obtain from \linsyshcsprei\ together with \eqonpbtrans\
the following system of equations
\eqn\linsyshcssdh{(\pi_2^*{\bar\partial}_\CA){\hat\psi}_\pm\ =\ 0, \quad
  D^+_{\a\ad_1\cdots\ad_{2b_\a-1}}{\hat\psi}_\pm\ =\ 0\quad{\rm and}\quad
  D^i_{+\ad_1\cdots\ad_{2f_i-1}}{\hat\psi}_\pm\ =\ 0,}
where the pulled back transition function is split according to 
$f_{+-}={\hat\psi}_+^{-1}{\hat\psi}_-$. In addition, we also assume the 
existence 
of a gauge for the solutions to the equations of motion of the hCS theory in 
which the components $\CA_{{\bar\l}_\pm}$ vanish identically. This is clearly 
equivalent to say that the bundle $\CE\to\CP^{3|\CN}_{b,f}$ is holomorphically 
trivial on the rational curves 
$\IC P^1_{x,\eta}\hookrightarrow\CP^{3|\CN}_{b,f}$.
As the correspondence space is the direct product
$$\CF^{2(b_1+b_2)+3|2(f_1+\cdots+f_\CN)+\CN}\ =\ 
  \IC^{2(b_1+b_2+1)|2(f_1+\cdots+f_\CN)+\CN}\times\IC P^1,$$
this then implies that the pulled back bundle is holomorphically trivial 
ensuring the existence of a gauge where $\pi^*_2\CA^{0,1}=0$. Therefore, the
system \linsyshcssdh\ can be gauge transformed to
\eqna\linsyshcssdhI
$$\eqalignno{{\bar\partial}\psi_\pm\ &=\ 0, 
       &\linsyshcssdhI a\cr
  (D^+_{\a\ad_1\cdots\ad_{2b_\a-1}}+\CA^+_{\a\ad_1\cdots\ad_{2b_\a-1}})
   \psi_\pm\ &=\ 0, &\linsyshcssdhI b\cr
  (D^i_{+\ad_1\cdots\ad_{2f_i-1}}+\CA^i_{+\ad_1\cdots\ad_{2f_i-1}})
   \psi_\pm   \ &=\ 0, &\linsyshcssdhI c\cr
}$$ 
which is nothing but the auxiliary linear system for the truncated 
$\CN$-extended self-dual SYM hierarchy \linsyssdh{}. Note that in 
\linsyshcssdhI{a} ${\bar\partial}$ denotes the anti-holomorphic part of the
exterior derivative on the correspondence space.

In summary, the moduli space of solutions to 
hCS theory defined on the enhanced supertwistor space can bijectively be
mapped onto the moduli space of solutions to the $\CN$-extended 
self-dual SYM hierarchy, i.e., we have established \otohbsdhhcs.

%--------------------------------------------------------------------------
\subsec{Self-dual super Yang-Mills hierarchies: An example}

Let us now exemplify our discussion. Consider the truncated $\CN=2$ 
self-dual SYM hierarchy of type $(b_1,b_2,f_1,f_2)=({1\over2},{1\over2},1,1)$.
Its field equations are given by 
\eqn\examplehierarchyI{\eqalign{\fc_{\ad\bd}\ &=\ 0,\cr
            \epsilon^{\a\b}\epsilon^{\ad\bd}\cnab_{\a\ad}\cnab_{\b\bd}
            \pc\ \!\!^i\ &=\ 0,\cr
            \epsilon^{\ad\bd}\cnab_{\a\ad}\Wc\ \!^{[ij]}_{\bd\gd}-
            2\{\pc\ \!\!^{[i},\cnab_{\a\gd}\pc\ \!\!^{j]}\}\ &=\ 0,\cr
}}
and follow from \superfieldeqhXVII. The $\CN=2$ self-dual SYM equations, 
which are the first three equations of \fieldeqnfour\ (with $i,j=1,2$), are by 
construction a ``subset'' of \examplehierarchyI. Namely, apply 
to the last equation of \examplehierarchyI\ the operator $\cnab_{\b\dd}$ and 
contract with $\epsilon^{\a\b}$ to obtain
$$    \epsilon^{\ad\bd}\epsilon^{\a\b}\cnab_{\b\dd}
      \cnab_{\a\ad}\Wc\ \!^{[ij]}_{\bd\gd}-2\epsilon^{\a\b}
      \{\cnab_{\b\dd}\pc\ \!\!^{[i},\cnab_{\a\gd}\pc\ \!\!^{j]}\}\ =\ 0,
$$
where we have used the identity \exreci{}. Equations \prepotentialI\ together 
with \exreci{} then imply
\eqn\examplehierarchyIII{\eqalign{\fc_{\ad\bd}\ &=\ 0,\cr
      \epsilon^{\a\b}\cnab_{\a\ad}\cc\ \!\!^i_{\bd\b}\ &=\ 0,\cr
      \epsilon^{\ad\bd}\epsilon^{\a\b}\cnab_{\a\ad}\cnab_{\b\bd}
      \Wc\ \!^{[ij]}_{\gd\dd}-\epsilon^{\a\b}
      \{\cc\ \!^{[i}_{\gd\a},\cc\ \!^{j]}_{\dd\b}\}\ &=\ 0,\cr
}}
which reduce to the $\CN=2$ self-dual SYM equations when the dotted indices of
$\cc\ \!\!^i_{\ad\a}$ and $\Wc\ \!^{[ij]}_{\ad\bd}$ are chosen to be ${\dot1}$.
Note that as $i,j$ run only from $1$ to $2$, the last equation of 
\examplehierarchyI\ can be rewritten in the form
\eqn\examplehierarchyII{\eqalign{\fc_{\ad\bd}\ &=\ 0,\cr
            \epsilon^{\a\b}\epsilon^{\ad\bd}\cnab_{\a\ad}\cnab_{\b\bd}
            \pc\ \!\!^i\ &=\ 0,\cr
            \epsilon^{\ad\bd}\cnab_{\a\ad}\Gc_{\bd\gd}-
            \sfrac{3}{4}\epsilon_{ij}\{\pc\ \!\!^{i},
            \cnab_{\a\gd}\pc\ \!\!^{j}\}\ &=\ 0,\cr
}}
where we have defined the anti-self-dual two-form according to
$\Gc_{\ad\bd}\equiv{3\over8}\epsilon_{ij}\Wc\ \!^{[ij]}_{\ad\bd}$.

On the other hand, hCS theory on the enhanced supertwistor space 
$$\CP^{3|2}[1,1|2,2],$$
where we again adopted the notation from the beginning of section 4,
has already been considered in \PopovNK. There, the analysis  
led to the same equations of motion. We note that 
$\CP^{3|2}[1,1|2,2]$ is a Calabi-Yau supermanifold (see below). Therefore,
we also have a well-defined action functional for the hCS theory and hence
for the truncated self-dual SYM hierarchy, 
\eqn\anaction{S\ =\ \int{\rm d}^4x\ {\rm tr}\left\{\Gc\ \!^{\ad\bd}\fc_{\ad\bd}
              +\sfrac{3}{8}\epsilon_{ij}\pc\ \!\!^i\ \!\cnab_{\a\ad}
              \cnab\ \!^{\a\ad}\pc\ \!\!^j\right\}}
as well.

%--------------------------------------------------------------------------
\subsec{Hierarchies and open topological $B$-model}

So far, we have been studying the (truncated) hierarchies of the $\CN$-extended
self-dual SYM theory from the field theoretic point of view. We will now argue 
that the open 
topological $B$-model describes certain corners of the hierarchies.
This should be not that surprising, however, as we know since Witten's work
\WittenNN\ that the $\CN=4$ self-dual SYM theory can be described by the 
open topological $B$-model defined on the supertwistor space.
Remember that for having a well defined $B$-model, we need to 
impose the Calabi-Yau condition on the target manifold $X$, whose complex 
dimension is assumed to be $(3|\CN)$, in the sequel.  
The Calabi-Yau condition is reflected 
in the requirement of the vanishing of the first Chern number of the target 
space. Furthermore, we have seen in \refs{\WittenFB,\WittenNN} that in this 
case the (cubic open string field theory of the) open topological
$B$-model reduces to hCS theory.
The action functional for the latter theory reads as
\eqn\BmodelHCS{S\ = \ \int_Y\,\Omega\wedge{\rm tr}
               \left(\CA^{0,1}\wedge
               {\bar\partial}\CA^{0,1}+\sfrac{2}{3}\CA^{0,1}\wedge\CA^{0,1}
               \wedge\CA^{0,1}\right),}
where $\Omega$ is the holomorphic measure on $X$, $\CA^{0,1}$ 
represents the 
$(0,1|0,0)$-part of a super gauge potential $\CA$ (that is, it is assumed that
$\CA^{0,1}$ does not contain anti-holomorphic fermionic components and 
does holomorphically depend on the fermionic coordinates) 
and ``${\rm tr}$'' denotes the trace over ${\frak g}{\frak l}(n,\IC)$.
In \BmodelHCS, $Y\subset X$ is the submanifold of $X$ which is obtained by the
requirement that all of 
the complex conjugated fermionic coordinates are put to zero \WittenNN.
Let now the target space be the supertwistor space $\CP^{3|\CN}$. Clearly,
this only works for $\CN=4$ since only then the supertwistor space is a 
Calabi-Yau manifold. In this case, the holomorphic measure $\Omega$
can be taken as
\eqn\hmeasure{\Omega\big|_{\CU_\pm}\ =\ 
              \pm\,{\rm d}z^1_\pm\wedge{\rm d}z^2_\pm
              \wedge{\rm d}\l_\pm{\rm d}\eta^\pm_1{\rm d}\eta^\pm_2
              {\rm d}\eta^\pm_3{\rm d}\eta^\pm_4.}
Since the equations of motion of hCS theory defined on 
$\CP^{3|\CN}$ are gauge equivalent to the $\CN$-extended self-dual SYM 
equations on $\IR^4$, $\CN=4$ self-dual SYM theory is described by the open
topological $B$-model. In sections 3 and 4, we have introduced a generalization
of the supertwistor space $\CP^{3|\CN}$, namely the enhanced supertwistor space
$\CP^{3|\CN}_{b,f}$. The question one needs to address is to clarify when this 
space becomes a Calabi-Yau supermanifold.\foot{For the discussion of super 
Ricci-flatness of Calabi-Yau supermanifolds, etc., see
\RocekBI.} The total first Chern number is given by
$$ c_1(\CP^{3|\CN}_{b,f})\ =\ 2(b_1+b_2)-2(f_1+\cdots+f_\CN)+2. $$
We therefore obtain a whole family of Calabi-Yau spaces which is parametrized
by the discrete parameters $b_\a,f_i$ subject to
$$2(b_1+b_2)-2(f_1+\cdots+f_\CN)+2\ =\ 0.$$
Note that in this case the holomorphic measure is given by a similar looking
expression as \hmeasure. Clearly, when we choose the parameters
to be $b_\a=f_i={1\over2}$, we have $\CN=4$ supersymmetries and the hierarchies
reduce to the standard $\CN=4$ self-dual SYM equations. In general, however, we
can conclude that the open topological $B$-model defined on the manifold 
$\CP^{3|\CN}_{b,f}$ for fixed $\CN>0$, with the parameters $b_\a,f_i$ subject 
to the above condition, describes certain truncated versions of the
$\CN$-extended self-dual SYM hierarchy. 
Of course, for fixed $\CN$ there is an infinite number of 
possibilities of choosing $b_\a$ and $f_i$. Note that one may also consider 
the formal limit $b_\a,f_i\to\infty$. This limit must be taken in such a 
way, however, that the Calabi-Yau condition $c_1=0$ holds. Thus, the $B$-model 
will also describe the full $\CN$-extended self-dual SYM hierarchy.

%--------------------------------------------------------------------------
%--------------------------------------------------------------------------
\newsec{Conclusions and outlook}

In this paper, we described the construction of hidden symmetry algebras of the
$\CN$-extended self-dual SYM equations on $\IR^4$ by means of the supertwistor 
correspondence. In particular, we exemplified our discussion by focusing
on super Kac-Moody and Kac-Moody-Virasoro type symmetries. However, the 
presented algorithm does not only apply for those algebras but for any kind of 
algebra one may define on the supertwistor side, i.e., given some infinitesimal
transformations of the transition function $f_{+-}$ of some holomorphic vector
bundle $\CE\to\CP^{3|\CN}$ generated by some (infinite-dimensional) algebra, 
one can map those via the linearized Penrose-Ward transform to a corresponding 
set of symmetries on the gauge theory side. The only thing one should require 
for these transformations is that they preserve the complex structure on 
$\CP^{3|\CN}$.

Furthermore, the affine extension of the superconformal algebra led us to a new
family of supertwistor spaces, which we denoted by $\CP^{3|\CN}_{b,f}$ and
called enhanced supertwistor spaces. The supertwistor correspondence for those 
spaces eventually gave us the $\CN$-extended self-dual SYM hierarchies, which
in turn describe graded Abelian symmetries of the $\CN$-extended self-dual SYM
equations. We have shown that the moduli spaces of solutions to those 
hierarchies can bijectively be mapped to the moduli spaces of solutions to hCS 
theory on $\CP^{3|\CN}_{b,f}$. As $\CP^{3|\CN}_{b,f}$ turned out to be a 
Calabi-Yau supermanifold for certain values of $b_\a$ and $f_i$, we have also 
seen that the open topological $B$-model on $\CP^{3|\CN}_{b,f}$ describes those
(truncated) hierarchies.

\bigskip\noindent
There are a lot of open issues which certainly deserve further investigations:

\noindent
$\bullet$ The first task is to generalize the above discussion to the full 
$\CN=4$ SYM theory. In restricting to the self-dual truncation, we have just 
given the first step. It is well known that $\CN=3\ (4)$ SYM theory in four
dimensions
is related via the supertwistor correspondence to a quadric of dimension 
$(5|6)$ which lives in  
$\CP^{3|3}\times\CP^{3|3}$ (see \refs{\WittenXX,\WittenNN,\PopovRB} 
and references therein). The latter space and hence the quadric need to be 
covered 
with at least four coordinate patches. Hence, due to this four-set (open) 
covering certain technicalities in constructing the symmetry algebras 
will appear, but nevertheless there will be no principal problems. One simply
needs to work out the details.

\noindent
$\bullet$ Having given the supertwistor construction of hidden symmetries of
the full $\CN=3\ (4)$ SYM theory, one needs to investigate -- using, e.g.,
twistorial methods -- the quantum corrections of the obtained symmetry 
algebras. For instance, by 
passing to the quantum regime, one will not obtain centerless Kac-Moody 
algebras such as \SConAVIII\  and \affineEVII, but rather
$$[\d^m_a,\d^n_b\}\ =\ f_{ab}^{\ \ c}\d^{m+n}_c+k\d_{ab}\d_{m,-n}, $$
where $k$ is the level of the Kac-Moody algebra. Such algebras can in turn 
be used to define representations of so-called quantum (super) Yangians (see, 
e.g., \SchoutensAU\ for the case of two-dimensional CFTs and 
\refs{\ChauRP,\ChauMP} for the case of quantum SDYM theory).\foot{For the 
definition of (super) Yangians see, e.g., references 
\refs{\DrinfeldRX-\BernardYA,\DolanUH,\Nazarov}.} Such quantum Yangians play an
increasing and important role in the investigation of quantum integrability of 
the superconformal gauge theory \DolanUH.

\noindent
$\bullet$ Another issue also worthwhile to explore is the construction
of hidden symmetry algebras (and hierarchies) of gravity theories, in 
particular of conformal supergravity (see, e.g., \FradkinAM\ for a review). The
description of conformal supergravity in terms of the supertwistor 
correspondence has been discussed in \refs{\WittenNN,\BerkovitsJJ,\AhnYU}. 
By applying similar techniques as those presented in this paper, one will 
eventually obtain infinite-dimensional symmetry algebras
(and correspondingly an infinite number of conserved nonlocal charges) of
the conformal supergravity equations, generalizing the 
results known for self-dual gravity (see, e.g., references \BoyerAJ). 

\noindent
$\bullet$ Finally, we address another point for further investigations.
It has been conjectured by Ward \WardGZ\ that all integrable models in $D<4$
dimensions can be obtained from the self-dual YM equations in four dimensions.
Typical examples of such systems are the nonlinear Schr\"odinger equation, the
Korteweg-de Vries equation, the sine-Gordon model, etc. All of them follow 
from the self-dual Yang-Mills equations by incorporating suitable algebraic 
ans\"atze for the self-dual gauge potential followed by a dimensional 
reduction. In a similar spirit, the Ward conjecture can be supersymmetrically 
extended in order to derive the supersymmetric versions of the above-mentioned 
models. Therefore, it would be of interest to take the $\CN$-extended 
self-dual SYM hierarchy presented in this paper and to derive the 
corresponding super hierarchies of these integrable systems in $D<4$ 
dimensions and to compare them with results known in the literature.

%--------------------------------------------------------------------------
\bigbreak\bigskip\bigskip\centerline{{\bf Acknowledgments}}\nobreak

I am thankful to Alexander Kling, Olaf Lechtenfeld, Stefan Petersen, 
Christian S\"amann, Sebastian Uhlmann and Robert Wimmer for discussions and
for commenting on the manuscript. I would also like 
to thank Niall J. MacKay for email correspondence. In particular, I
am deeply grateful to Alexander Popov for numerous discussions 
and valuable comments. This work was done within the
framework of the DFG priority program (SPP 1096) in string theory.

%--------------------------------------------------------------------------
%--------------------------------------------------------------------------
\appendix{A}{Hidden symmetries and sheaf cohomology}

In this appendix, we shall present a more formal approach to the symmetry 
algebras by using the sheaf cohomology machinery. In particular, we talk about 
super Kac-Moody and super Virasoro symmetries.

%--------------------------------------------------------------------------
\subsec{\v Cech description of holomorphic vector bundles}

First, let us recall some basic definitions.  Consider a complex 
(super)manifold $M$ with the (open) covering ${\frak U}=\{\CU_m\}$. 
Furthermore, we are interested in smooth maps from open subsets of 
$M$ into the non-Abelian group $GL(n,\IC)$ 
as well as in a sheaf ${\frak G}$ of such $GL(n,\IC)$-valued functions. 
A $q$-cochain of the covering ${\frak U}$ with values in ${\frak G}$ is a 
collection $\psi=\{\psi_{m_0\cdots m_q}\}$ of sections of the sheaf ${\frak G}$
over nonempty intersections $\CU_{m_0}\cap\cdots\cap\CU_{m_q}$. We will denote
the set of such $q$-cochains by $C^q(\fU,\fG)$. We stress that it has a
group structure, where the multiplication is just pointwise multiplication.

We may define the subsets of cocylces $Z^q(\fU,\fG)\subset C^q(\fU,\fG)$. For
example, for $q=0,1$ these are given by
$$\eqalign{Z^0(\fU,\fG)\ &\equiv\ \{\psi\in C^0(\fU,\fG)\ |\ \psi_m=\psi_n
           ~~{\rm on}~~\CU_m\cap\CU_n\neq\emptyset\},\cr
           Z^1(\fU,\fG)\ &\equiv\ \{\psi\in C^1(\fU,\fG)\ |\ \psi_{nm}=
           \psi_{mn}^{-1}~~{\rm on}~~\CU_m\cap\CU_n\neq\emptyset\cr
           &\kern2.5cm{\rm and}~~\psi_{mn}\psi_{np}\psi_{pm}=1~~
           {\rm on}~~\CU_m\cap\CU_n\cap\CU_p\neq\emptyset\}.\cr
}$$
These sets will be of particular interest later on. We remark that from the 
first of these two definitions it follows that $Z^0(\fU,\fG)$ coincides with 
the group 
$$H^0(M,\fG)\ \equiv\ \fG(M)\ =\ \Gamma(M,\fG),$$
which is the group of global sections of the sheaf $\fG$. Note that in general
the subset $Z^1(\fU,\fG)\subset C^1(\fU,\fG)$ is not a subgroup of the group 
$C^1(\fU,\fG)$.

We say that two cocycles $f,f'\in Z^1(\fU,\fG)$ are equivalent if 
$f'_{mn}=\psi_m^{-1}f_{mn}\psi_n$ for some $\psi\in C^0(\fU,\fG)$. The set
of equivalence classes induced by this equivalence relation is the first
cohomology set and denoted by $H^1(\fU,\fG)$. If the $\CU_m$ are all Stein
manifolds we have the bijection
$$H^1(\fU,\fG)\ \cong\ H^1(M,\fG).$$
Furthermore, we shall also need the sheaf of holomorphic sections of the 
trivial bundle $M\times GL(n,\IC)$, which we denote by $\CO_{GL}$, in the 
sequel.

Let us now stick to the supertwistor space $\CP^{3|\CN}$ and the correspondence
space $\CF^{5|2\CN}$ with their two-set open coverings $\fU=\{\CU_+,\CU_-\}$ 
and ${\tilde\fU}=\{{\tilde\CU}_+,{\tilde\CU}_-\}$, respectively. We note that 
$\CU_\pm$ and  ${\tilde\CU}_\pm$ are indeed Stein manifolds. Then any cocyle 
$f=\{f_{+-}\}\in Z^1(\fU,\fG)$ defines uniquely a complex rank $n$ vector 
bundle $\CE$ over $\CP^{3|\CN}$ and via $\pi_2$ over $\CF^{5|2\CN}$. Equivalent
cocyles define isomorphic complex vector bundles and hence, since $\CU_\pm$ are
Stein manifolds, the isomorphism class of complex rank $n$ vector bundles is 
parametrized by $H^1(\CP^{3|\CN},\fG)$. Choosing $\fG$ to be $\CO_{GL}$, we see
that holomorphic vector bundles over the supertwistor space are parametrized by
$H^1(\CP^{3|\CN},\CO_{GL})$, i.e.,
$\CM_{\rm hol}(\CP^{3|\CN})\subset H^1(\CP^{3|\CN},\CO_{GL}).$

%--------------------------------------------------------------------------
\subsec{Super Kac-Moody symmetries}

Let us study small perturbations of the transition function $f_{+-}$ of a 
holomorphic vector bundle $\CE\to\CP^{3|\CN}$ and its pull-back bundle 
$\pi_2^*\CE\to\CF^{5|2\CN}$. Consider $\CF^{5|2\CN}$ with the covering 
${\tilde{\frak U}}=\{{\tilde\CU}_+,{\tilde\CU}_-\}$. Then we define an action 
of $C^1({\tilde{\frak U}},\CO_{GL})$ on $Z^1({\tilde{\frak U}},\CO_{GL})$ by
\eqn\defcz{h\,:\, f_{+-}\ \mapsto\ h_{+-}f_{+-}h_{-+}^{-1},}
where $h=\{h_{+-},h_{-+}\}\in C^1({\tilde{\frak U}},\CO_{GL})$ and 
$f=\{f_{+-}\}\in Z^1({\tilde{\frak U}},\CO_{GL})$. Obviously, the group 
$C^1({\tilde{\frak U}},\CO_{GL})$ acts transitively on
$Z^1({\tilde{\frak U}},\CO_{GL})$, since for an arbitrary \v Cech $1$-cocycle
$f\in Z^1({\tilde{\frak U}},\CO_{GL})$ one can find an 
$h\in C^1({\tilde{\frak U}},\CO_{GL})$ such that $f_{+-}=h_{+-}h_{-+}^{-1}$ and
$f_{-+}=h_{-+}h_{+-}^{-1}$. The stabilizer of the trivial \v Cech $1$-cocycle, 
given by $f=1$, is simply the subgroup 
$$ C_\Delta^1({\tilde{\frak U}},\CO_{GL})\ \equiv\ 
\{h\in C^1({\tilde{\frak U}},\CO_{GL})
   \,|\,h_{+-}=h_{-+}\}$$
implying that $Z^1({\tilde{\frak U}},\CO_{GL})$ can be identified with the 
coset
$$ Z^1({\tilde{\frak U}},\CO_{GL})\ \cong\ C^1({\tilde{\frak U}},
   \CO_{GL})/C_\Delta^1({\tilde{\frak U}},\CO_{GL}). $$

Now we are interested in infinitesimal deformations of the transition function
$f_{+-}$. Consider the sheaf $\CO_{{\frak g}{\frak l}}$ of holomorphic sections
 of the bundle $\CF^{5|2\CN}\times{\frak g}{\frak l}(n,\IC)$ and the sheaves
$\CS_{GL}$ and $\CS_{{\frak g}{\frak l}}$ consisting of those 
$GL(n,\IC)$-valued and ${\frak g}{\frak l}(n,\IC)$-valued smooth functions, 
respectively, that are annihilated by $\partial_{{\bar\l}_\pm}$. We have a
natural infinitesimal action of the group $C^1({\tilde{\frak U}},\CO_{GL})$
on the space $Z^1({\tilde{\frak U}},\CO_{GL})$ 
which is induced by the linearization of \defcz. Namely, we get
\eqn\defczi{\d f_{+-}\ =\ \d h_{+-}f_{+-}-f_{+-}\d h_{-+},}
where $\d h=\{\d h_{+-},\d h_{-+}\}\in C^1({\tilde{\frak U}},
\CO_{{\frak g}{\frak l}})$ and $f=\{f_{+-}\}\in 
Z^1({\tilde{\frak U}},\CO_{GL})$. As in subsection 3.1, we introduce the
${\frak g}{\frak l}(n,\IC)$-valued function
\eqn\defofphi{\varphi_{+-}\ \equiv\ \psi_+(\d f_{+-})\psi_-^{-1},}
where $\psi_\pm\in C^0({\tilde{\frak U}},\CS_{GL})$ and $f_{+-}=\psi_+^{-1}
\psi_-$, as before. Here, $C^0({\tilde{\frak U}},\CS_{GL})$ is the group of
$0$-cochains. From equation \defczi\ it is then immediate that
$$\varphi_{+-}\ =\ -\varphi_{-+}.$$
Moreover, $\varphi_{+-}$ is annihilated by $\partial_{{\bar\l}_\pm}$. Thus, it 
defines an element $\varphi\in Z^1({\tilde{\frak U}},\CS_{{\frak g}{\frak l}})$
with $\varphi=\{\varphi_{+-}\}$. Any $1$-cocycle with values in 
$\CS_{{\frak g}{\frak l}}$ is a $1$-coboundary since 
$H^1(\CF^{5|2\CN},\CS_{{\frak g}{\frak l}})=0$. Therefore, we have
\eqn\eqphii{\varphi_{+-}\ =\ \phi_+-\phi_-,}
where 
$\phi=\{\phi_+,\phi_-\}\in C^0({\tilde{\frak U}},\CS_{{\frak g}{\frak l}})$.

Linearizing $f_{+-}=\psi_+^{-1}\psi_-$, we obtain
$$\d f_{+-}\ =\ f_{+-}\psi_-^{-1}\d\psi_-
           -\psi_+^{-1}\d\psi_+f_{+-}.$$
and hence, by virtue of the equations \defofphi\ and \eqphii\ we obtain
\eqn\eqforhpmi{\d\psi_\pm\ =\ -\phi_\pm\psi_\pm.}
In summary, given some $\d h=\{\d h_{+-},\d h_{-+}\}$ one derives via
\defczi-\eqphii\ the perturbations $\d\psi_\pm$.\foot{In our example of
subsection 3.2, $\d h$ was given by $\d h_{+-}=\d h_{-+} =\l_+^m X_a$.}
Now one may proceed as in subsection 3.1 to arrive at the formulas \infpwtrans\
for the nontrivial symmetries. Symmetries obtained this way are called super 
Kac-Moody symmetries.

%--------------------------------------------------------------------------
\subsec{Super Virasoro symmetries}

Above we have introduced Kac-Moody symmetries of the $\CN$-extended super
SDYM equations which were generated by the algebra
$ C^1({\tilde{\frak U}},\CO_{{\frak g}{\frak l}})$. In this subsection
we focus on symmetries, which are related with the group of local
biholomorphisms of the supertwistor space $\CP^{3|\CN}$.

Consider the supertwistor space $\CP^{3|\CN}$. Remember that it comes 
with the local coordinates $(Z^a_\pm)=(z_\pm^\a,\l^\pm_\ad,\eta_i^\pm)$. 
On the intersection $\CU_+\cap\CU_-$, they are related by the transition 
function $t_{+-}^a$, i.e.,
$$Z_+^a\ =\ t^a_{+-}(Z_-^b).$$
Let us denote the group of local biholomorphisms of the supertwistor space 
$\CP^{3|\CN}$ by $\fH_\CP$, which is the subgroup of the diffeomorphism
group of $\CP^{3|\CN}$ which consists of those maps from
$\CP^{3|\CN}\to\CP^{3|\CN}$ which preserve the complex structure of 
$\CP^{3|\CN}$. To $\fH_\CP$ corresponds the algebra
$$C^0(\fU,\CV_\CP)$$
of $0$-cochains of $\CP^{3|\CN}$ with values in the sheaf $\CV_\CP$ of
holomorphic vector fields on the supertwistor space. Let us now consider the
algebra $C^1(\fU,\CV_\CP)$ whose elements are collections of vector fields 
\eqn\elementsofc{\chi\ =\ \{\chi_{+-},\chi_{-+}\}\ =\ 
               \left\{\chi^a_{+-}\partial_a^+,\chi^a_{-+}\partial_a^-
                 \right\},}
where we have abbreviated $\partial^\pm_a\equiv\partial/\partial Z^a_\pm$.
In particular, the $\chi_{+-}$ and $\chi_{-+}$ are elements of the algebra
$\CV_\CP(\CU_+\cap\CU_-)$ of holomorphic vector fields on the intersection
$\CU_+\cap\CU_-$. Thus, $C^1(\fU,\CV_\CP)$
can be decomposed according to
$$C^1(\fU,\CV_\CP)\ \cong\ \CV_\CP(\CU_+\cap\CU_-)\oplus
    \CV_\CP(\CU_+\cap\CU_-).$$

Kodaira-Spencer deformation theory then tells us that the algebra 
$C^1(\fU,\CV_\CP)$ acts on the transition functions $t_{+-}^a$ according to
\eqn\kodspe{\d t^a_{+-}\ =\ \chi_{+-}^a-\chi^b_{-+}\partial_b^-t^a_{+-}}
which can equivalently be rewritten as
\eqn\kodspetwo{\d t_{+-}\ \equiv\ \d t^a_{+-}\partial^+_a\ =\ 
               \chi_{+-}-\chi_{-+}.}

Let us consider a subalgebra
$$ C^1_\Delta(\fU,\CV_\CP)\ \equiv\ \{\chi\in C^1(\fU,\CV_\CP)\,|\,
   \chi_{+-}=\chi_{-+}\} $$
of the algebra $C^1(\fU,\CV_\CP)$. Then the space 
$$ Z^1(\fU,\CV_\CP)\ =\ \{\chi\in C^1(\fU,\CV_\CP)\,|\, \chi_{+-}=-\chi_{-+}\}
$$
is given by the quotient
$$ Z^1(\fU,\CV_\CP)\ \equiv\ C^1(\fU,\CV_\CP)/C^1_\Delta(\fU,\CV_\CP).$$
We stress that the
transformations \kodspe\ change the complex structure of $\CP^{3|\CN}$
if $\chi_{+-}\neq\chi_+-\chi_-$, where 
$\{\chi_+,\chi_-\}\in C^0(\fU,\CV_\CP)$. Therefore,
$$H^1(\fU,\CV_\CP)\ =\ Z^1(\fU,\CV_\CP)/C^0(\fU,\CV_\CP)$$
is the tangent space (at a chosen complex structure)
of the moduli space of deformations of the complex structure on $\CP^{3|\CN}$.

Consider now a holomorphic vector bundle $\CE$ over the supertwistor space.
We may define the following holomorphic action (thus preserving the complex 
structure on $\CP^{3|\CN}$) of the
algebra $C^0(\fU,\CV_\CP)$ on the transition functions $f_{+-}$ of the
holomorphic vector bundle $\CE$: 
\eqn\defofactonfks{\d f_{+-}\ =\ \chi_+(f_{+-})-\chi_-(f_{+-}),}
with $\chi_+$ and $\chi_-$ restricted to $\CU_+\cap\CU_-$. 
In a similar manner, 
$C^0(\fU,\CV_\CF)$ acts on the pulled-back transition functions of
the pull-back bundle $\pi_2^*\CE\to\CF^{5|3\CN}$. Here, $\CV_\CF$ is the
sheaf of holomorphic vector fields on the correspondence space $\CF^{5|3\CN}$.

After this little digression, we may now follow the lines presented 
in subsection 3.1 to arrive at the formulas \infpwtrans\ for the nontrivial 
symmetries. We refer to symmetries obtained in this way as super Virasoro 
symmetries since they are related with the group of local biholomorphisms.

%--------------------------------------------------------------------------
%--------------------------------------------------------------------------
\appendix{B}{Remarks on almost complex structures}

Consider the superspace $(\IR^{2m|2n},I)$, where $I$ is the canonical metric
on $\IR^{2m|2n}$,
\eqn\ospmnI{I\ \equiv\ \left(\matrix{{\bf 1}_{2m} & 0\cr 0&\o_{2n}\cr}\right)
            \qquad{\rm and}\qquad
         \o_{2n}\ \equiv\ \left(\matrix{0&-{\bf1}_n\cr{\bf 1}_n&0\cr}\right).}
 The group of isometries of $\IR^{2m|2n}$
is the matrix supergroup
$$OSp(2m|2n)\ \subset\ GL(2m|2n),$$ 
where $GL(2m|2n)$ is the group of all invertible matrices of dimension 
$(4m^2+4n^2|8mn)$. In particular,
\eqn\ospmn{OSp(2m|2n)\ \equiv\ \{g\in GL(2m|2n)\ |\  
             ^{\rm st}\!g\,I\,g\,=\,I\}.}
The corresponding Lie superalgebra ${\frak o}{\frak s}{\frak p}(2m|2n)$ is 
given by
\eqn\ospmnLie{{\frak o}{\frak s}{\frak p}(2m|2n)\ =\ \{X\in 
           {\frak g}{\frak l}(2m|2n)\ |\  ^{\rm st}\!X\,I+X\,I\,=\,0\}.}
In \ospmn\ and \ospmnLie, the superscript ``st'' denotes the supertranspose 
which is defined as
\eqn\ospmnII{X\ =\ \left(\matrix{A&B\cr C&D}\right)
             \qquad\Longrightarrow\qquad
             ^{\rm st}\!X\ \equiv\ \ 
             \left(\matrix{ ^{\rm t}\!A&-\ \!^{\rm t}C\cr 
                           \!^{\rm t}\!B& ^{\rm t}\!D}\right).}
Here, $A$ and $D$ are even $2m\times 2m$ and $2n\times 2n$ matrices, while 
$B$ and $C$ are odd $2m\times 2n$ and $2n\times 2m$ matrices, respectively.
Explicitly, \ospmnLie\ reads as
\eqn\ospmnLieI{^{\rm t}\!A+A\ =\ 0,\qquad B-\ \!\!^{\rm t}\!C\o_{2m}\ =\ 0
               \qquad{\rm and}\qquad ^{\rm t}\!D\o_{2m}+\o_{2m}D\ =\ 0,}
where $X$ is given by \ospmnII.

An almost complex structure on the space $\IR^{2m|2n}$ is an endomorphism 
$\CJ$ from $\IR^{2m|2n}\to \IR^{2m|2n}$ with $\CJ^2=-{\bf 1}$. In 
matrix representation we have
\eqn\ospmnIII{\CJ\ =\ \left(\matrix{J_{2m}&0\cr 0&J_{2n}}\right), 
             \qquad{\rm with}\qquad
         J_{2k}\ \equiv\ \left(\matrix{0&-{\bf1}_k\cr{\bf 1}_k&0\cr}\right).}
As we are interested in the moduli space of all (constant) almost complex
structures, we need the subgroup $H\subset OSp_{\rm c}(2m|2n)$ (the subscript
``c'' stands for the connected component of $OSp(2m|2n)$, i.e., 
${\rm sdet}(g)=+1$ for $g\in OSp(2m|2n)$) such that
\eqn\ospmnIV{H\ \equiv\ \{g\in OSp_{\rm c}(2m|2n)\ |\ 
             g^{-1}\,\CJ\,g\,=\,\CJ\}.}
The corresponding algebra is
\eqn\ospmnV{{\frak h}\ =\ \{X\in {\frak o}{\frak s}{\frak p}(2m|2n)
             \ |\  [X,\CJ]\,=\,0\}.}
Writing out explicitly the condition $[X,\CJ]=0$
together with the explicit form of $X$ given in \ospmnII, one realizes
that $A\in {\frak s}{\frak o}(2m)\cap {\frak s}{\frak p}(2m,\IR)$, that is, 
$A\in {\frak u}(m)$. Similarly, both $^{\rm t}\!D\o_{2n}+\o_{2n}D=0$ and 
$[X,\CJ]=0$ show that 
$D\in {\frak s}{\frak o}(2n)\cap {\frak s}{\frak p}(2n,\IR)
\cong{\frak u}(n)$. The remaining condition given in \ospmnLieI\ together with
$[X,\CJ]=0$ yield that $B$ and $C$ are of the form
\eqn\ospmnVi{B\ =\ \left(\matrix{B_1&B_2\cr -B_2&B_1}\right)\qquad
           {\rm and}\qquad C\ =\ -\left(\matrix{^{\rm t}\!B_2&^{\rm t}\!B_1\cr 
                  -\ \!^{\rm t}\!B_1&^{\rm t}\!B_2}\right),}
where the $B_{1,2}$ are real $m\times n$-matrices. Then, we may identify $B$ 
and $C$ according to
\eqn\ospmnVii{ B\ \mapsto\ \i(B_1+\i B_2)\qquad{\rm and}\qquad
               C\ \mapsto\  ^{\rm t}\!B_2+\i\ \!\!^{\rm t}\!B_1,}
i.e., $C$ is the negative Hermitian adjoint of $B$. In summary, the algebra
${\frak h}$ can therefore be identified with ${\frak u}(m|n)$.\foot{Recall,
the matrices $X\in{\frak u}(m|n)$ are defined by the condition that $A$ and
$D$ are skew-Hermitian while $C$ is the negative Hermitian adjoint of $B$
(cf., e.g., \DolanUH).} Hence, the supercoset
\eqn\ospmnVII{OSp_{\rm c}(2m|2n)/U(m|n)}
parametrizes all almost complex structures on $\IR^{2m|2n}$. Note that as
the (real) dimension of $OSp_{\rm c}(2m|2n)$ is $(m(2m-1)+n(2n+1)|4mn)$
and of $U(m|n)$ is $(m^2+n^2|2mn)$, respectively, the dimension of the 
supercoset \ospmnVII\ is $(m(m-1)+n(n+1)|2mn)$. 

Let us now stick to the case when $m=2$ and $n=\CN$. As in the purely bosonic
case, we may introduce 
\eqn\ospmnIX{\CP_\CN\ \equiv\ P(\IR^{4|2\CN},OSp_{\rm c}(4|2\CN))
              \times_{OSp_{\rm c}(4|2\CN)} (OSp_{\rm c}(4|2\CN)/U(2|\CN)),}
where $P(\IR^{4|2\CN},OSp_{\rm c}(4|2\CN))$ is the principal 
$OSp_{\rm c}(4|2\CN)$-frame bundle on $\IR^{4|2\CN}$. Since the supermanifold
$\IR^{4|2\CN}$ is trivial, \ospmnIX\ becomes
\eqn\ospmnX{\CP_\CN\ =\ 
              \IR^{4|2\CN}\times(OSp_{\rm c}(4|2\CN)/U(2|\CN)).}
Clearly, when $\CN=0$ \ospmnX\  reduces to 
$\IR^4\times\IC P^1$, which is diffeomorphic to the bosonic twistor space
$\CP^3=\IC P^3\setminus\IC P^1$. Therefore, one may regard \ospmnX\ as another
super extension of the twistor space.

%--------------------------------------------------------------------------
%--------------------------------------------------------------------------
\appendix{C}{Superconformal algebra}

In this appendix, we shall give the (anti)commutation relations of the 
generators \scgen\ of the superconformal algebra. They are
\eqn\superconformalalgebraI{\eqalign{
    [P_{\a\ad},K_{\b\bd}]\ &=\ 2(\epsilon_{\a\b}J_{\ad\bd}+\epsilon_{\ad\bd}
                               J_{\a\b})-\epsilon_{\a\b}\epsilon_{\ad\bd}D,\cr
    \{Q_{i\a},Q^j_\ad\}\ &=\ \d^j_iP_{\a\ad},\qquad
    \{K^{i\a},K^\ad_j\}\ =\ -\d_j^iK^{\a\ad},\cr
    \{Q_{i\a},K^{j\b}\}\ &=\ 2\d^j_i({J_\a}^\b+\sfrac{1}{4}\d_\a^\b D)
                      +\sfrac{1}{2}\d_\a^\b\d_i^j(1-\sfrac{4}{\CN})A-\d_\a^\b
                             T^j_i,\cr                               
   [T^i_j,K^{k\a}]\ &=\ -(\d^k_j K^{i\a}-\sfrac{1}{\CN}\d_j^iK^{k\a}),\cr
   [A,K^{i\a}]\ &=\ -\sfrac{1}{2}K^{i\a},\qquad
   [D,K^{i\a}]\ =\ -\sfrac{1}{2}K^{i\a},\cr
   [J_{\a\b},K^{i\g}]\ &=\ \sfrac{1}{2}\epsilon_{\d(\a}\d^\g_{\b)}K^{i\d},
   \qquad[P_{\a\ad},K^{i\b}]\ =\ -\d^\b_\a Q^i_\ad\cr
   [T^j_i,Q_{k\a}]\ &=\ \d^j_k Q_{i\a}-\sfrac{1}{\CN}\d^j_iQ_{k\a},\cr
   [A,Q_{i\a}]\ &=\ \sfrac{1}{2}Q_{i\a},\qquad
   [D,Q_{i\a}]\ =\ \sfrac{1}{2}Q_{i\a},\cr
   [J_{\a\b},Q_{i\g}]\ &=\ -\sfrac{1}{2}\epsilon_{\g(\a}Q_{i\b)},\qquad
   [Q_{i\b},K^{\a\ad}]\ =\ \d^\b_\a K^\ad_i,\cr
   [T^j_i,T^l_k]\ &=\ \d^j_k T^l_i-\d^l_i T^j_k,\cr
   [D,P_{\a\ad}]\ &=\ P_{\a\ad},\qquad[D,K^{\a\ad}]=-K^{\a\ad},\cr
   [J_{\a\b},K^{\g\gd}]\ &=\ \sfrac{1}{2}\epsilon_{\d(\a}\d^\g_{\b)}K^{\d\gd},
   \qquad
   [J_{\a\b},P_{\g\gd}]\ =\ -\sfrac{1}{2}\epsilon_{\g(\a}P_{\b)\gd},\cr
   [J_{\a\b},J^{\g\d}]\ &=\ -\d^{(\g}_{(\a}{J_{\b)}}^{\d)}.\cr
}}

 \listrefs

\end